\pdfoutput=1

\documentclass{aa}

\usepackage{graphicx}
\usepackage[varg]{txfonts}
\usepackage{amsmath}
\usepackage{verbatim}
\usepackage{latexsym}
\usepackage{multirow}
\usepackage{fixltx2e}
\usepackage{color}

\usepackage{caption}

\newcommand{\ind}[1]{_{\mathrm{#1}}}

\newcommand{\espadon}{ESPaDonS\xspace}
\newcommand{\espadons}{ESPaDonS\xspace}

\begin{document}


\title{Atmospheric circulation of Venus measured with visible imaging spectroscopy at the THEMIS observatory}


\author{Patrick Gaulme\inst{\ref{inst1},\ref{inst2},\ref{inst3}} 
\and Fran\c cois-Xavier Schmider\inst{\ref{inst4}} 
\and Thomas Widemann\inst{\ref{inst5},\ref{inst6}} 
\and Ivan Gon\c{c}alves\inst{\ref{inst4}}
\and Arturo L\'opez Ariste\inst{\ref{inst7}}
\and Bernard Gelly\inst{\ref{inst8}}
}

\institute{Max-Planck-Institut f\"{u}r Sonnensystemforschung, Justus-von-Liebig-Weg 3, 37077, G\"{o}ttingen, Germany \email{gaulme@mps.mpg.de}\label{inst1}
\and 
Department of Astronomy, New Mexico State University, P.O. Box 30001, MSC 4500, Las Cruces, NM 88003-8001, USA \label{inst2}
\and
Physics Department, New Mexico Institute of Mining and Technology, 801 Leroy Place, Socorro, NM 87801, USA \label{inst3}
\and
Laboratoire Lagrange, Universit\'e de Nice Sophia Antipolis, UMR 7293, Observatoire de la C\^ote d'Azur (OCA), Nice, France \label{inst4}
\and 
LESIA, Observatoire de Paris, PSL, CNRS, Sorbonne Universit\'e, U. Paris-Diderot, France \label{inst5}
\and
DYPAC, Universit\'e Versailles-Saint-Quentin, Universit\'e Paris Saclay, France \label{inst6}
\and
IRAP, Universit\'e de Toulouse, CNRS UMR 5277, CNES, UPS.  14, Av. E. Belin. 31400 Toulouse, France \label{inst7}
\and
THEMIS Observatory, La Laguna, Tenerife, Spain \label{inst8}
}

\titlerunning{Venus winds with visible imaging spectroscopy}
\authorrunning {Gaulme et al.}
\abstract{
Measuring the atmospheric circulation of Venus at different altitudes is important for understanding its complex dynamics, in particular the mechanisms driving super-rotation. Observationally, Doppler imaging spectroscopy is in principle the most reliable way to measure wind speeds of planetary atmospheres because it directly provides the projected speed of atmospheric particles. However, high-resolution imaging spectroscopy is challenging, especially in the visible domain, and most knowledge about atmospheric dynamics has been obtained with the cloud tracking technique. The objective of the present work is to measure the global properties of the atmospheric dynamics of Venus at the altitude of the uppermost clouds, which is probed by reflected solar lines in the visible domain. Our results are based on high-resolution spectroscopic observations with the long-slit spectrometer of the solar telescope THEMIS. We present the first instantaneous ``radial-velocity snapshot'' of any planet of the solar system in the visible domain, i.e., a complete radial-velocity map of the planet obtained by stacking data on less than 10\,\% of its rotation period. From this, we measured the properties of the zonal and meridional winds, which we unambiguously detect. We identify a wind circulation pattern that significantly differs from previous knowledge about Venus. The zonal wind reveals a ``hot spot'' structure, featuring about 200 m s$^{-1}$ at sunrise and 70 m s$^{-1}$ at noon in the equatorial region. Regarding meridional winds, we detect an equator-to-pole meridional flow peaking at 45 m s$^{-1}$ at mid-latitudes, i.e., about twice as large as what has been reported so far.}

\maketitle


\section{Introduction}
\label{sect_intro}
The atmosphere of Venus is well known for its super-rotation in a retrograde direction. The atmospheric zonal rotation period strongly varies with altitude, from a corotation with the surface of the planet of 243.02 Earth days at ground level, down to 4.4 days at cloud tops, where it peaks at about 100 m s$^{-1}$ at equator. First evidenced from the ground \citep{Boyer_Guerin_1969}, the atmospheric super-rotation has been extensively studied both from space and ground-based telescopes \citep{Gierasch_1997, Limaye_1988, Rossow_1990}. The cloud top region is important as it constrains the global mesospheric circulation in which zonal winds generally decrease with height while thermospheric subsolar-to-antisolar (SSAS)\ winds increase \citep{Lellouch_1997, Widemann_2007, Widemann_2008}. It also shows important variability at various spatial and temporal scales \citep{Sanchez_Lavega_2008,Hueso_2012, Hueso_2015, Patsaeva_2015, Khatuntsev_2013,Machado_2012, Machado_2014}. 

Characterizing the meridional circulation is also important for understanding the maintenance of the super-rotation, by determining the global mean and eddy circulations and the associated meridional transport of angular momentum and energy \citep[e.g.,][]{Limaye_Rengel_2013}. The role of thermal tides to transport angular momentum vertically in low latitudes has been confirmed \citep{Lebonnois_2010, Takagi_Matsuda_2007}. It has also been noticed that the latitudinal distribution of zonal wind at cloud tops may result from an equilibrium between the impact of thermal tides and the angular momentum transport by the meridional circulation \citep{Lebonnois_2010}, providing grounds for systematic and simultaneous observations of both zonal and meridional regimes. 

For Venus, as for dense atmospheres in the solar system, most atmospheric dynamics measurements come from the cloud tracking technique. 
The method consists of following cloud features at specific wavelengths taken on image pairs obtained at various times. Although the clouds are almost featureless in visible light, there are prominent features in UV and infrared wavelengths \citep{Titov_2008, Titov_2012}. Cloud motions are considered to be a good proxy for true atmospheric motions and are capable of providing a systematic long-term monitoring of the atmospheric winds \citep[e.g.,][]{Sanchez_Lavega_2008, Peralta_2008, Moissl_2009, Peralta_2012}. Outstanding measurements were obtained by the ESA Venus Express mission \citep[hereafter VEx;][]{Svedhem_2007}, whose main goal was a better understanding of the atmospheric circulation, with a specific attention to the origin of the super-rotation. Cloud tracking measurements were provided by the Venus Monitoring Camera \citep[VMC;][]{Markiewicz_2007} and the Visible and InfraRed Thermal Imaging Spectrometer \citep[VIRTIS;][]{Drossart_2007}.

Despite the exquisite quality of VEx measurements, cloud tracking indicates the motion of large cloud structures (limited by spatial resolution), which is an indication of the speed of iso-pressure regions rather than the speed of the actual cloud particles. In the case of Venus, cloud stuctures may represent the phase speed of a condensation wave, possibly associated with vertical mixing or chemical processes associated with the UV absorber \citep{Widemann_2007,Hueso_2015,Machado_2014,Machado_2017}. A complementary solution to access direct wind speed measurement is  Doppler spectrometry because it measures the actual speed of cloud particles and has different but complementary limitations in local time, latitudinal, and temporal coverage than an orbiting spacecraft. The idea of comparing cloud tracking and Doppler spectroscopic measurements has emerged with the idea of supporting the VEx,  and eventually Venus Climate Orbiter (Akatsuki), which entered the Venus orbit in 2015 \citep{Nakamura_2016}.

Characterizing the atmospheric circulation of Venus based on high-resolution spectroscopy is actually an old idea, but it was logically considered to be challenging \citep[e.g.,][]{Moreux_1928}\footnote{``Ainsi, nos id\'ees sur la climatologie de V\'enus sont li\'ees \`a la d\'etermination de la dur\'ee de la rotation de la plan\`ete et, en pr\'esence de l'incertitude des observations, il semble que le mieux est de r\'eserver nos conclusions. Le doute cependant aurait pu \^etre lev\'e, pensait-on, au moyen du spectroscope. Cet instrument, en effet, par le d\'eplacement des raies sombres en un sens ou dans l'autre, nous indique si l'objet lumineux vis\'e par l'appareil se rapproche ou s'\'eloigne de l'observateur : c'est le principle de la m\'ethode Doppler-Fizeau [...]. Malheureusement, lorsqu'on photographie simultan\'ement le spectre des deux bords oppos\'es de la plan\`ete, les diff\'erences de vitesses relatives doivent \^etre si faibles que le d\'eplacement des raies est \`a peine sensible. Il n'est donc pas \'etonnant que le proc\'ed\'e, aux mains de diff\'erents astronomes, ait donn\'e des solutions contradictoires.'' \citep{Moreux_1928}}. The first reliable measurements with modern spectrographs started in the 1970s \citep{Traub_Carleton_1975, Young_1979}. 
Starting in 2007, the ground-based support to VEx kicked off many observational projects \citep{Lellouch_Witasse_2008}, including the present project. 
Significant results on the upper mesospheric dynamics were obtained using mid-infrared heterodyne spectroscopy \citep{Sornig_2008, Sornig_2012} and millimeter and submillimeter wave spectroscopy \citep{Clancy_2008,Clancy_2012,Lellouch_2008,Moullet_2012}, but Doppler spectroscopy is more challenging at shorter wavelengths. Visible observations of solar Fraunhofer lines scattered by Venus clouds were performed by \citet{Widemann_2007, Widemann_2008, Gabsi_2008, Gaulme_2008, Machado_2012, Machado_2014, Machado_2017}. 
Most results regard average zonal wind profiles as a function of latitude at the cloud-top level. \citet{Widemann_2007, Widemann_2008} and \citet{Machado_2014, Machado_2017} also reported to have measured instantaneous zonal and meridional wind circulation from data obtained with the \espadon \'echelle spectrometer of the Canada France Hawaii Telescope (CFHT), which we discuss in this paper (Sect. 2). 

Back in 2007, we proposed to use the THEMIS\footnote{T\'elescope H\'eliographique pour l'\'etude du Magn\'etisme et des Instabilit\'es Solaires} solar telescope  to get Doppler maps of Venus by scanning the planet with the 100 arcsec long-slit spectrometer MulTiRaies (MTR), whose resolution ranges from 100,000 to 1,000,000 in the visible \citep{Mein_Rayrole_1985}. Two factors motivated this choice. First, the use of a long slit allows for reconstructing complete radial-velocity (RV) maps of the planet, which is not possible with single-fiber spectrographs, as \espadon, or spectrographs with shorter slits, as that of the Ultraviolet and Visual Echelle Spectrograph (UVES) of the Very Large Telescope (VLT)\footnote{In \citet{Machado_2012}, the slit length of UVES\ was 11 arcsec versus $\approx$20-arcsec for the diameter of Venus.}. Secondly, a solar telescope can easily observe during the daytime, which is convenient for Venus.
We led two test campaigns in 2007 and 2008, and an ``actual'' campaign in 2009, when the observing setup and data processing technique were fully developed. The 2007 test did not meet good weather conditions and the results were extensively discussed in a previous publication \citep[][hereafter G08]{Gaulme_2008}. Only a portion of the visible phase of Venus was scanned in the 2007 observations. A rough estimate of the zonal wind was obtained, i.e., $151\pm16$ m s$^{-1}$, which was significantly larger than previously measured. In 2008, the observing setup consisted of repeatedly scanning Venus from west to east. This observing protocol showed that the mean RV would drift with time, for instrumental reasons, which introduced an uncontrollable bias in the zonal wind estimate, as we could not disentangle RV variations due to zonal winds from instrumental drifts. During the next run, in September 2009, we opted to alternatively scan Venus from north to south (NS) and south to north (SN) with the slit parallel to Venus equator to circumvent this issue. That way, as described later, we could ensure reliable RV values at each latitude and erase possible drifts by combining  NS and SN scans.
 
In this paper, we report the results of the 2009 campaign, where we obtain for the first time an instantaneous and spatially resolved RV map of any planet of the solar system. We first review the previous works dealing with Venus RV measurements, and we bring into question some aspects of what was done so far in relation with our own observing approach (Sect. 2). We then detail the data acquisition and processing techniques that lead to photometric and RV image reconstruction (Sect. 3). Next, we present the RV maps obtained on different days and our empirical model, which involves zonal and meridional winds as well as a local-time dependence of the zonal component (Sect. 4). Finally, we discuss our results and compare them to both cloud tracking and recent spectroscopic observations (Sect. 5). 

\begin{figure}[t!]
\center
\includegraphics[width=6cm]{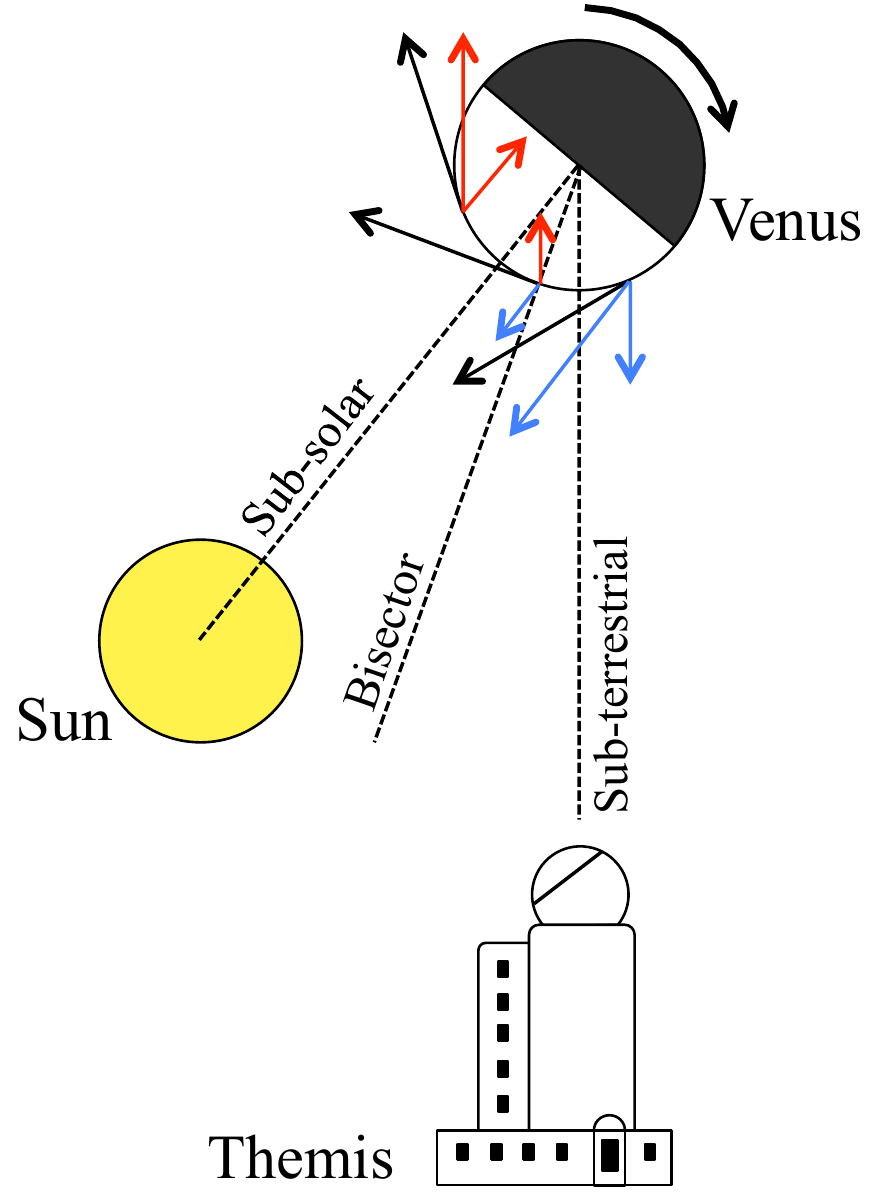} 
\caption{Schematic view of the Doppler shift of a prograde rotating planet as seen from the THEMIS Observatory. The Doppler effect measured in the visible on the reflecting cloud deck is the sum of the motion relative to the Sun and the Earth. Radial velocities are zero along the bisector meridian, located halfway in between the subsolar and subterrestrial meridian.}
\label{fig_themis}
\end{figure}
\begin{figure}[t!]
\center
\includegraphics[width=7cm]{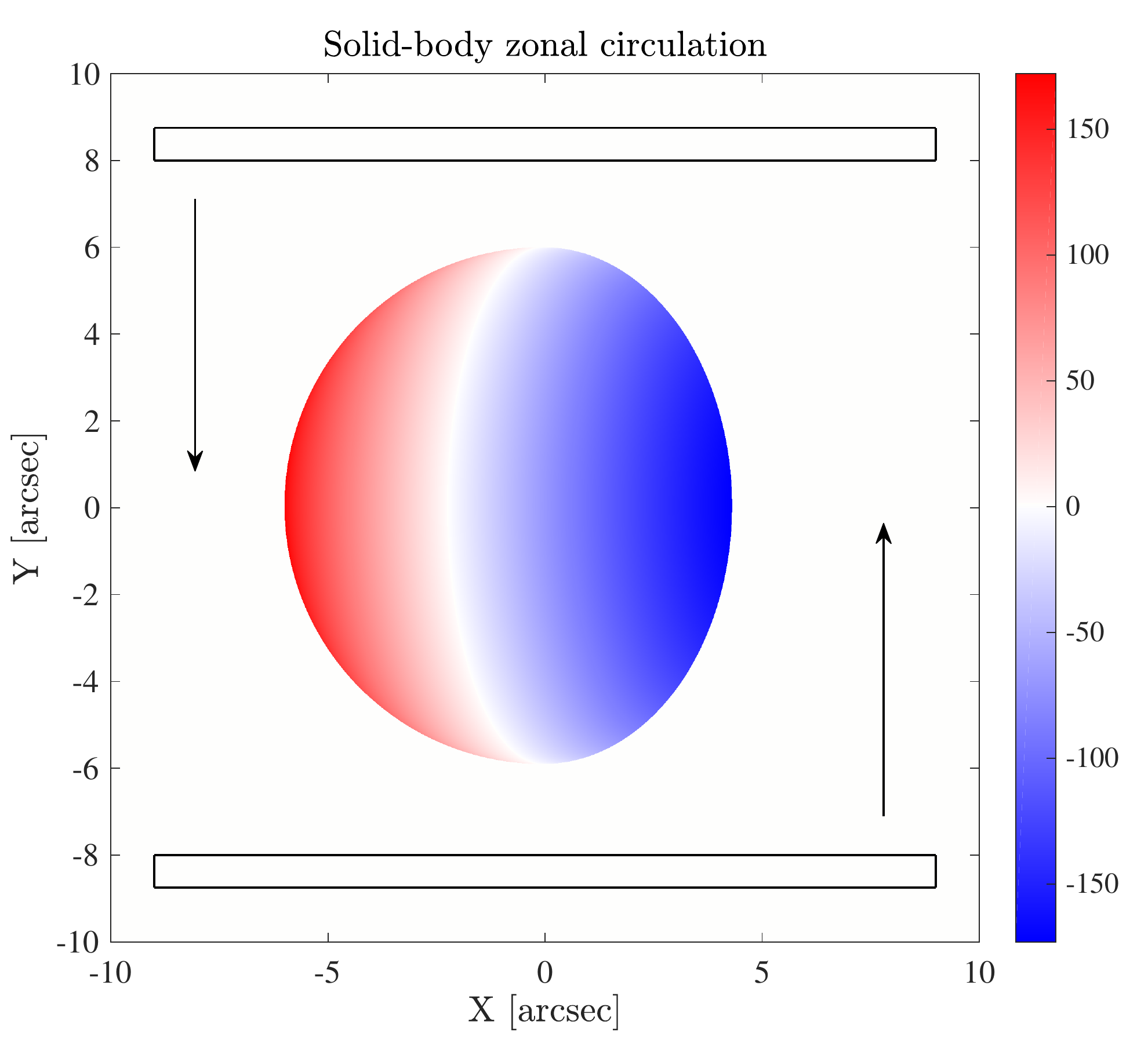}
\includegraphics[width=7cm]{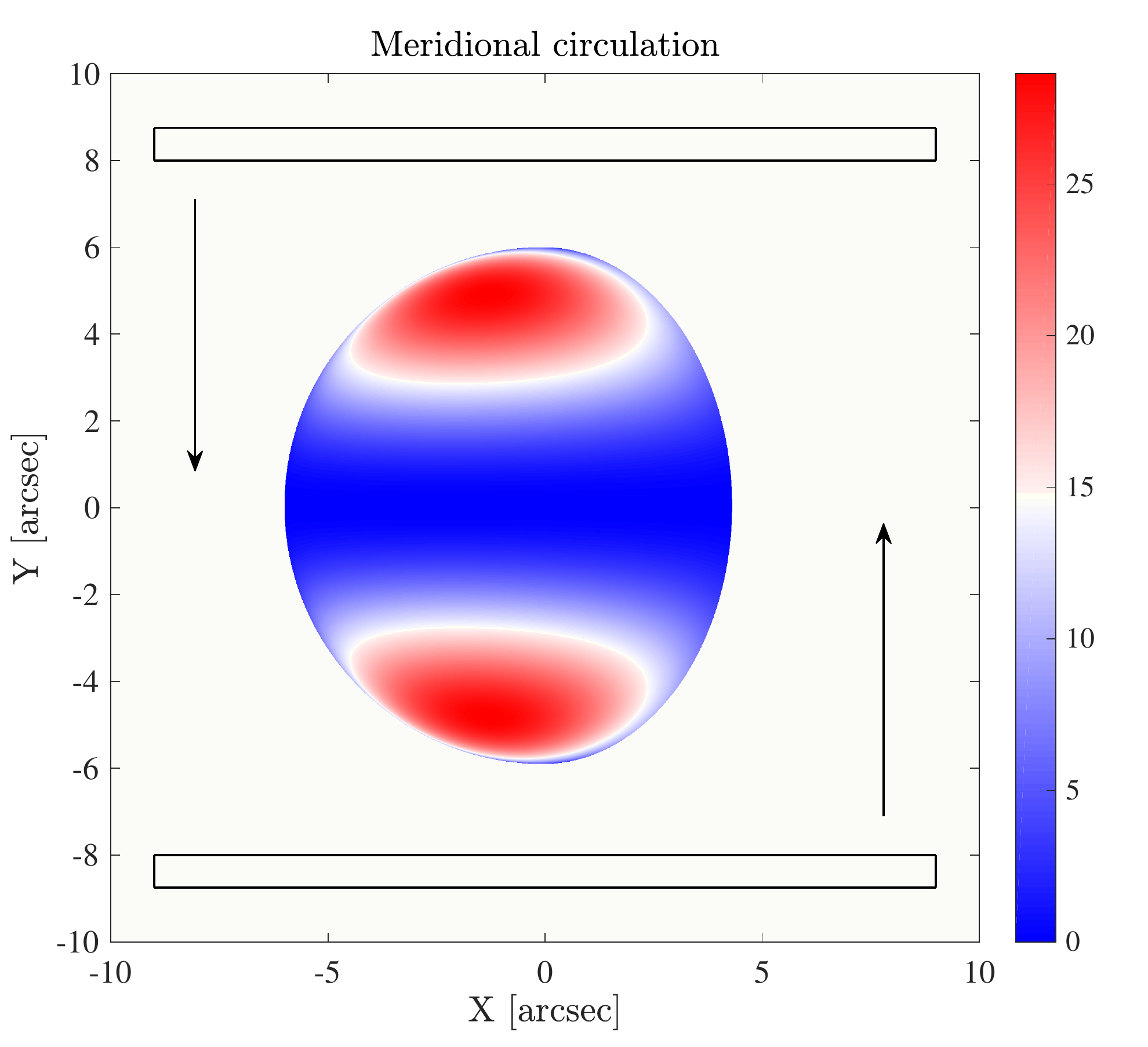}
\caption{Simulation of RV maps corresponding to a model of solid-body rotation, i.e., $V=V\ind{equator} \cos(\lambda)$ (top), and a meridional circulation, i.e., $V=V\ind{\lambda=45^\circ} |\sin(2\lambda)|$ (bottom), where $\lambda$ stands for the latitude. For the zonal wind $V\ind{equator} = 100$ m s$^{-1}$, for the meridional wind $V\ind{\lambda=45^\circ} = 20$ m s$^{-1}$. The Doppler shift assumes the use of reflected solar lines and a phase angle of 43$^\circ$. The black boxes indicate the orientation of the entrance slit of the spectrometer and the arrows show the scanning direction. The slit width corresponds with the exact width (0.75 arcsec) relative to Venus, but the length was reduced from 100 to 18 arcsec for illustration purposes.}
\label{fig_RV_simu}
\end{figure}

\section{Two decades of efforts}
\label{sect_review}
\subsection{Expected Doppler map}
The Doppler technique has been extensively used from the ground with a variety of observational techniques involving millimeter, infrared, and visible spectroscopy. Spectroscopic tracers of the atmospheric dynamics of Venus include: CO and isotopic $^{13}$CO rotational lines in the millimeter wave range to probe at 105 km and 95 km, respectively; infrared emission of CO$_2$ in nonlocal thermodynamical equilibrium (110 km level); and solar Fraunhofer lines and Venus CO$_2$ molecular lines in the visible probing cloud top level (70 km) and a few kilometers above, respectively. 
Heterodyne spectroscopic measurements have been interpreted as a combination of SSAS and zonal winds \citep[e.g.,][]{Lellouch_1994, Clancy_2008, Lellouch_2008, Sornig_2008}, with variable relative contributions. The cross-terminator SSAS velocity is usually inferred to be on the order of 100 m s$^{-1}$, while equatorial zonal velocity varies between 0 and 200 m s$^{-1}$. Poleward meridional winds may also have been marginally detected \citep{Lellouch_1994, Lellouch_2008}.

From visible observations, the atmospheric circulation of Venus is mostly zonal at the top of cloud layers, peaking at about 100 m s$^{-1}$ at the equator \citep[e.g.,][]{Sanchez_Lavega_2017}. The zonal wind was originally considered to follow either a solid body rotation, i.e., the same rotation period at all latitudes, or cylindrical, i.e., the same velocity at all latitudes, except in the polar regions where it should drop to zero. Cloud tracking and RV measurements led to a slightly different picture; an average zonal wind profile as a function of latitude looks like the letter ``M'': wind rising from 0 to 120 m s$^{-1}$ from poles to 45$^\circ$ latitude and decreasing to 100 m s$^{-1}$ around the equator \citep[e.g.,][]{Machado_2017}. 
A meridional circulation corresponding to equator-to-pole Hadley cells was measured both from cloud tracking and RV measurements, peaking at about 20 m s$^{-1}$ at 45$^\circ$ latitude \citep[e.g.,][]{Machado_2017}.

The Doppler shift of solar Fraunhofer lines reflected on a planet is the sum of the RV of the planet relatively to the Sun and to the observer. In case of a planet at exact opposition, the Doppler effect of reflected solar lines is then doubled. In the case of a non-zero phase angle, i.e., the angle Sun-planet-observer, the Doppler shifts cancel each other on the meridian located at the bisector of the subsolar and subterrestrial points \citep[e.g.,][]{Gabsi_2008}. A retrogradely rotating zonal circulation therefore displays a blueshift in the morning and a redshift in the afternoon (Fig. \ref{fig_themis}). In our case, observations were performed during Earth morning elongation, which means that we were seeing the morning terminator of Venus. In Fig. \ref{fig_RV_simu}, we represent theoretical RV maps of a solid-body rotator and a meridional circulation based on two Hadley cells at the phase angle corresponding to the 2009 campaign. On top of this, uniform RV offsets are expected, corresponding to the relative motion of Venus with respect to the Sun and mostly of Venus with respect to the observer on the rotating Earth. The motions are well documented in the ephemeris database and must be taken into account. 

Beyond real RV fields, the rotation of the Sun as seen from Venus introduces a bias in RV measurements, as originally introduced by \citet{Young_1975} and subsequently completed by \citet{Gaulme_2018}. Rays from a different part of the Sun, which show different RVs, reach the planet with (slightly) different incidence angles. Regions of the Sun that are closer to the horizon contribute less to the reflected solar spectrum than regions closer to zenith. Thus, the RV integrated over the whole solar disk is not zero at a given point of Venus. In other words, even if Venus were not rotating, we would still measure a Doppler shift near Venus terminator, and that Doppler shift would mimic a retrograde rotation because the solar rotation is prograde. \citet{Gaulme_2018} demonstrated that the RV field $\Delta V\ind{Y}$ associated with the so-called ``Young'' effect is expressed as  
\begin{equation}
\Delta V\ind{Y}(\gamma,\theta) =  Y(\Lambda) \tan\gamma \sin\theta
\label{eq_young}
,\end{equation}
where $\gamma$ is the solar-zenith angle, $\theta$ the inclination of the solar spin axis with respect to local horizon, and $Y(\Lambda)$ a coefficient that is about 2.9 m s$^{-1}$ at $\Lambda = 550$ nm. 
The analytical expression by \citet{Young_1975} of the artificial Doppler shift $\Delta V\ind{Y}$ on Venus was calculated for the equator and did not include the $\sin\theta$ term. In addition, his expression did not include the solar limb darkening and the Sun's differential rotation, which leads the coefficient by Gaulme et al. to be smaller by about 10\,\% than that of Young. 
We note that the new expression of the Young effect extended to all latitudes by \citet{Gaulme_2018} questions the wind measurements that have been done so far near Venus's terminator, which made use of the \citet{Young_1975} analytical formulation \citep{Widemann_2007, Widemann_2008, Gabsi_2008, Machado_2012, Machado_2014, Machado_2017}. 

The observed RV map is not only the sum of all the above-listed contributions, it is also strongly biased by the atmospheric seeing. Firstly, as extensively studied by \citet{Gaulme_2018}, the atmospheric seeing modifies the apparent location of the planet in the sky whenever the planet is not observed at full phase (opposition). This leads to biases when inverting RV maps because it affects the position of longitudes and latitudes on the Venus image. Secondly, the seeing convolves regions of variable RV and photometry, which tends to reduce the apparent amplitude of atmospheric motions. 
As originally pointed out by \citet{Civeit_2005}, the resulting RV map is the convolution of the RV signal with the photometric map of the considered object, including its degradation by seeing. The mean Doppler $\Delta V\ind{obs}$ measured in a given pixel $(x,y)$ on the detector can be expressed as 
\begin{equation} 
\Delta V\ind{obs}(x,y)  =  \displaystyle{\frac{(\Delta V\ F* P) (x,y)}{(F * P)(x,y)}} \label{eq_RV_convol_0}
,\end{equation}
where $\Delta V$ and $F$ are the real RV and photometric maps prior to seeing degradation, $P$ is the point spread function (PSF) of the atmospheric seeing, and the asterisk sign $*$ indicates the convolution product. This effect tends to reduce the amplitude of RV variations from west to east. Besides, it cancels out most of the Young  effect, making it almost negligible \citep{Gaulme_2018}. In this paper, we take into account the data degradation by atmospheric seeing to extract zonal and meridional wind circulations, for the first time with RV measurements of Venus performed in the visible. 

\subsection{Review of previous works}
In this section, we review the recent measurements of the atmospheric circulation of Venus with Doppler spectroscopy in the visible domain. This includes the works that started after 2000, slightly before or together with the VEx ground-based support, which were shortly described above. 
Techniques used for Doppler velocimetry in the visible solar spectrum on the dayside are mostly high-resolution \'echelle spectroscopy with single optical fiber feeding, i.e., single aperture measurements \citep{Widemann_2007,Widemann_2008,Gabsi_2008,Machado_2014,Machado_2017}, but also long-slit spectrometry \citep{Gaulme_2008,Machado_2012}. In the framework of coordinated campaigns to support VEx science investigations, the major breakthrough in terms of observational techniques for visible high-resolution spectroscopy of Venus has been led by Widemann and Machado \citep{Widemann_2007, Widemann_2008,Machado_2012,Machado_2014,Machado_2017}. We review all of these works, from the pioneering observations of \citet{Widemann_2007}, \citet{Gaulme_2008}, and \citet{Gabsi_2008}  to the robust observational protocols of \citet{Machado_2017}.

\subsubsection{Bushwhacking: Early works}
Among pioneering works, we consider \citet{Widemann_2007,Gabsi_2008,Gaulme_2008} apart from other works, as they constitute the first attempts of the past two decades to measure the winds of Venus with high-resolution spectrometers in the visible wavelength. 
As for any new project, their initial observational protocols and early analysis assumptions made their results significantly discrepant with respect to later studies. These three works nevertheless contributed toward renewing interest in the technique and kicking off the ground based-support to VEx and Akatsuki, which were fundamental steps toward reaching later successful measurements. 

\citet{Widemann_2007} reported measurements of the average global winds of Venus with the AURELIE high-resolution spectrometer at the 1.52 m telescope of Observatoire de Haute Provence, France. These authors measured the Doppler effect on Venus CO$_2$ absorption lines and a few reflected solar Fraunhofer lines in the range 8600-8800 \AA. Their results confirmed the existence of a zonal retrograde flow, even though the measured mean equatorial velocity of $75\pm15$ m s$^{-1}$ was relatively low with respect to posterior measurements, and strong day-to-day variations ($\pm65$ m s$^{-1}$) were identified. By combining the results from all data, these authors also reported the possible detection of a SSAS circulation component of amplitude of about $40$ m s$^{-1}$ at the terminator. 
Given the consistency of observational results reported since then, it is possible that the daily variations of zonal winds were artifacts because most observations realized later on with improved protocols never displayed anything similar. Nevertheless, this work had the great merit of paving the way for the methods -- especially the sequential pointing -- used in \citet{Widemann_2008,Machado_2014,Machado_2017}.

The \citet{Gabsi_2008} observations were performed with the EMILIE high-resolution, cross-dispersed spectrograph and its associated calibrating instrument the Absolute Astronomical Accelerometer (AAA), at Observatoire de Haute-Provence, France. From their three best observing nights, they reported zonal wind values of $75\pm6$, $85\pm3$ and $91\pm6$ m s$^{-1}$ without considering the Young effect. These authors reported a better stability by introducing the Young effect according to the \citet{Young_1975} formula and retrieved a mean zonal circulation of 48, 47 and $51\pm3$ m s$^{-1}$, which is far from what has been measured otherwise. No other observations of Venus with this instrument have been published since then.

\citet{Gaulme_2008} reported a test of the long-slit MTR spectrometer of the THEMIS solar telescope to scan Venus and retrieve a complete RV map of it, which had never been done before. The system planned for scanning Venus did not work and the weather was poor. The result is a partial RV map that was modeled with a global zonal circulation pattern ($151\pm16$ m s$^{-1}$) without considering the \textit{Young} effect. The analysis of these preliminary data was promising enough to justify further observations, which constitutes the base of the present work.

\subsubsection{Stable radial-velocity measurements}
The series of papers by \citet{Widemann_2008,Machado_2014,Machado_2017} shows a remarkable consistency both in terms of methods and results. 
All observations were done with the \espadons spectro-polarimeter at the 3.6 m CFHT, in coordination with VEx/VIRTIS-M in 2011 and 2014 \citep{Machado_2014,Machado_2017}.
The spectrometer \espadons covers the whole visible range (3700-10,500 \AA) at an average resolution of 80,000 and is fed by an optical fiber whose field of view (FOV) is 1.6 arcsec. The observation protocol was sequential, like in most other ground-based wind measurement techniques, either with CO rotational lines or infrared (IR) CO$_2$ non-LTE emission lines. Only \citet{Moullet_2012} performed Doppler mapping with interferometric observations based on CO lines at the Plateau de Bure interferometer.

The sequential acquisition consists of selecting a set of positions on the Venus dayside and making one observation at a time per position to cover all positions. The whole scan of the dayside hemisphere is then repeated once or twice, according to observing conditions. Centering and guiding is manually controlled with the help of a Venus template that is taped on the display of the guiding camera (Fig. 1 of \citealt{Machado_2017}). To compensate RV drifts of about 100 m s$^{-1}$, the spectrometer's wavelength calibrations were carried out using both Thorium-Argon (ThAr) lamps and a set of telluric lines following the standard protocol developed at CFHT \citep{Donati_1997}. Final RV measurements were corrected from the various motions (Venus-Sun, Earth-Venus), as well as the Young effect according to the \citet{Young_1975}  expression. 

In all three papers, the complete RV dataset is interpreted as a horizontal zonal circulation, where two possible regimes are considered: a solid-body or a cylindrical regime. Except for the preliminary observations at CFHT by \citet{Widemann_2008}, which suffered of a lack of stability (zonal winds in between 92 and 155 m s$^{-1}$), the results by Machado et al. provide very consistent values of the mean zonal wind at equator from day to day and in between 2011 and 2014; i.e., $117.3\pm18.0$ and $117.5\pm14.5$ in \citet{Machado_2014} and $119.6\pm16.5$, $122.6\pm31.3$,  $119.6\pm26.0$, and $118.1\pm19.5$ in \citet{Machado_2017} with 2$\sigma$ errors. Beyond mean zonal winds, \citet{Widemann_2008} introduced a meridional wind component as part of their measurements, while \citet{Machado_2014, Machado_2017} looked for the presence of equator-to-pole meridional circulation on the spectra that were taken -- on purpose -- along the bisector meridian, where zonal RV signal is canceled. \citet{Machado_2014,Machado_2017} reached a conclusion on the detection of a meridional circulation peaking at about 20 m s$^{-1}$ at mid-latitudes, thus confirming the results obtained with Venus Express cloud tracking measurements \citep{Sanchez_Lavega_2008, Hueso_2012, Hueso_2015, Khatuntsev_2013, Machado_2014, Machado_2017}.

\begin{figure*}[t!]
\includegraphics[width=6cm]{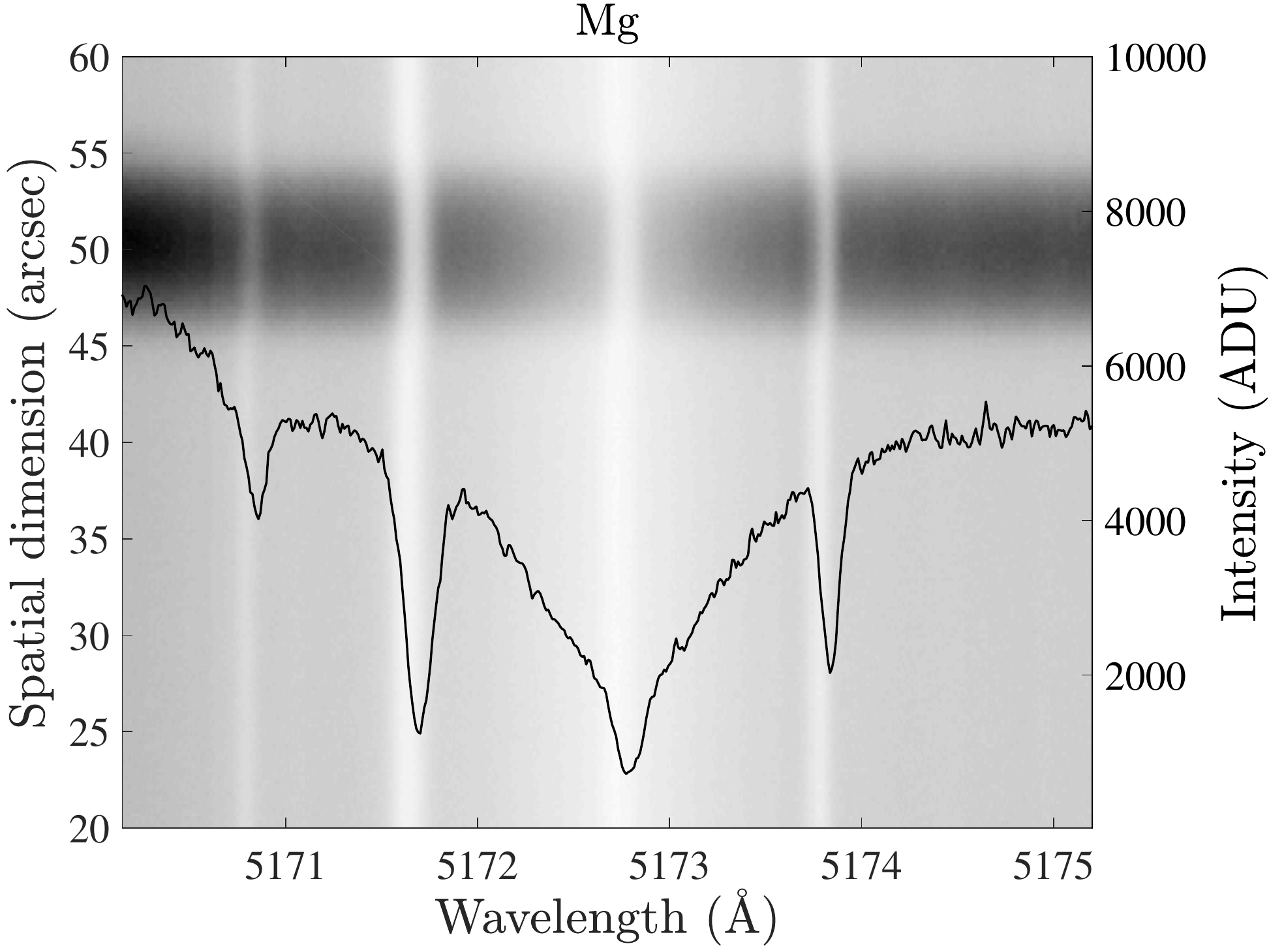}
\includegraphics[width=6cm]{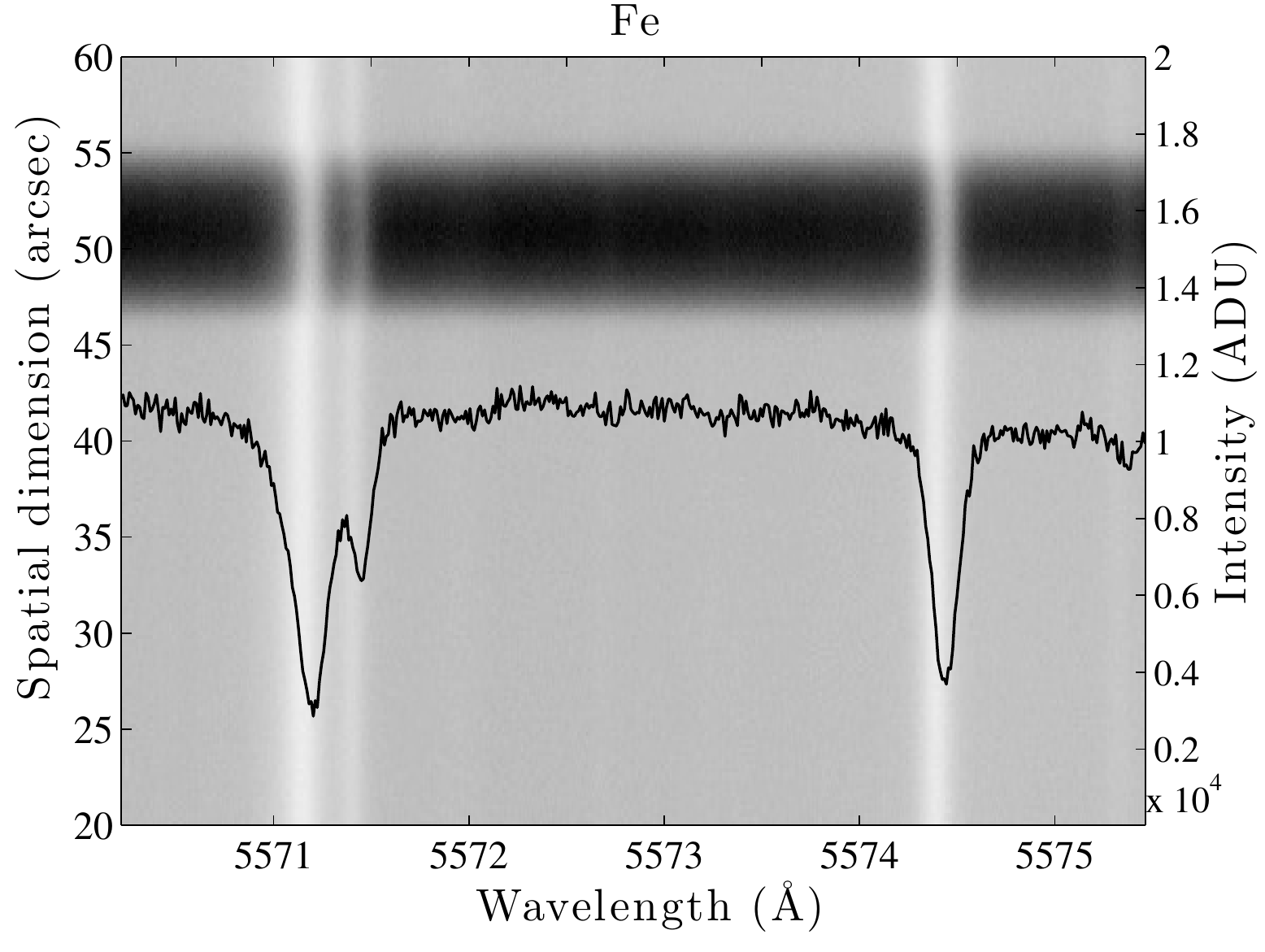}
\includegraphics[width=6cm]{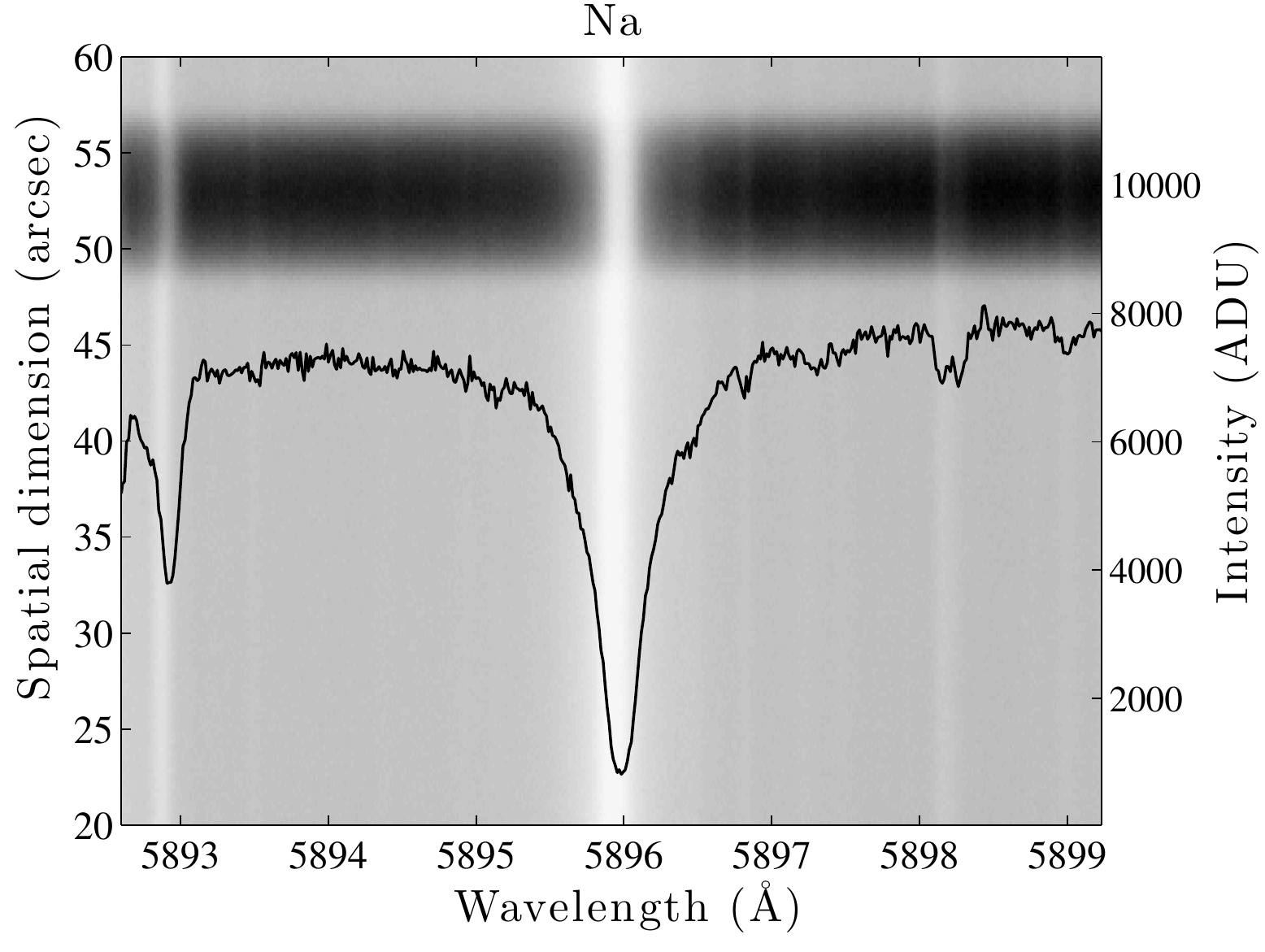}
\caption{Example of spectral images obtained with THEMIS in 2008 while scanning Venus. Color scale is inverted, i.e., dark is bright and white is faint. The three spectral ranges are centered on Mg, Fe, and Na Fraunhofer lines in the mid-visible domain. The 2D image is a zoom of the actual image where the dark area corresponds to Venus. The Fraunhofer lines are tthe vertical white structures. Since this spectra were taken during daytime, the absorption lines are visible both on Venus and the Earth sky. For all of the deepest lines, we clearly distinguish the Doppler shift in between Venus and Earth atmosphere. In these figures, the slit was positioned parallel to Venus' equator close to the sub-Earth meridian. The black line, corresponding to the y-axis on the right side, is the Venus spectrum at about zero latitude.}
\label{fig_spectres}
\end{figure*}

\citet{Machado_2012} reported wind measurements performed with the long-slit spectrometer UVES at the VLT in Paranal, Chile. The UVES spectrometer covers the whole visible range (3000-10,000 \AA) with an average spectral resolution of 100,000 and a spatial resolution of 0.2 arcsec per pixel. The slit size was set to $11\times 0.3$ arcsec, while the diameter of the planet was in between 20 and 22 arcsec with phase angles of 82.9$^\circ$ and 87.8$^\circ$ during two runs in May and June 2007, respectively. 
The advantages of this instrument are its large band pass (like \espadons), large collecting area (8.2 m telescope), small pixel FOV, and mapping capability. 
These authors focused on a few localized positions on the planet: three that were parallel to the rotation axis and six parallel to the equator. The positions along the rotation axis aimed at putting into light any asymmetry of zonal rotation in between both hemispheres. 
However, such configuration does not allow for determining the absolute speed of zonal circulation. To the contrary, the slit parallel to the equator aimed at providing absolute measurements of the amplitude of zonal circulation and possible longitudinal variations. Intentionally, no meridional circulation was considered at the time. 

For each spectrum, the Doppler shift was computed by correlating the spectrum with the spectrum at the center of the slit, as previously done by \citet{Luz_2005, Luz_2006} and \citet{Civeit_2005} with observations of Titan. In other words, the actual reference spectrum used for spectral calibration therefore changed from position to position of the slit. 
To check for instrument stability and to correct for optical slit curvature, exposures of the built-in ThAr lamp were taken after the VLT/UVES science exposures. To this set of corrections, the additional subtraction of the Young effect from the \citet{Young_1975} expression was performed near the terminator. 

From the spectra obtained with the slit parallel to the equator, \citet{Machado_2012} reported an average zonal circulation in between 106 and $127\pm14$ m s$^{-1}$. In addition, they investigated possible variations of zonal wind amplitude as a function of local time. They identified a slight increase of the wind speed near the terminator, which corresponded to the evening side during Spring 2007, at about 150 m s$^{-1}$ at 10$^\circ$ away from it. From the spectra obtained with the slit parallel to the rotation axis, they detected a slight asymmetry of zonal circulation by measuring that winds are faster by $6\pm5$ m s$^{-1}$ in the southern hemisphere. 
The M-shaped latitudinal profile of mean zonal circulation is later compared with other published methods (e.g., Fig. 14 of \citealt{Machado_2017}).

\subsubsection{A few concerns and the big picture}

We base our concerns on a recent study by \citet{Gaulme_2018}, who in particular have shown that atmospheric seeing introduces biases regarding both the localization of the planet on the detector and the RV field. None of the cited works fully takes into account these effects. As regards observations by \citet{Machado_2014, Machado_2017}  it is unlikely that it significantly influenced the final wind determination, thanks to the good seeing conditions. 
Nevertheless, we consider that neglecting the bias on RV by atmospheric seeing has likely led to underestimating the speed of zonal circulation next to the terminator. In the same way, making use of the \citet{Young_1975} expression of the Young effect is erroneous for data taken out of the equatorial region and may have biased part of the results. However, we note that the CFHT observations were done relatively far from the terminator, where the seeing effects on both RVs and the Young effect are small relative to the amplitude of zonal circulation. To the contrary, both UVES/VLT and MTR/THEMIS observations included the terminator region.

More specifically, because of this potential seeing effect, we question an aspect of the Machado et al. (2012) analysis, where the detection of small-scale longitudinal wave structure as a function of local time on Venus is considered (RV variations of about 10-30 m s$^{-1}$ over spatial scales of 216 km). Firstly, a scale of 216 km corresponds to two pixels and is at the very limit of the Shannon criterion on signal sampling. Secondly, with a pixel FOV of 0.2 arcsec and an atmospheric seeing larger than 1.2 arcsec, detecting such small-scale variations sounds optimistic. 

Regarding the CFHT observations \citep{Machado_2014, Machado_2017}, we identify two minor issues. Firstly, the fit of a pure zonal circulation intentionally neglects the fraction of the RV signal attributed to the meridional component. This choice could alter the retrieved speed of the zonal wind. However, considering the relatively low amplitude of meridional wind reported so far ($\sim20$ m s$^{-1}$ at mid latitudes), this approximation likely introduced a marginal bias. Secondly, Figs. 4 of both papers show local variations of the zonal wind as a function of longitude and latitude, which were obtained from the sum of the model plus the residuals (Machado, priv. comm.). This result is questionable because local values of the zonal wind should be obtained by dividing the residuals by the zonal-circulation projection factor, which is null at the bisector meridian. This would result in divergent error bars on zonal wind speed along that meridian. Still, we note a good agreement in between the \citet{Machado_2014,Machado_2017} results with the simultaneous cloud tracking data of VIRTIS-M/VEx.
In the present paper, we follow the approach proposed by \citet{Gaulme_2018} in which we perform a global fit of the observed RVs, including a zonal plus a meridional component. This does not allow for a direct comparison of our measurements with the local wind values of \citet{Machado_2014, Machado_2017}.


Overall, we retain that both observations conducted at CFHT with the fiber-fed spectrometer \espadons and at VLT with the UVES long-slit spectrometer led to identifying a mean zonal circulation of about 120 m s$^{-1}$ in between $\pm45^\circ$ parallels with an M-shaped profile. The UVES deprojected RV measurements per latitude indicated a slight increase of the zonal circulation toward terminator. The CFHT measurements along the bisector meridian show the presence of a double Hadley-cell equator-to-pole circulation with amplitude peaking at about 20 m s$^{-1}$ at mid-latitudes.


\section{Telescope, instrument, and methods}
\label{sect_telescope}
\subsection{THEMIS telescope with the MTR spectrometer}

The THEMIS observatory is a solar telescope dedicated to accurate measurement of polarization of solar spectral lines with high spatial, spectral, and temporal resolutions  \citep{Mein_Rayrole_1985}. It is a 90 cm diameter Ritchey - Chr\'etien telescope. As for G08, it has been operating in the MTRmode \citep{Mein_Rayrole_1985} with no polarimetric analysis, which simultaneously permits spectral observations in up to four different spectral domains. Spectrometry with a slit produces 2D images,  whose horizontal component is the optical spectrum and vertical component the spatial dimension (Fig. \ref{fig_spectres}). The slit is 100 arcsec long, 0.75 arcsec wide, and the spectral resolution was set to $R = 150,000$. The guiding and positioning on the planet was controlled by a tip-tilt mechanism.

Using this solar observatory instead of a classical night-time telescope has two advantages. Firstly, a high-resolution spectrometer with such a long slit is not common at all. This allows us to get a whole cut of any planet at once; all planets have an apparent diameter lower than 60 arcsec - including a significant fraction of Earth skylight, which is helpful to monitor the stability of the spectrometer, as we see later. 
Secondly, it allows us to not be limited by the sunlight since it is designed to stare at the Sun. We could for instance observe Venus at phase angles that are not accessible with night-time telescopes, as in 2008, where the phase was about 13$^\circ$, i.e., the elongation 9$^\circ$.

As in G08, we worked with solar lines that are reflected on the cloud decks of Venus. In the present observations, three 10 \AA\ broad spectral domain were considered, all in the central part of the visible spectrum and each centered around one or several deep and sharp Fraunhofer lines: the magnesium line at 5173 \AA, the iron doublet at 5573 \AA, and the sodium D1 at 5896 \AA . The fourth detector was tested to get Doppler shifts on CO$_2$ molecular lines, but the S/N of these data was not good enough to go further. The theoretical velocity sensitivity of a single line is roughly proportional to the slope of the line and the total amount of photons. We refer G08 for the detailed estimate. The maximum theoretical velocity sensitivities of each line are 32, 17, and 42 m s$^{-1}$ arcsec$^{-1}$ minute$^{-1}$  for Mg, Fe, and Na respectively.  

The observing protocol consisted of repeatedly scanning the planet with the slit, either from NS or from SN. We would place the slit at a given location on Venus and acquire several exposures in a row. Then we would move to another position, and so on until the planet was completely covered. Steps from one position to the next were set to 0.8 arcsec, which roughly corresponded to the slit width. Technically, it was actually the opposite, as we positioned Venus with respect to the slit and not the other way around. In practice, the position of the slit on the planet was done by tilting the ``M5'' mirror, which sits at the pupil plane of the telescope and whose angle is controlled by a piezo-electric device. The rotation of the planet with respect to the slit was obtained by rotating both the derotator and the entrance slit by $180^\circ$. As we show in Sect. \ref{sect_RV_maps}, we found out that the steps increased while moving on the planet from SN, which means that this occurs as a function of the elongation. This is not fully understood but can be explained if one of the key components was not perfectly aligned (e.g., M5 not exactly at pupil plane, or derotator not perfectly aligned). In the following, we indicate as a ``scan'' a set of observations that includes spectra from one edge of the the planet to the opposite (either from NS or the reverse). Scans were typically composed of 16 consecutive positions on sky -- enough to account for the blurring caused by atmospheric seeing -- with 10 exposures per position.

\begin{figure}[t!]
\center
\includegraphics[width=4.3cm]{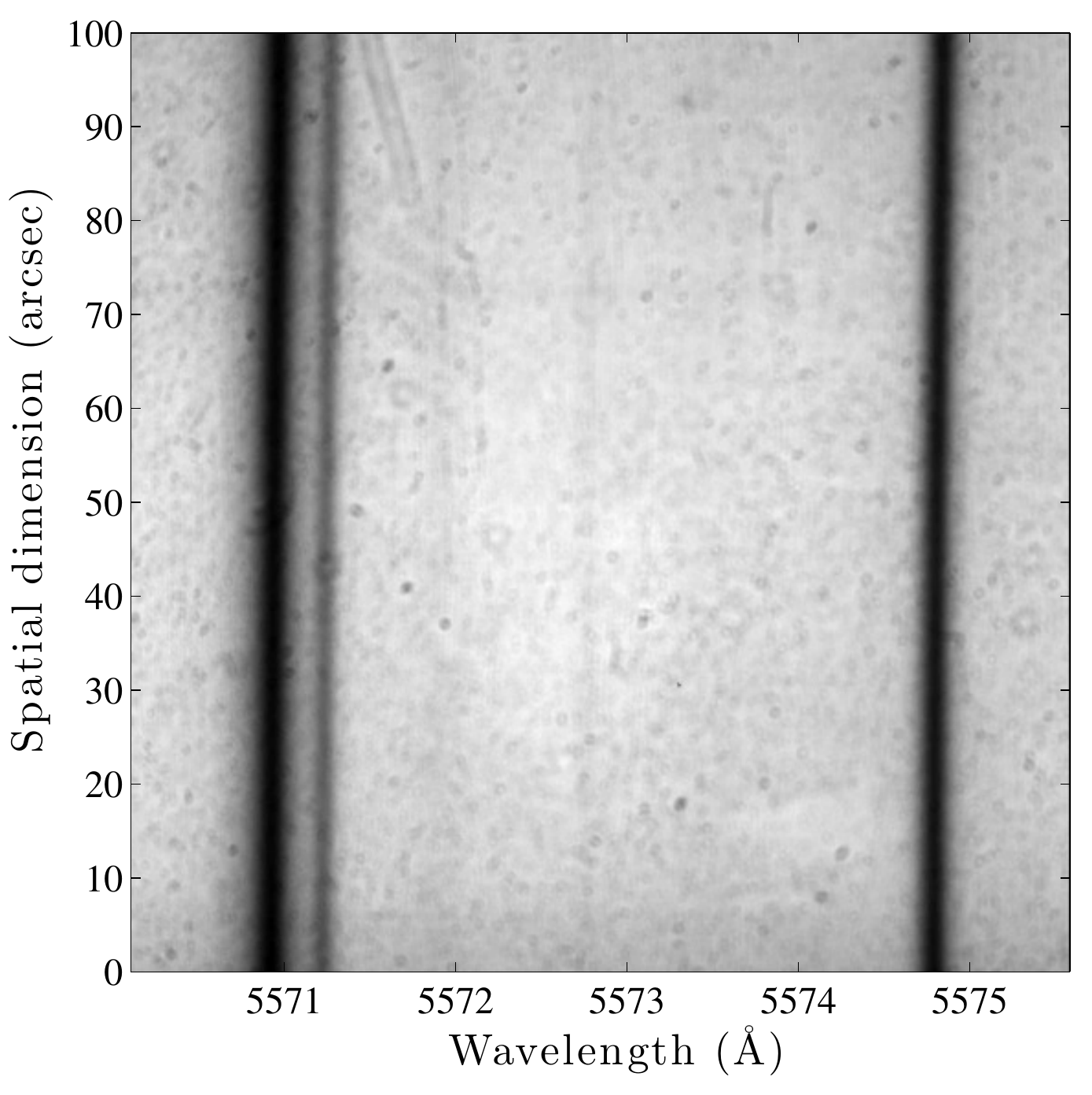} 
\includegraphics[width=4.3cm]{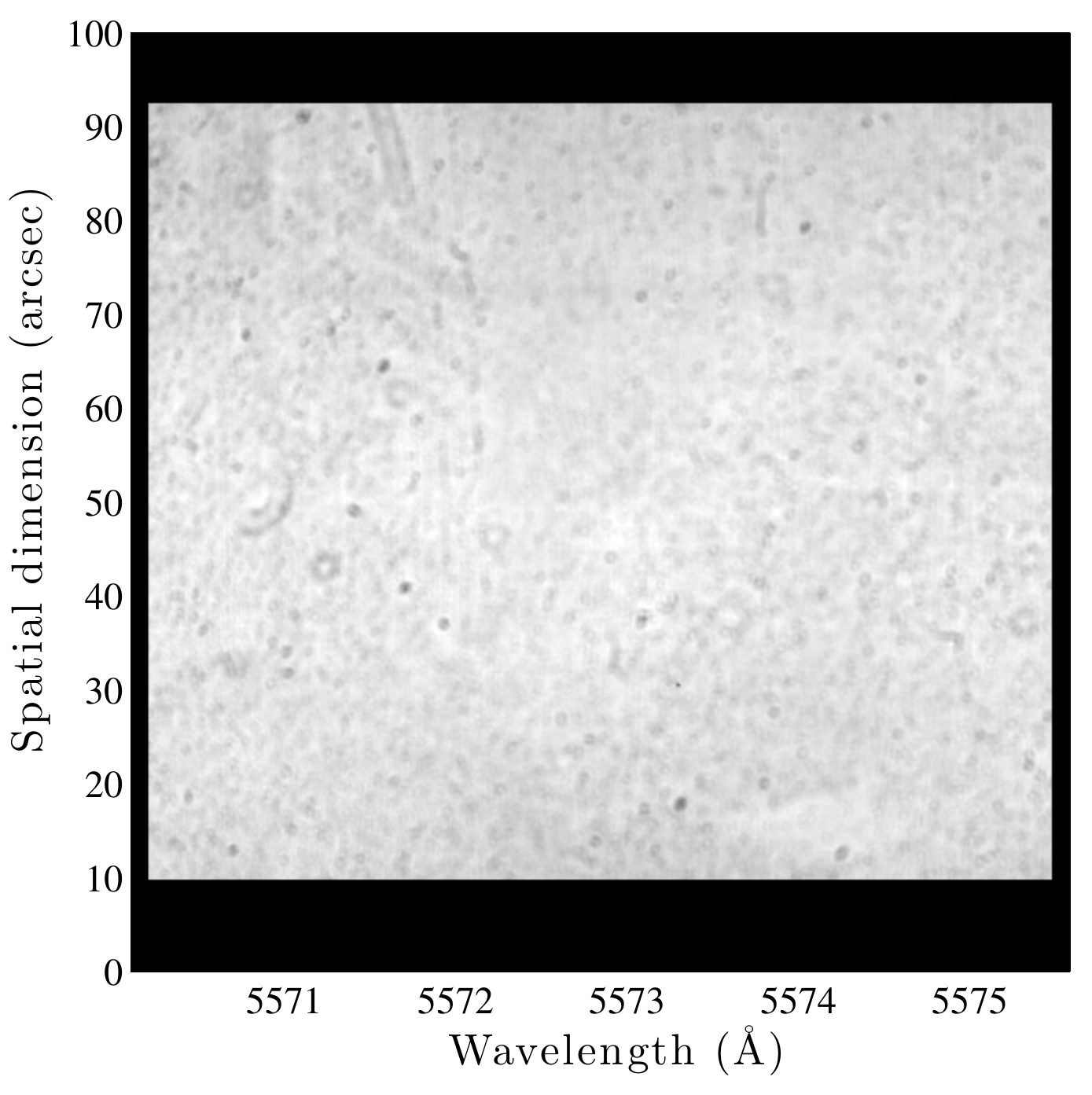}
\caption{Example of mean solar spectrum (left panel) and resulting flat field (right panel) on the detector dedicated to the Fe line during the 2009 campaign. A mask was applied to the redressed flat field. Vertically, it corresponds to regions that Venus never crosses, while horizontally, the image quality was affected by the rectification/interpolation.}
\label{fig_solar_flat}
\end{figure}

\subsection{Calibration data and preprocessing steps}
Calibration files are fundamental to ensure high quality Doppler measurements. In particular, errors on flat fields can induce distortions of the shape of spectral lines along the spatial dimension. In other words, it can produce fake Doppler shift across the planet. A quasi-absence of flat fields was one of the main limitations of the data quality of G08, where a fixed Doppler pattern on the detector was identified and removed a posteriori from the RV map. We considerably improved our protocol during the 2008 and 2009 campaigns. 

On the one hand, dark fields were obtained before each scan, and a master dark field was obtained by averaging them together. Using individual dark fields or the master dark field did actually not make any significant difference on RV measurements. On the other hand, obtaining a flat field is somewhat delicate with a slit spectrometer, given that daylight presents spectral lines. Using incandescent lights into the dome was not possible because of the specific configuration of this solar telescope. We chose to acquire about 400 spectra by directly pointing at the Sun to get a very high S/N spectral image on the whole detector by averaging them all (Fig. \ref{fig_solar_flat}). We note that the 400 spectra were taken at random locations on the solar disk to average out irregular Doppler shifts in the solar photosphere. Indeed, spatial and spectral resolutions of THEMIS is high enough to clearly distinguish by eye distortions of the Fraunhofer lines dues to the granulation and presence of p-waves close to the surface. 

These three spectral images were then used to measure and correct the geometrical distortions on the detector: instead of being vertical, absorption lines appeared to be slightly bent (Fig. \ref{fig_solar_flat}). We determined row by row the center of each absorption line from the image derivative with respect to the spectral axis ($x$-axis). A third-order polynomial fitting was used to fit the measured position of the bottom of the lines and to rectify the images with a cubic spline interpolation algorithm. We note that distortion is different for each detector and was estimated and corrected independently. The optical distortions were stable enough to not repeat this calibration in between observations. However, for cautiousness, we repeated the process every day  and did not notice any significant variation. An average line profile could be computed by vertically collapsing the rectified image of the average solar spectrum. Each row of the solar spectral image was finally divided by the average line profile to obtain a flat-field image (see Fig. \ref{fig_solar_flat}). 

\begin{figure}[t]
\center
\includegraphics[width=4.3cm]{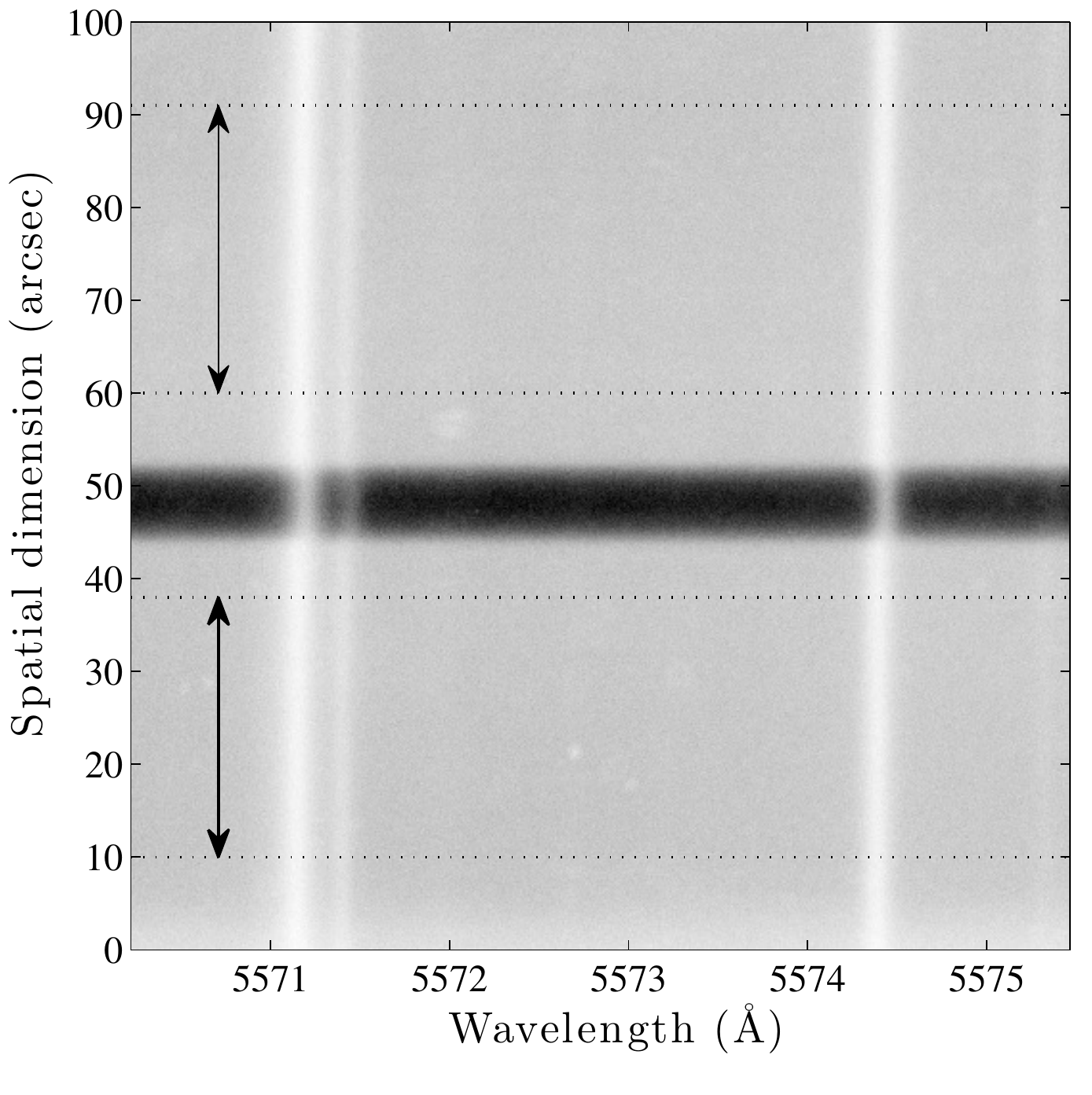}
\includegraphics[width=4.3cm]{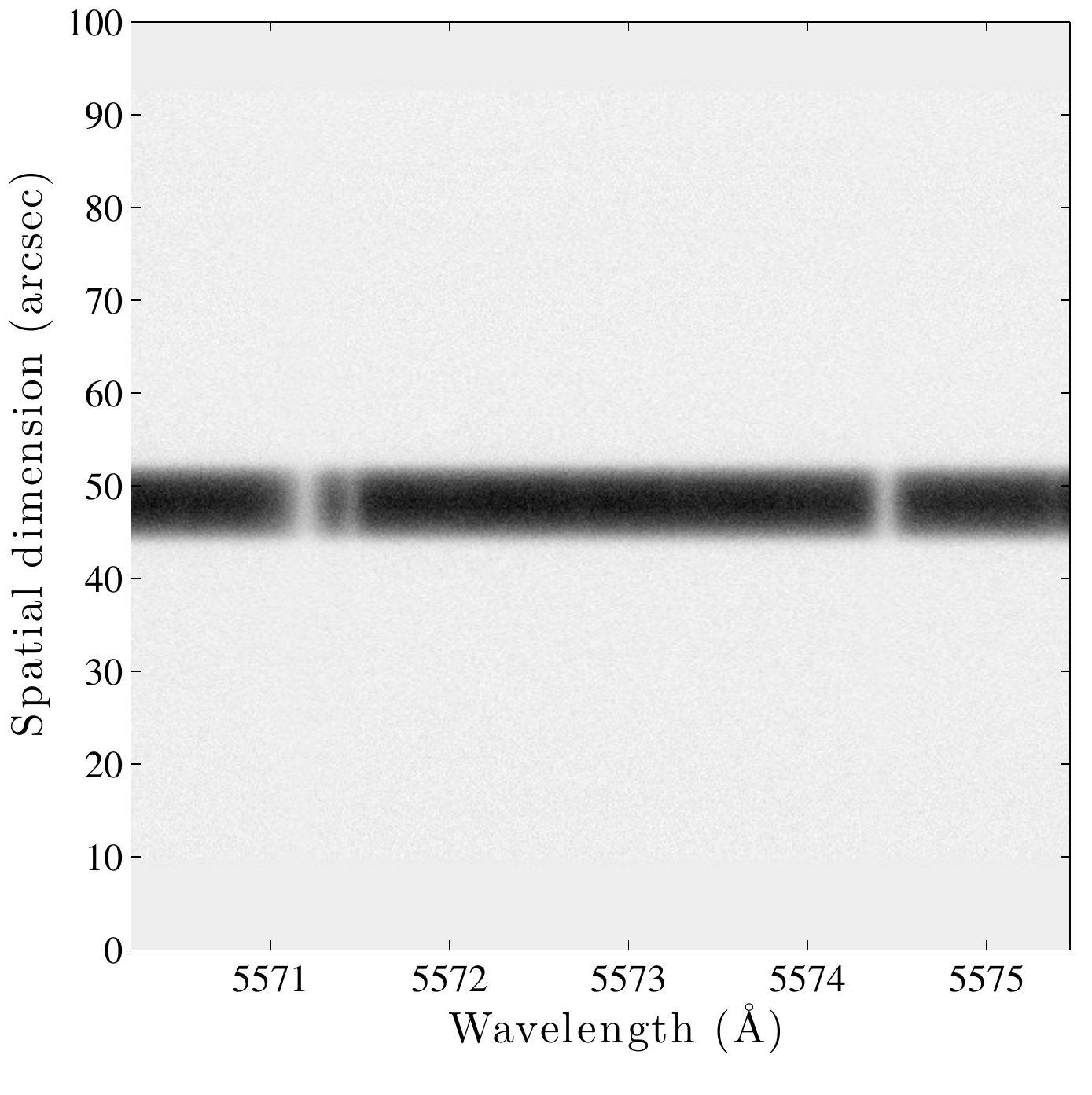}
\caption{Example of data preprocessing on an image taken during the 2008 campaign on May 5. Color scale is inverted, i.e., dark is bright and white is faint. The image is centered on the Fe doublet at 557 nm. The left panel shows the raw image, which is the result of a 1 minute exposure on Venus during daytime. The dark area corresponds to the region occupied by Venus spectrum, while the Fraunhofer lines scattered by Earth's sky are clearly visible out of it. The sky lines are tilted because of the geometric distortions mentioned in the text. The areas delimited by dotted lines and indicated by two vertical arrows are used to compute the average sky spectrum, which is then subtracted from the whole image. The right panel shows the ``cleaned'' image after dark field removal, rectification of geometric distortion, flat-field division, and mean sky lines subtraction. }
\label{fig_data_proc}
\end{figure}

The dark and flat fields therefore obtained were subtracted and divided, respectively, from each spectral image of Venus.
One last important item to preprocess the data taken during daytime, i.e., most of the data, consists of removing spectral lines that are scattered by the Earth's sky. For this, we selected the area of the detector where Venus is not present and collapsed all rows to get an average background line profile, as we did for the flat field. This contribution was then subtracted from the spectral image. This process had to be repeated for each spectral image because the solar lines scattered by Earth's atmosphere drift with Earth's rotation. The whole process was demonstrated to work adequately as illustrated in Fig. \ref{fig_data_proc}. 

To ensure a good data quality at each position of the slit, several exposures were taken every time, six in 2008 and ten in 2009. Thanks to the excellent pointing stability due to the tip-tilt mechanism, and after inspecting the images, we considered that Venus was not moving during the six or ten exposures at each slit position. In other words, shifts along the spatial dimension could happen while translating the slit on Venus from one position to another, but not otherwise, except if a cloud hid Venus in the meantime. The six or ten images taken at the same position were then averaged. The final product, i.e., ready to be used for measuring Doppler shifts, consisted of rebinning each of these mean spectral images along the spatial dimension. Indeed, the pixel FOV  (0.20 arcsec) of the MTR is well below the seeing - from 2 to 4 arcsec in 2009 - so that we could gain a factor two in S/N by rebinning four pixels along the $y$-axis, leading to a rebinned pixel resolution of 0.8 arcsec.

\subsection{Measuring Doppler shifts}
\label{sect_doppler}
\begin{figure}[t!]
\center
\includegraphics[width=7cm]{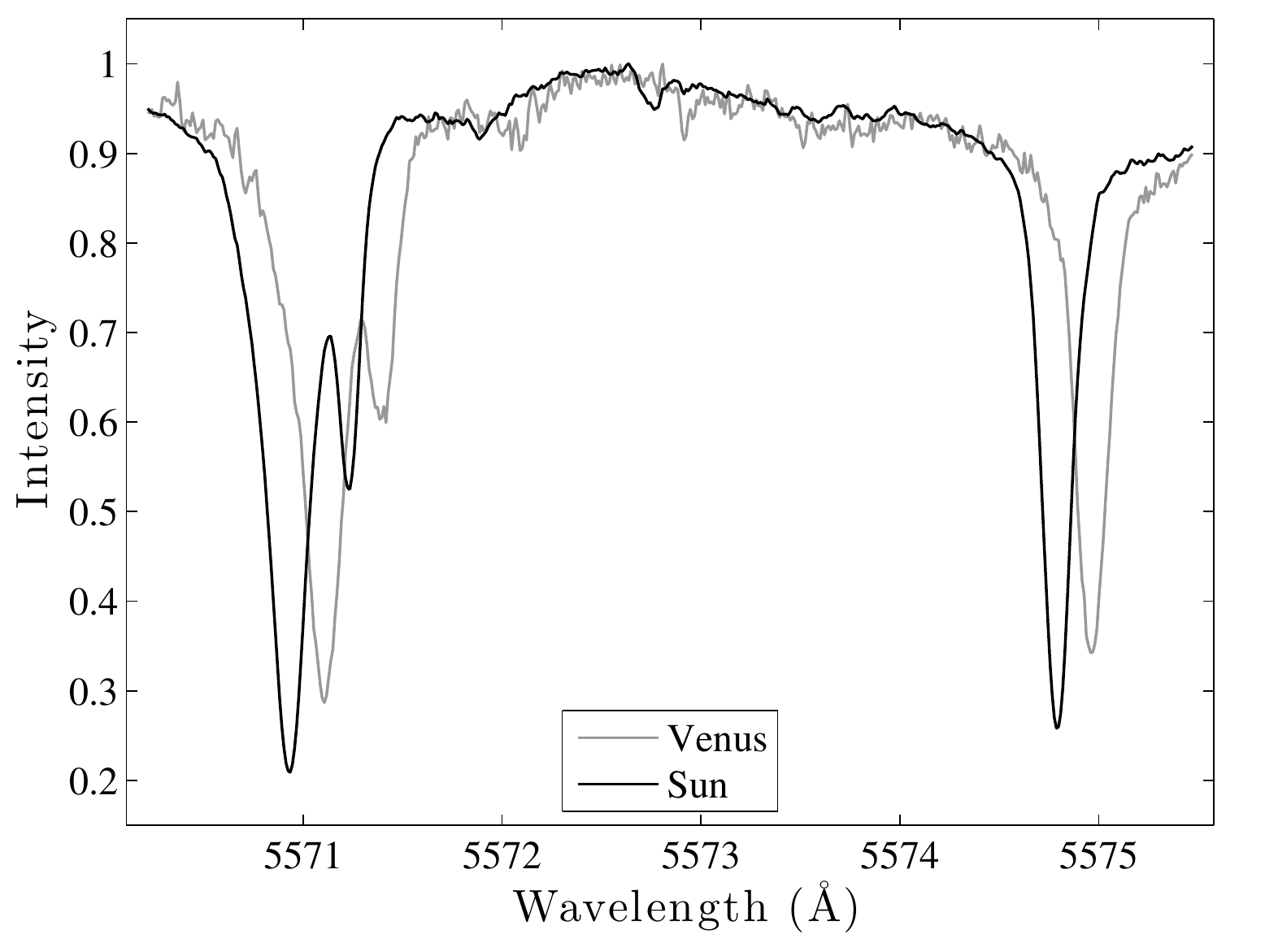}
\includegraphics[width=7cm]{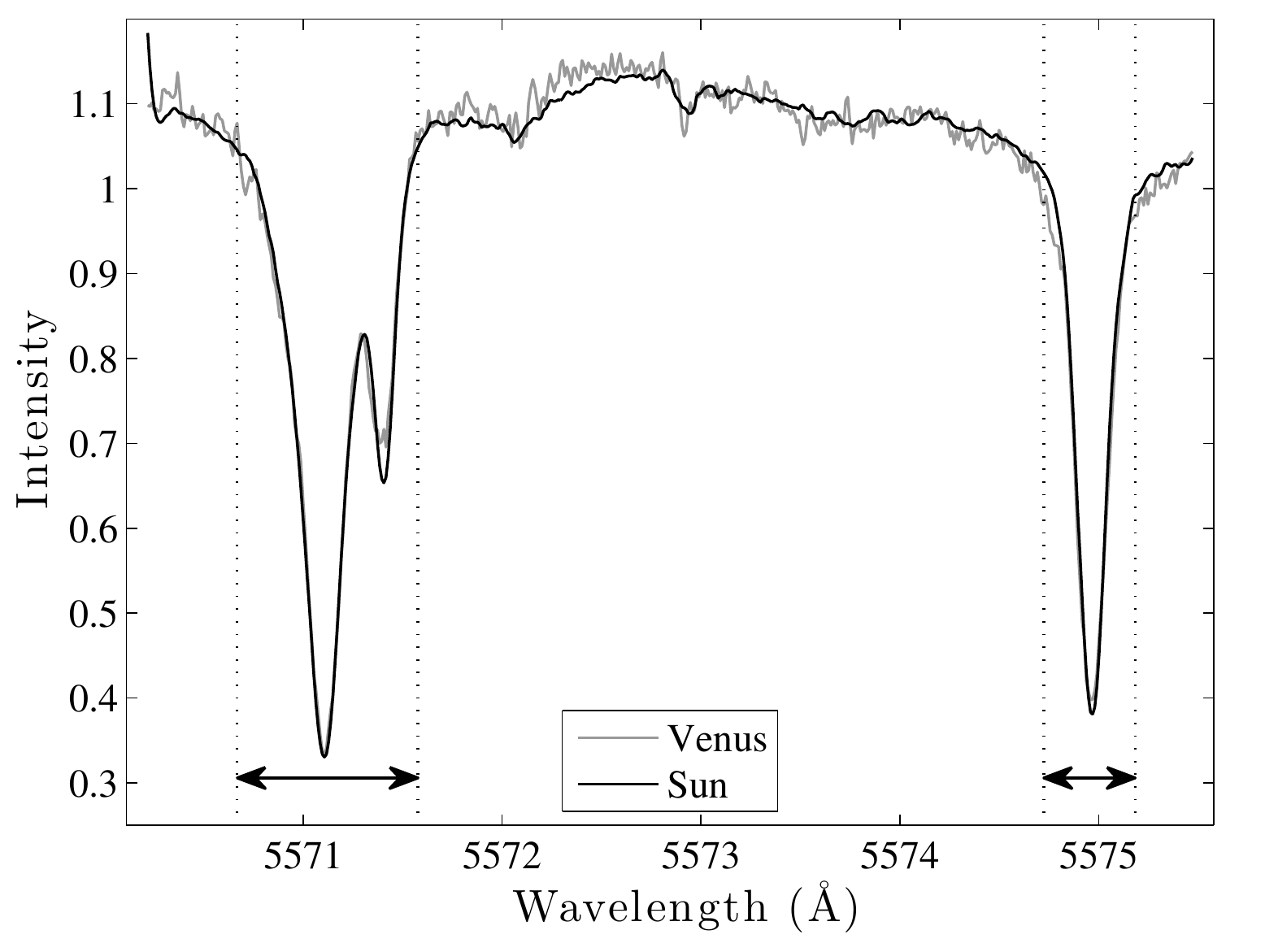}
\caption{Two-step process used to measure Doppler shifts. Top panel: the reference spectrum is the average of 400 spectra obtained on the Sun, while spectrum of Venus corresponds to a single location on Venus disk. Bottom panel: the same spectra after shifting the reference spectrum at the same location of Venus and after normalizing them. The wavelength ranges indicated by two arrows and vertical dotted lines indicate the selected areas where cross-correlation is applied (see text). Spectra were obtained on September 14, 2009. }
\label{fig_doppler_estimator}
\end{figure}
\begin{table*}[t]
\small
 \caption{Observation properties of Venus during the 2009 campaign at Teide Observatory. Columns 1 to 3: observation days, start, and end dates. Start and end times are in modified Julian date (MJD), i.e., Julian date JD - 2,400,000.5 days. Columns 4 to 10: ephemerides from the IMCCE database. Ephemerides are calculated every day of both campaign at 06:00:00 UTC. The last two columns indicate the number of scans done each day and the atmospheric seeing. 
}
\center
\begin{tabular}{l c c c c c c c c c c c }\hline
 Date &    Start & End & R.A          & Dec.           & Dist.     & $V$     & Phase  &Elong.     &$R$    &Scan \# & Seeing \\
           & [MJD]  & [MJD] & [ h  m  s]     & [o  '  "]         & [AU]     &[km s$^{-1}$] & [$^\circ$] &[$^\circ$] &[arcsec] &    & [arcsec]\\
\hline
2009.09.14 & 55088.300350 & 55088.633799 & 9 39 05     & 14 44 35   & 1.403    & 8.688 & 43.10   & 29.12   & 5.95   & 12        & 2.2-3.5\\
2009.09.16 & 55090.359586 & 55090.445014 & 9 48 40     & 14 01 20   & 1.413    & 8.529 & 42.29   & 28.64   & 5.91   & 2          & 3.0-3.7\\
2009.09.17 & 55091.357399 & 55091.502237 & 9 53 26     & 13 39 05   & 1.418    & 8.449 & 41.89   & 28.40   & 5.88   & 6          & 3.1-5.2\\
\hline
\end{tabular}
\label{tab_obs_cond}
\end{table*}

Measuring absolute RVs on any astronomical object is very challenging, and we seek these RVs for tens of m s$^{-1}$. This has been discussed in the case of Venus wind measurements in \citet{Young_1979} and \citet{Widemann_2007, Widemann_2008} with attempts of making absolute RV measurements using visible lines. These authors concluded the need of a reference point on Venus that would serve as a relative velocity reference, and they used such a point to perform differential velocity measurements on the disk. 

In our case, we measured Doppler shifts by comparing the position of a given spectrum on Venus with a high S/N reference spectrum, which was not one of Venus spectra. Instead, the reference spectrum is the mean solar spectrum used to build a reference flat field. Indeed, the solar spectrum scattered by Venus atmosphere is the result of integrating all upwelling and downwelling flows on the solar surface. The average of 400 spectra randomly located on the solar disk makes it representative of what Venus receives and reflects, and has the advantage of displaying a much larger S/N ($\approx 500$) than any individual spectrum obtained on Venus. With that method, the estimated shifts need to be corrected from the motion of Venus with respect to observer and the motion of the observer with respect to the Sun. All these components are well known and can be subtracted with the help of ephemeris data retrieved from the website of the \textit{Institut de M\'ecanique C\'eleste et de Calcul des Eph\'emerides} (http://www.imcce.fr). The RV differences between Venus and the Sun with respect to the Earth were about 3 and 9 km s$^{-1}$ in 2008 and 2009, respectively, i.e.,  0.06 and 0.18 \AA\  in the mid-visible. We note that a typical Fraunhofer line is about 0.3 \AA\ wide at half maximum. 

We employed a two-step process (Fig. \ref{fig_doppler_estimator}) to estimate the Doppler shift. Let us consider a spectrum obtained on Venus. We first computed the cross-correlation of the Venus spectrum with the reference solar spectrum. To give more weight to the spectral ranges that are sensitive to Doppler shifts, i.e., those with the steepest spectral slope, we multiplied each spectrum by the absolute value of its derivative before computing the cross-correlation. Then, we interpolated the reference spectrum on a grid shifted by this first estimate of the Doppler shift. We note that we interpolated the reference spectrum and \textit{not} the Venus spectrum because it shows a good enough S/N to not be altered by interpolation. Interpolation was performed with a spline algorithm.

A second step is necessary because even though we give more weight to the sensitive parts of the spectral range, the contribution of the noisy continuum still alters the measurement. Now that the two spectra are almost overlapping, we computed the cross-correlation of the two spectra on the cores of the Fraunhofer lines, instead of the complete spectral range. In the case of the Fe doublet, we selected two regions on the detector (Fig. \ref{fig_doppler_estimator}), then cross-correlated the spectra on each region, and took the average. 

We note that we considered using the Connes method to measure the Doppler shifts, which is commonly used in exoplanetary science \citep{Connes_1985}. However, this method is best suited when working with thousands of lines, while we are working with a maximum of nine lines. We still tested it instead of performing a second cross-correlation on the core of the lines but the results were clearly noisier, as seen by the presence of many outliers.

Last but not least, we estimated the error on Doppler velocity values from a mix of measurements and simulated data. For a given spectral range (Mg, Fe, or Na), we first built a model spectrum from the solar spectrum used for making the flat fields, which we smoothed over three spectral bins. Secondly, we simulated 100,000 of simulated spectra with a given S/N (either 10, 50, 100, or 1000) based on our model spectrum and a normally distributed noise, with a given Doppler shift (either 10, 50, 100, 200, or 1000 m s$^{-1}$). For each simulated spectrum, we ran our Doppler shift estimation routine. The histograms of the estimated Doppler shifts with respect to their exact values gave us an estimate of the measurement error. We observed that the error was independent from the absolute value of the Doppler shift and is -- as expected -- a linear function of the inverse of the S/N. For a S/N of 100, the average error on Doppler velocity estimates was written as
\begin{eqnarray}
\sigma\ind{v, Mg, ref} &=& 17\ \mbox{m s}^{-1} \\
\sigma\ind{v, Fe, ref} &=& 18\ \mbox{m s}^{-1} \\
\sigma\ind{v, Na, ref} &=& 32\ \mbox{m s}^{-1} 
.\end{eqnarray}
for the three spectral ranges we consider in this paper. Then, for each actual spectrum, we measured its S/N from the standard deviation of the data minus the model spectrum in the continuum regions, and we retrieved the velocity error by comparing the S/N to the reference simulated values, i.e.,
\begin{equation}
\sigma\ind{v} = \sigma\ind{v,ref}\ \frac{100}{\mathrm{S/N}}
\end{equation}
For the data taken on September 17, 2009, which include Mg, Fe, and Na spectra, the velocity error $\sigma\ind{v, avg}$ on an average map is written as\begin{equation}
\sigma\ind{v, avg} = \frac{1}{3}\ \sqrt{\sigma\ind{v, Mg}^2 + \sigma\ind{v, Fe}^2 + \sigma\ind{v, Na}^2},
\end{equation}
where $\sigma\ind{v, Mg}$, $\sigma\ind{v, Fe}$, and $\sigma\ind{v, Na}$ are the error maps in each spectral band.

\section{Observations}
\label{sect_obs}

\subsection{Observation setting}
As mentioned in the introduction, we were awarded observing time from May 2 to May 8, 2008 to extend the test reported in G08. The planet diameter was about 9.8 arcsec and the phase angle $\phi \approx 13^\circ$. This campaign ended up being a test campaign as it put into light some difficulties in reaching our goal, as we identified a spurious Doppler shift in the measurements that was not from astrophysical origin. For this run, we opted for an exclusive west-east (WE) scanning of the planet in the Venus coordinate frame, i.e., slit parallel to rotation axis, which was repeated over and over as long as the weather permitted.  The reason to choose a WE scanning was the possibility to construct a map easily from individual spectral images. By building a map, we actually mean two maps: a ``photometric'' image and an RV image of Venus. We intend  by photometric maps an image of Venus in the visible, but we do not aim to quantify the exact photon flux. 

With a WE scan, the photometric map can be obtained in a straightforward manner. Firstly, at each position of the slit, we considered the mean spectral image (result of averaging ten images),  which we projected along the $x$-axis to get a 1D vector that contains the  north-south (NS) photometric profile at the slit position along the equator. Knowing that slit positions are spaced by 0.8 arcsec, we just placed each vector one after the other into a 2D table and produced a map. However, this is not enough because the slit sometimes moves along the $y$-axis in between two positions, introducing vertical shifts on the detector. The fact of having a WE scan makes the vertical adjustment easy because all photometric profiles must be symmetrically centered on the apparent equator of Venus. Both the photometric and RV maps were then interpolated according to the shift measured from the photometric profiles. However, this choice, which was motivated by its simplicity for reconstructing the map, ended up being terrible. Indeed, the RV measurements appeared to be dominated by a signal that was not related to Venus but to the instrument configuration. The spectrometer had never been tested for measuring such small RV shifts, so the drift was a surprise.  We had no way to a posteriori disentangle the instrumental drift from the Venus wind circulation because both the instrumental drift and the zonal circulation were oriented along the WE direction. This is why we rotated the slit by 90$^\circ$ and scanned Venus along the NS direction during the following campaign.

The second campaign occurred from September 11 to 17, 2009, when Venus was displaying a phase of about 43$^\circ$ and an apparent diameter of 11.9 arcsec. The weather allowed us to scan Venus 12, 2, and 6 times on September 14, 16, and 17, respectively (Table \ref{tab_obs_cond}). On September 14 and 16, only spectra of the Fe doublet were taken, while spectra in the three ranges were taken the other day. In addition to scanning the planet along the NS direction we also scanned it along the reverse direction (SN) to help characterize the RV bias induced by the spectrometer. If truly deterministic, averaging the NS and SN maps would have allowed us to get rid of the bias without needing to model it. 

\subsection{Long path to assemble radial velocity maps}
\label{sect_RV_maps}
\begin{figure}[t!]
\center
\includegraphics[width=8cm]{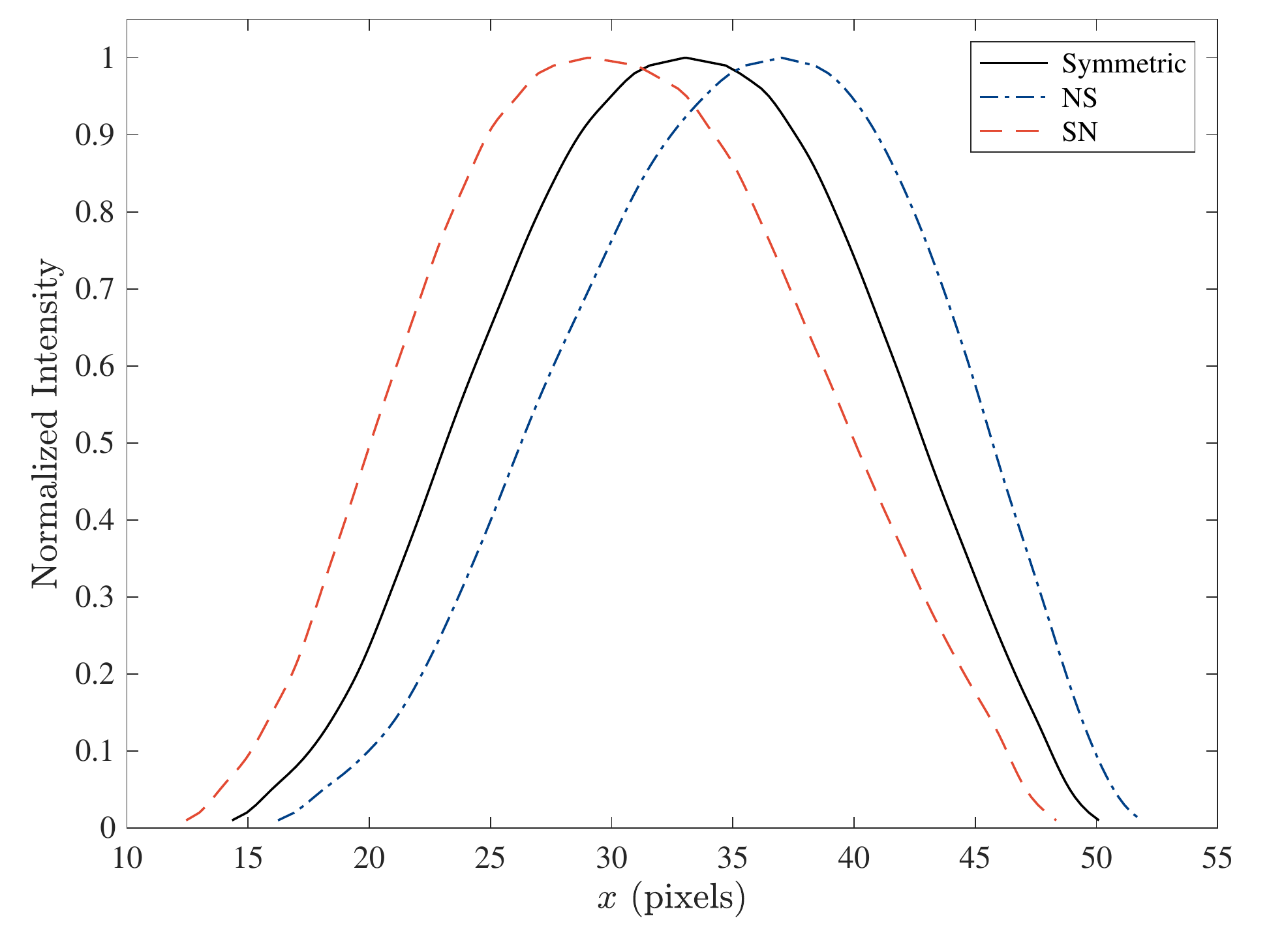}
\caption{Mean photometric profile from September 14, 2009 data. A so-called photometric profile is the sum of each photometric map projected along the $y$-axis on the $x$-axis.  Intensity was normalized such as maximum value is 1. The ``NS'' and ``SN'' photometric profiles are the average profiles for each scanning direction (NS or SN). The symmetric profile is the bisector of both profiles.  }
\label{fig_photo_profile}
\end{figure}
\begin{figure}[t]
\center
\includegraphics[width=7cm]{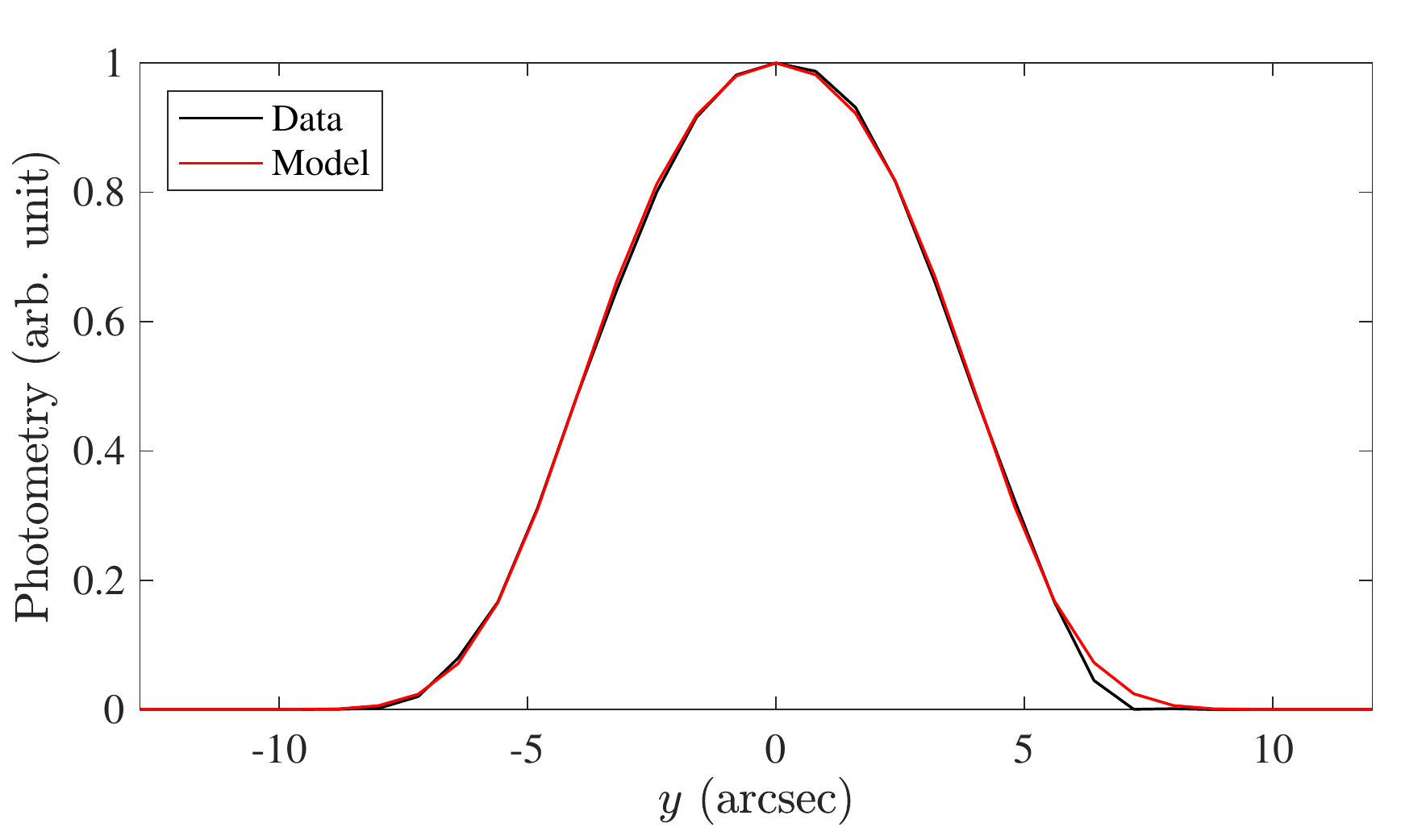}
\includegraphics[width=7cm]{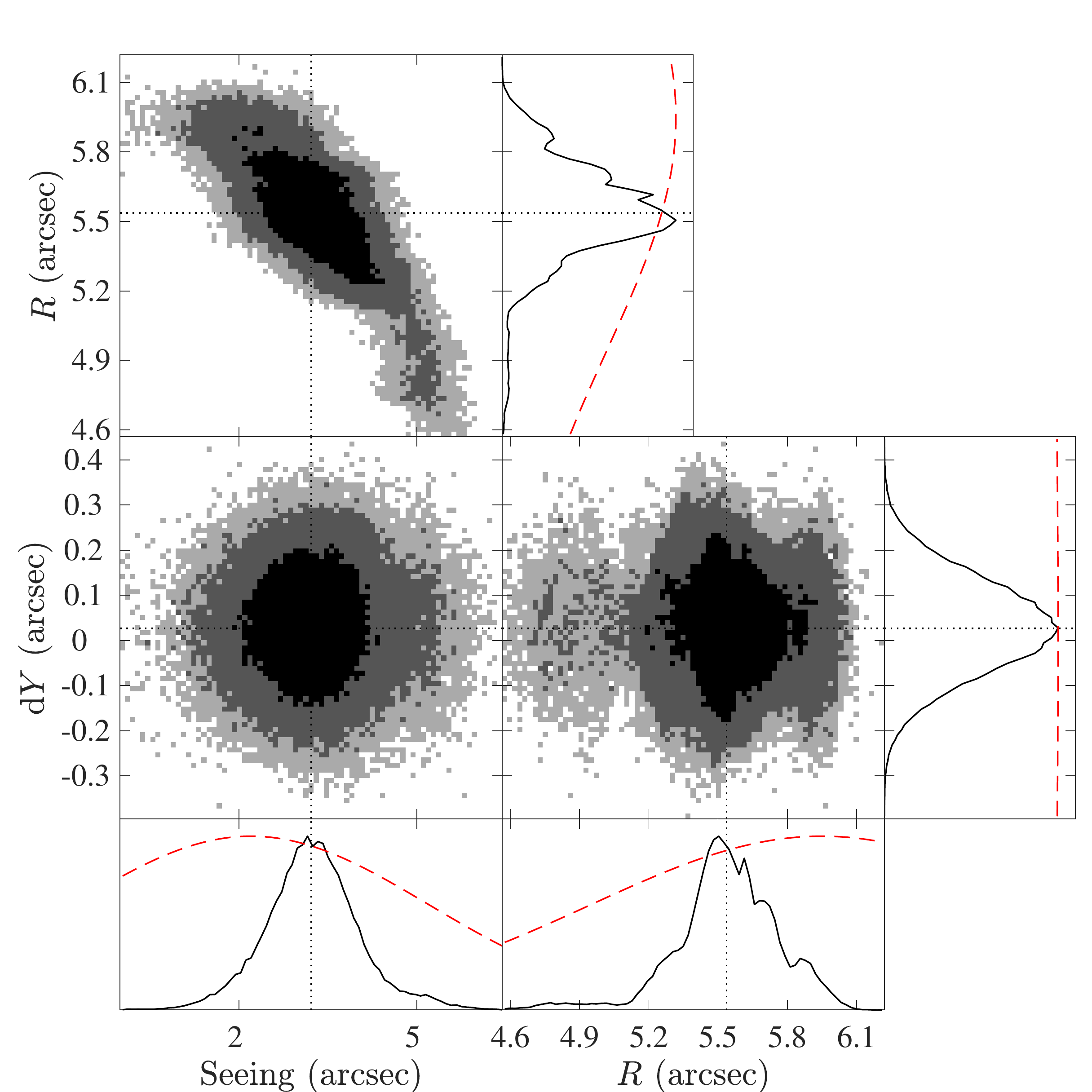}
\caption{Modeling Venus photometric profile with a Lambert law. Top panel: the black line is the mean photometric profile of Venus along the NS direction, and the red line is the best-fit profile. Bottom panel: the above fitting was obtained by maximizing the likelihood of a Lambert-sphere model  to the THEMIS observations with an MCMC algorithm.  Each subpanel represents the two-parameter joint posterior distributions of all free parameters. The 68.3\,\%, 95.5\,\%, and 99.7\,\% confidence regions are denoted by three different gray levels. The PDFs are plotted at the bottom and right of the panels. The dotted lines indicate the maximum probability value of the PDF of each parameter. After $500,000$ iterations, the fitted values are $3.21_{-0.78}^{+0.67}$ arcsec for the seeing, $5.54_{-0.22}{^+0.17}$ arcsec for the apparent radius, and $0.03_{-0.11}^{+0.10}$ arcsec for the planet decentering with respect to the center of the detector. The red curve on the top panel indicates the result of the fitted values.}
\label{fig_MCMC_fit_NS}
\end{figure}
The drawback of the NS/SN scanning approach is a challenging map construction. With a WE scan, the middle of the photometric profile at each 
position of the slit corresponds to the equator. With an NS scan, the middle of the photometric profile has no specific meaning for a planet that is not observed at opposition. It is even impossible to connect the location of the maximum of intensity to a given longitude, as its position is a function of the atmospheric seeing, which we do not know \citep{Gaulme_2018}. To be able to associate a latitude and a longitude on Venus to a given position along the slit, we must consider that the photometric profile is the result of the actual photometric profile, the latitude on the planet, the pixel FOV, and the atmospheric seeing. We note that for our specific instrumental configuration, the pixel FOV was known with an accuracy of about 5\,\%, which we aimed at refining for accurate wind measurements. If we had a complete photometric map of the planet, we could estimate both the atmospheric seeing and the apparent diameter (i.e., pixel FOV), by fitting a photometric profile degraded by a PSF, assuming the true photometric limb-darkening profile to be known.

Unfortunately, we do not have yet a photometric map and we need to know the seeing and the pixel FOV to build it. Where to start then? The first observational input consists of computing the photometric profile along the NS and the SN directions, directly from the data. For each position of the slit, we already have a photometric profile along the EW direction. If we collapse these photometric profiles, we get a total intensity as a function of the projected latitude. We can retrieve a proxy of the pixel FOV and the atmospheric seeing from this simple profile. In principle, modeling this profile should be enough to extract the parameters.

However, our planned processing strategy needed to be tuned because of an unexpected issue. In Fig. \ref{fig_photo_profile}, we plot the NS and SN photometric profiles of Venus, which are the average of the 12 scans done on September 14, 2009. They appear to be both asymmetric and in opposite directions, which means that the interval between each position of the slit increase as we move it. The farther the slit is from its original position, the larger is the step. This stretching along the NS direction thus prevents us from simply modeling the photometric profile and recombining NS and SN maps. We first need to symmetrize the photometric profile and then interpolate the maps on an evenly sampled grid of slit position. We compute the median photometric profile as the bisector of the NS and SN profiles (plain black line on Fig. \ref{fig_photo_profile}), which is symmetrical and we can model as the result of a limb-darkening law modulated by atmospheric seeing and pixel FOV. We note that the atmospheric seeing was not monitored by any device at the observatory, especially because most observing time was during daytime.

\begin{figure}[t!]
\center
\includegraphics[width=7cm]{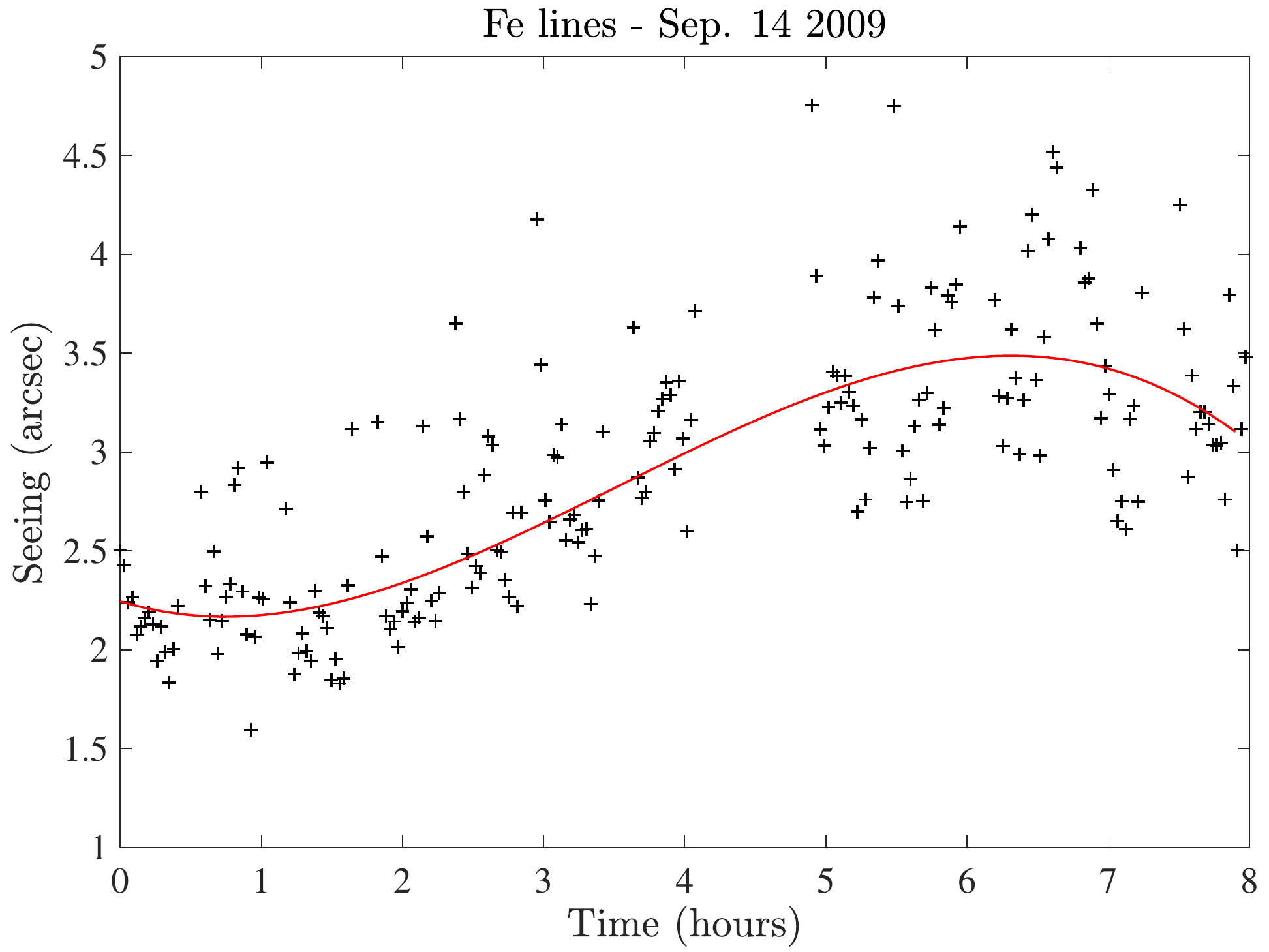}
\includegraphics[width=7cm]{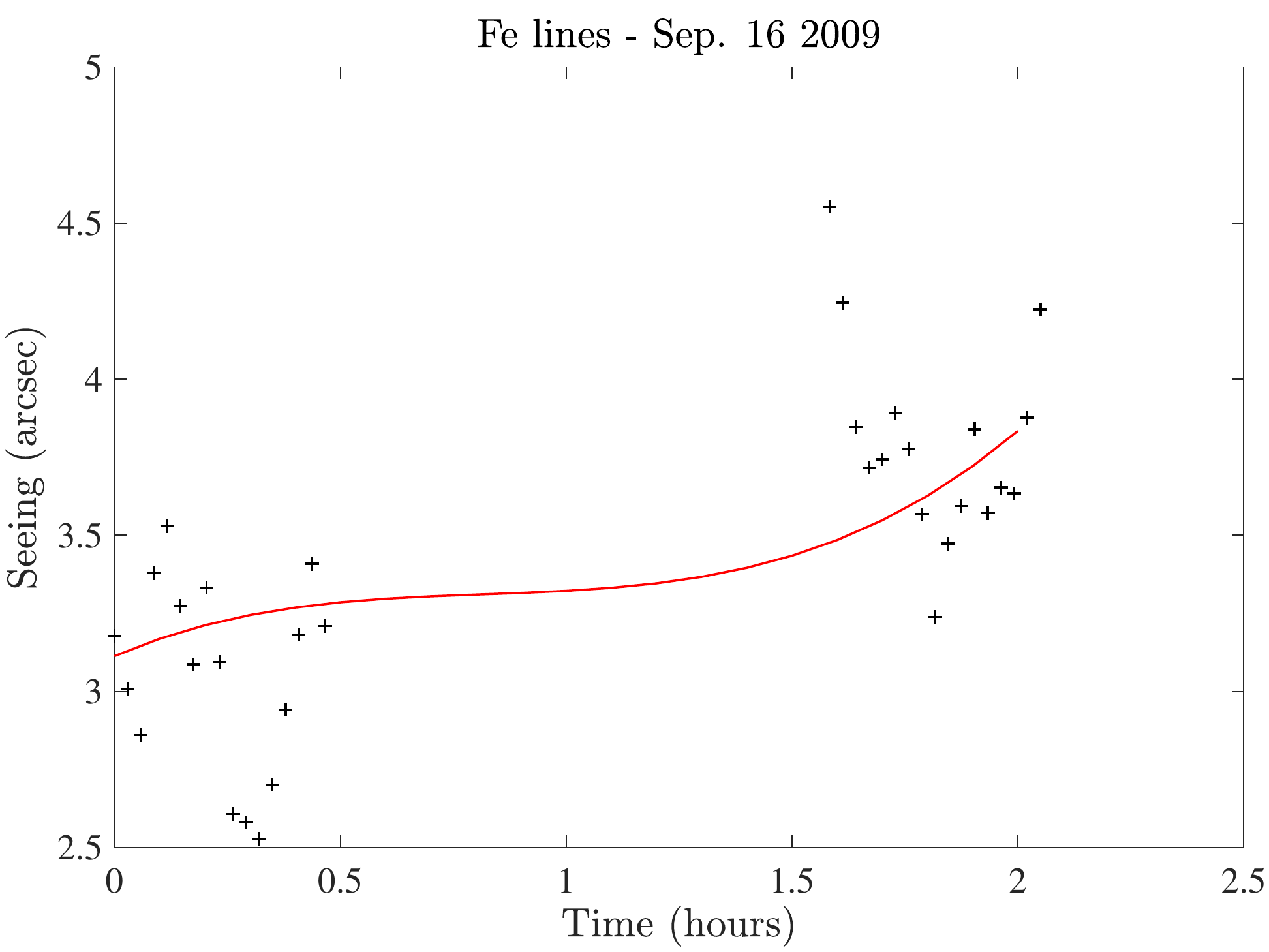}
\includegraphics[width=7cm]{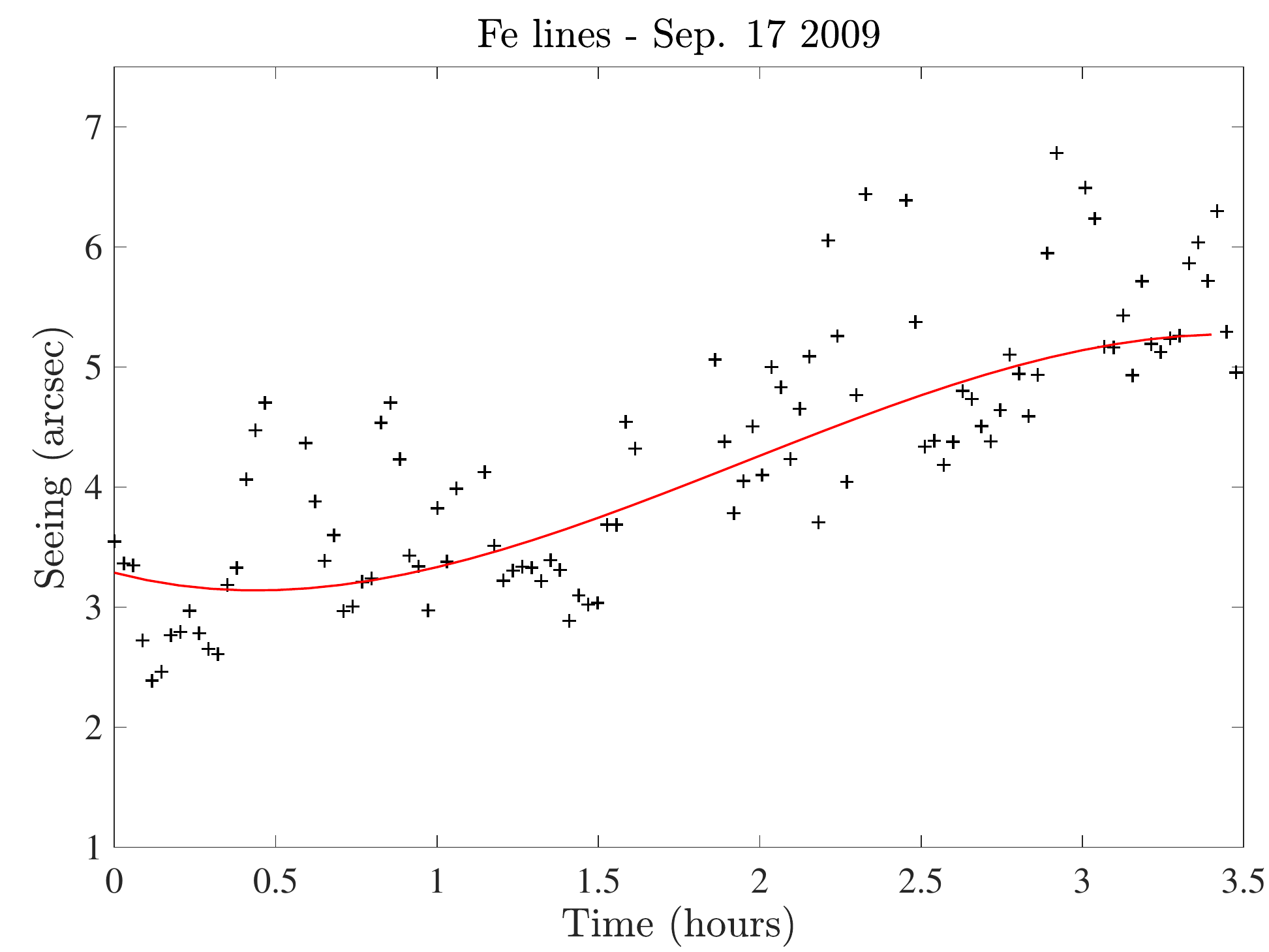}
\caption{Atmospheric seeing measured by fitting a Lambert limb-darkening law degraded by atmospheric seeing, as function of time. Time origin is set at the first exposure of the day. The red curve indicates a fit of it with a third order polynomial.}
\label{fig_seeing}
\end{figure}

\begin{figure}[t!]
\center
\includegraphics[width=8cm]{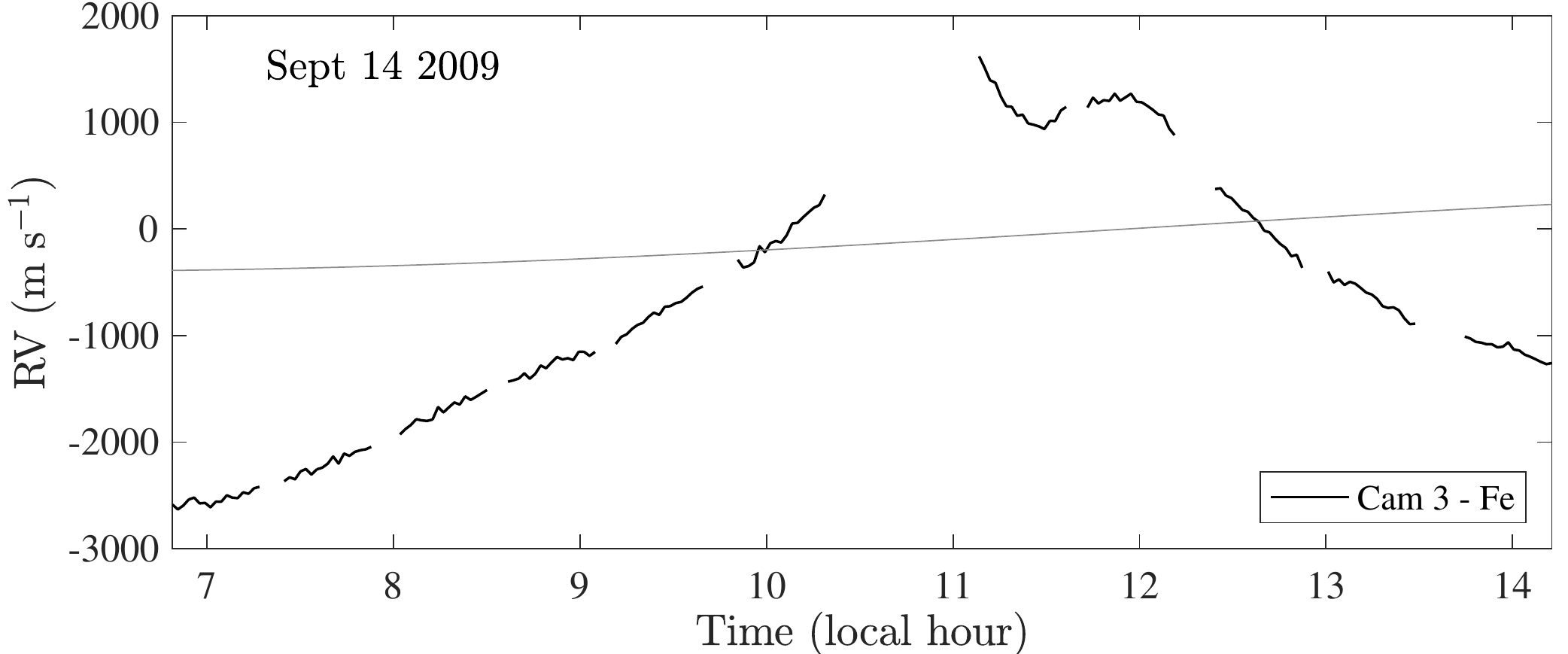}
\includegraphics[width=8cm]{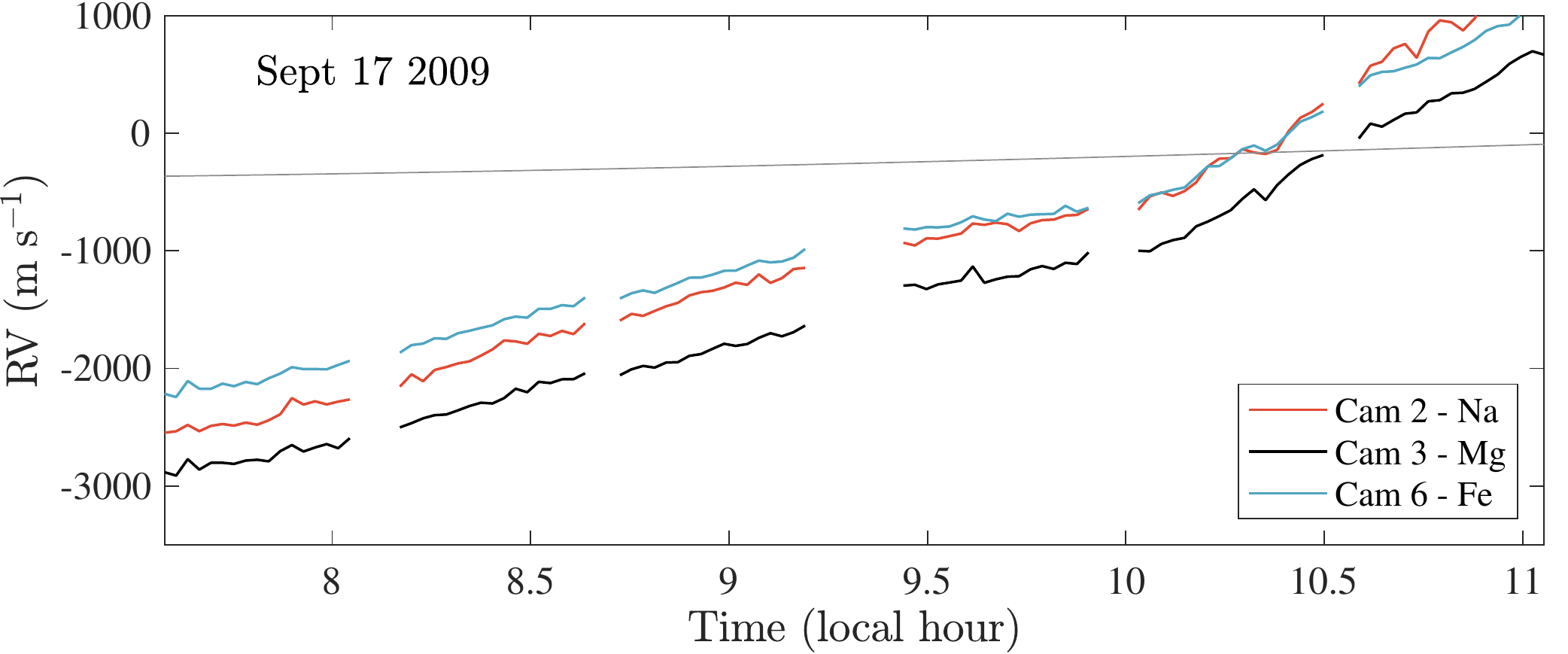}
\caption{Mean RV of the Fraunhofer lines scattered by the Earth's sky as functions of time. Each chunk (separated by empty spaces) corresponds to a scan. The first scan taken on September 14, 2009 is not shown as it was taken before sunrise. In both plots, the thin gray line represents the theoretical variation of the sky-scattered solar lines according to the ephemeris (Earth's rotation essentially). The abbreviation ``cam'' refers to the CCD camera's number.}
\label{fig_drifts_2009}
\end{figure}

\begin{figure}[t!]
\center
\includegraphics[width=8cm]{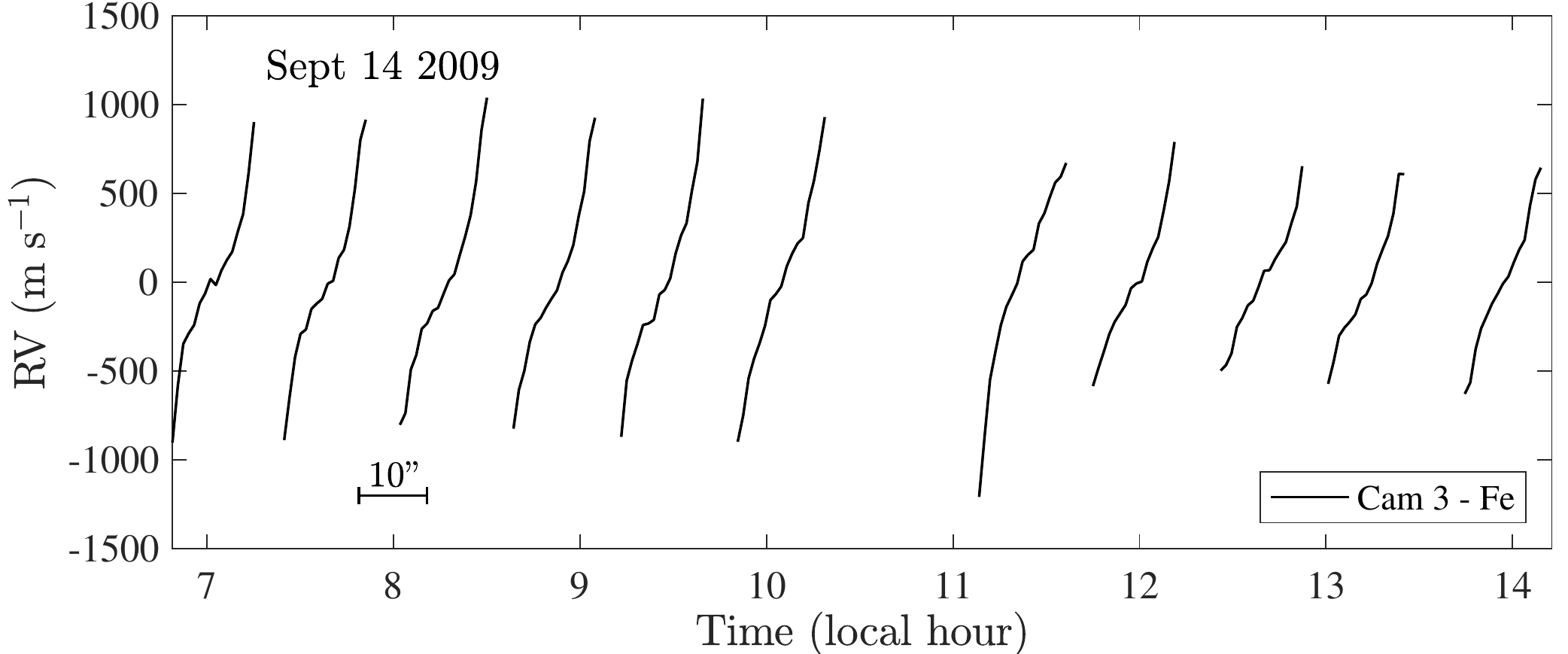}
\includegraphics[width=8cm]{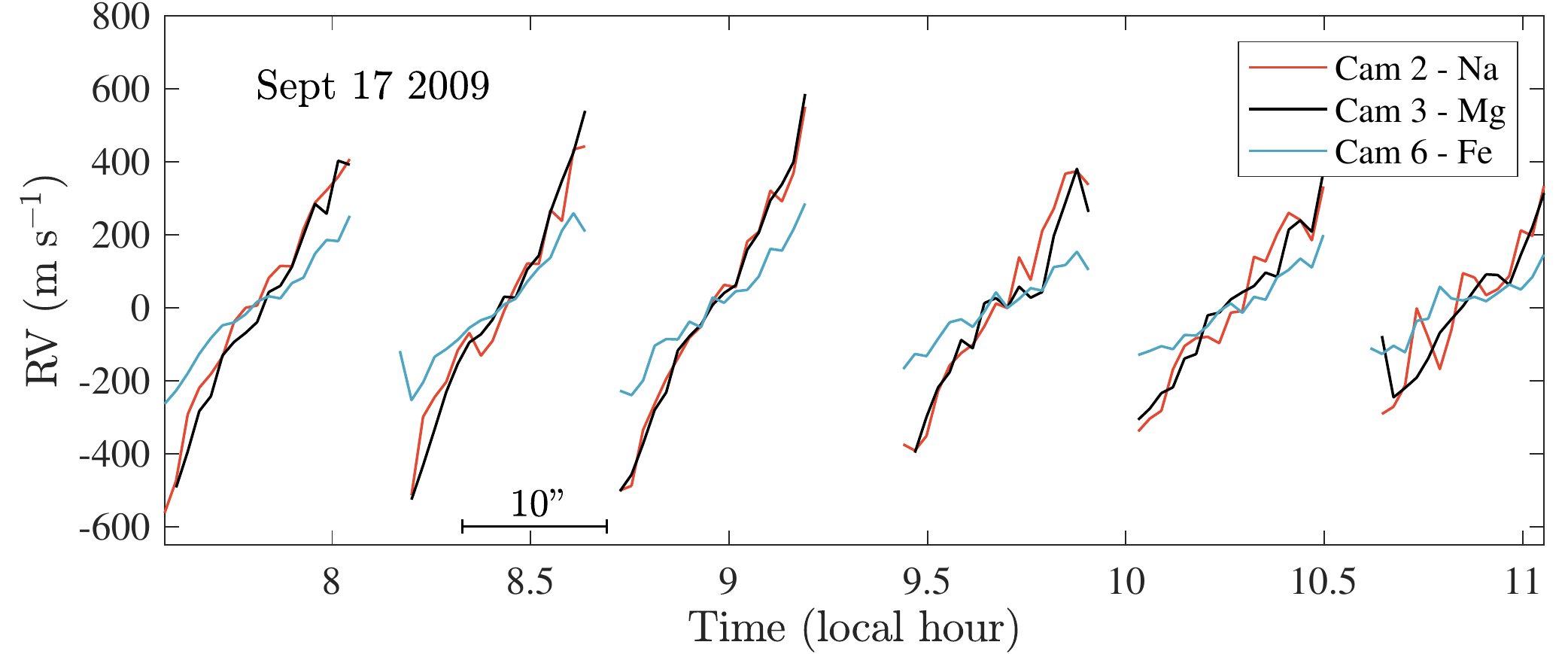}
\caption{Difference of RVs in between Venus and the Earth's sky. The horizontal line on the bottom panel corresponds to a spatial coverage of 10 arcsec. The abbreviation ``cam'' refers to the CCD camera's number. }
\label{fig_drifts_venus_2009}
\end{figure}

\begin{figure}[h!]
\includegraphics[width=8.5cm]{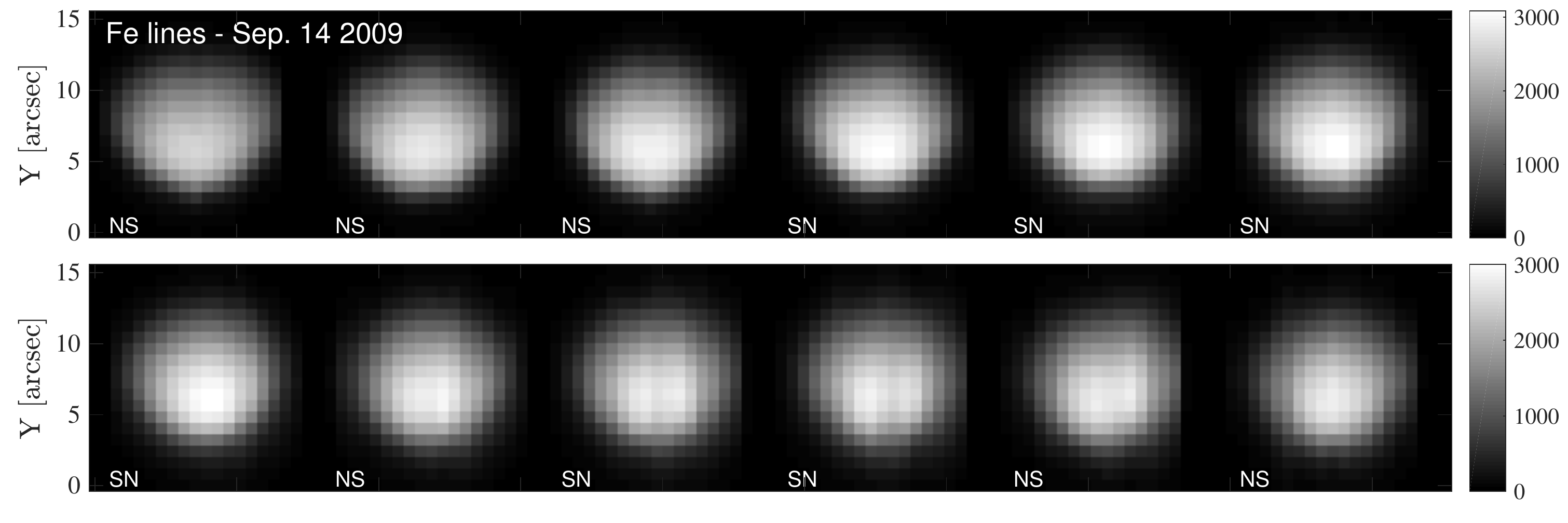}
\includegraphics[width=8.5cm]{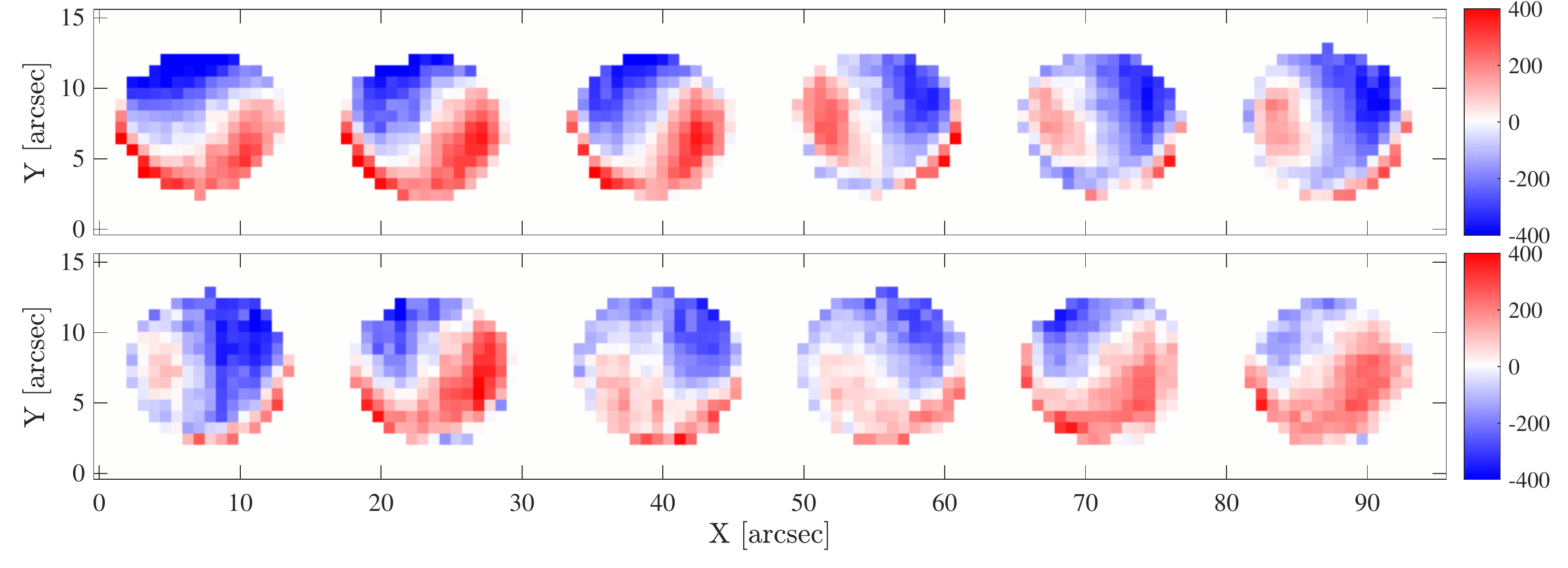}
\includegraphics[width=8.5cm]{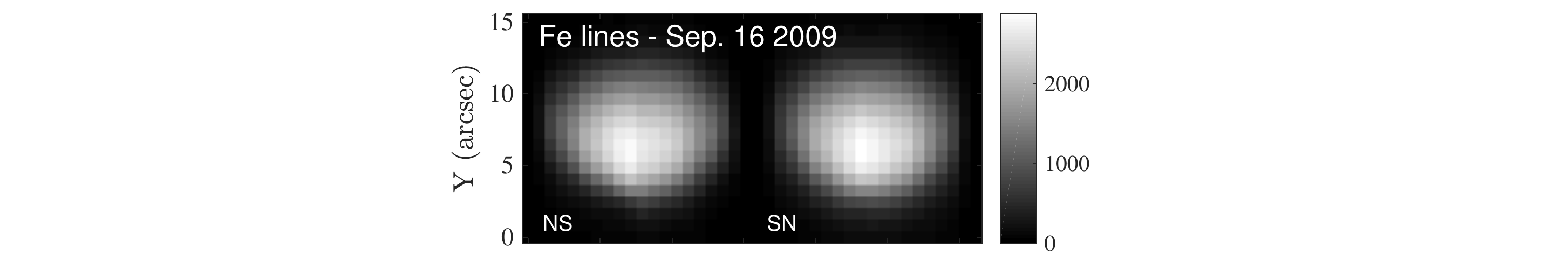}
\includegraphics[width=8.5cm]{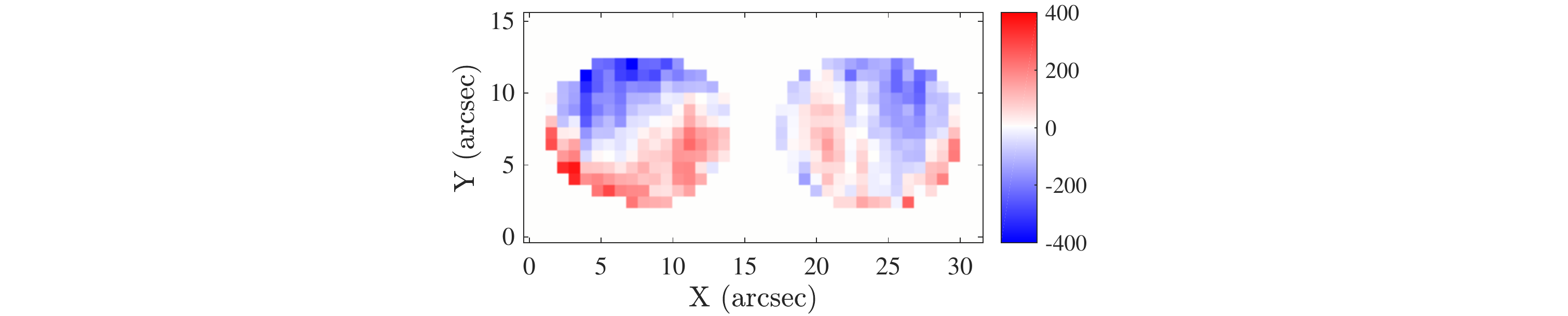}
\includegraphics[width=8.5cm]{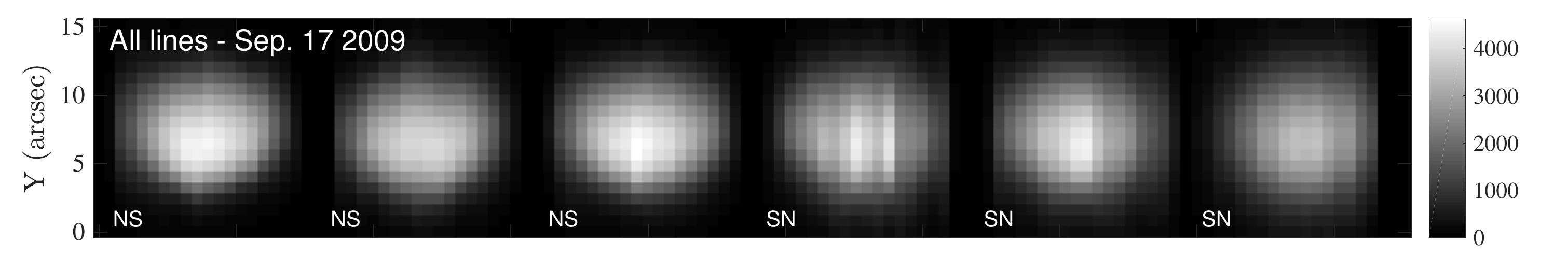}
\includegraphics[width=8.5cm]{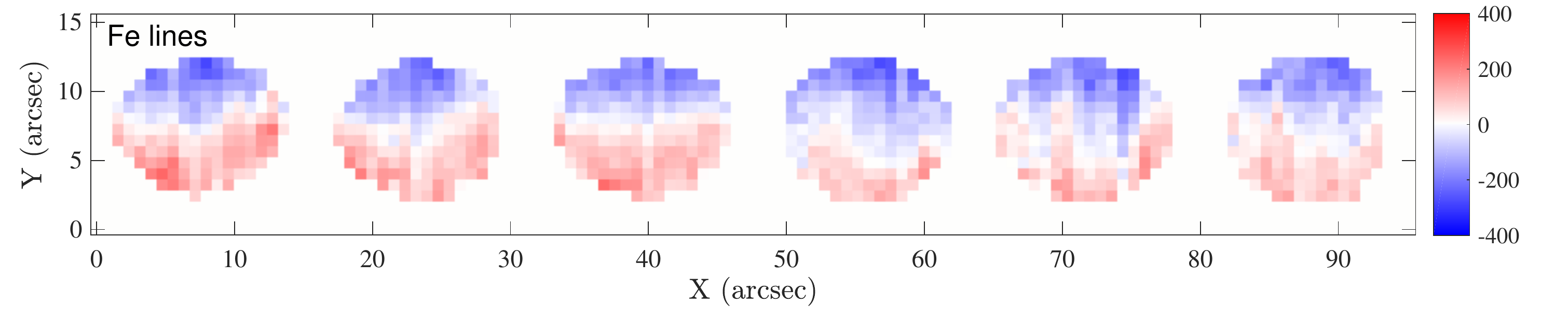}
\includegraphics[width=8.5cm]{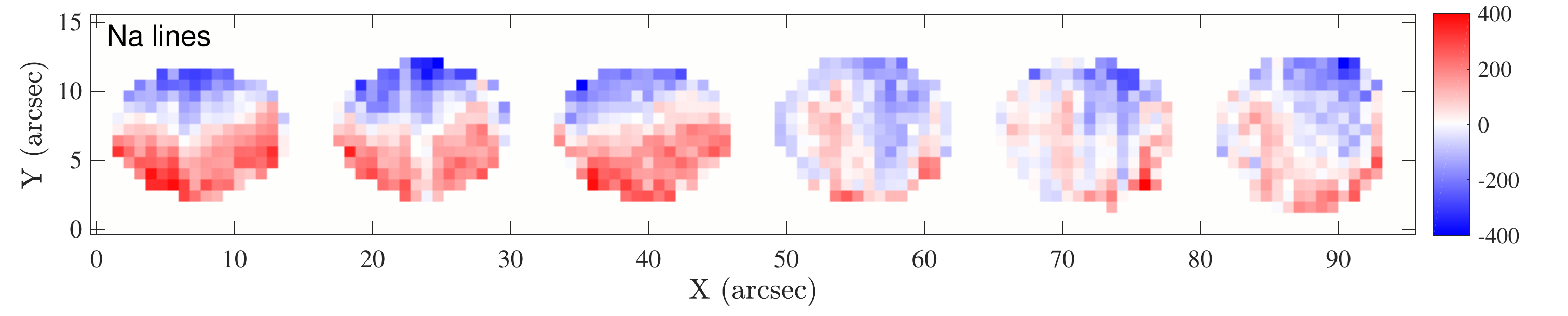}
\includegraphics[width=8.5cm]{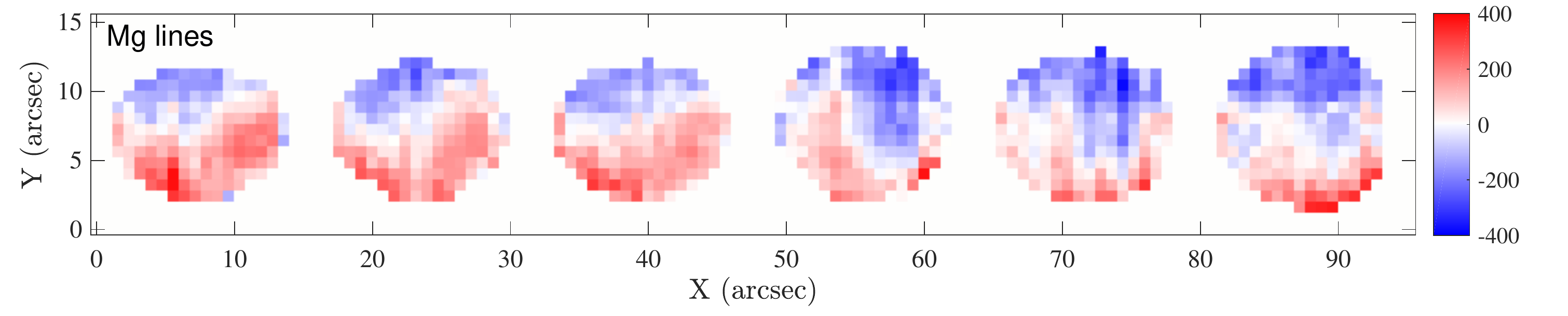}
\caption{Photometric and RV maps for September 14, 16, and 17, 2009. Data were taken on Fe lines only on September 14 and 16, and on Fe, Na, and Mg lines on September 17. Each RV map was ``flattened'' by subtracting a fitted plane along the NS direction, i.e., $x$-axis. Photometric intensity is expressed in ADU and RV in m s$^{-1}$.}
\label{fig_scans_2009}
\end{figure}


\begin{figure*}[ht!]
\center
\includegraphics[height=4.1cm]{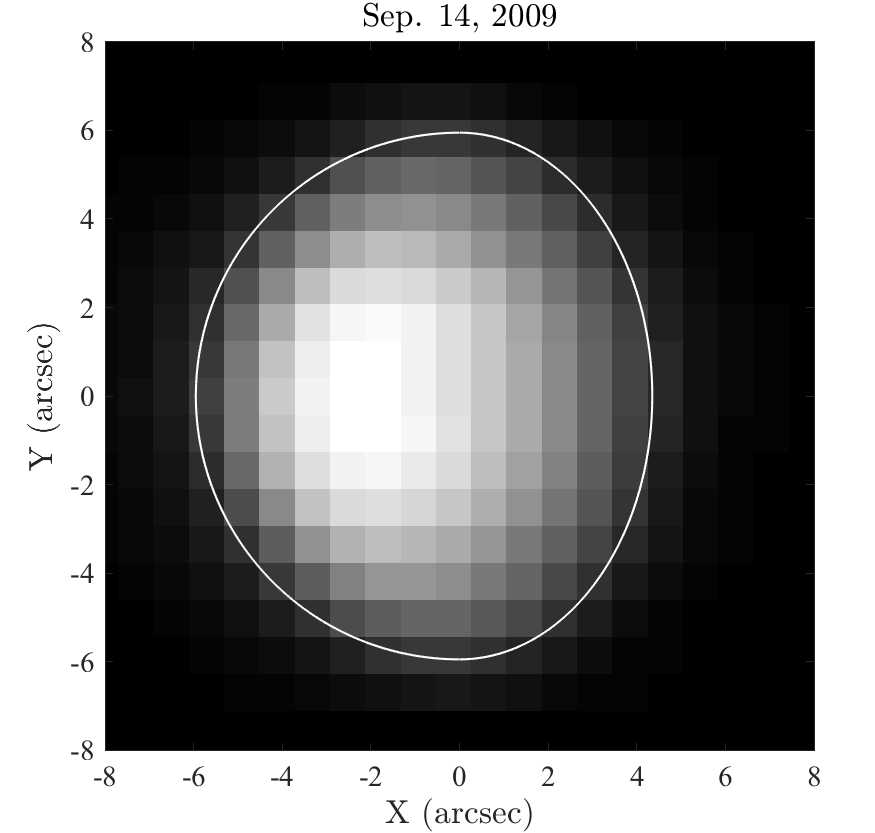}\includegraphics[height=4.1cm]{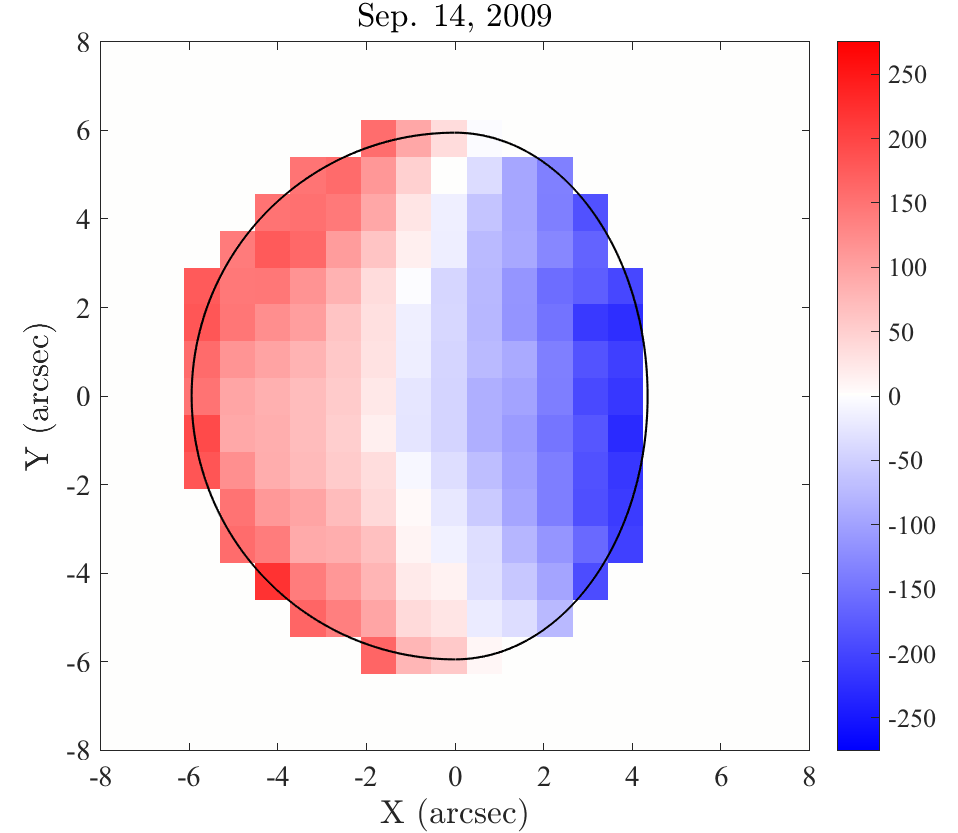}\includegraphics[height=4.1cm]{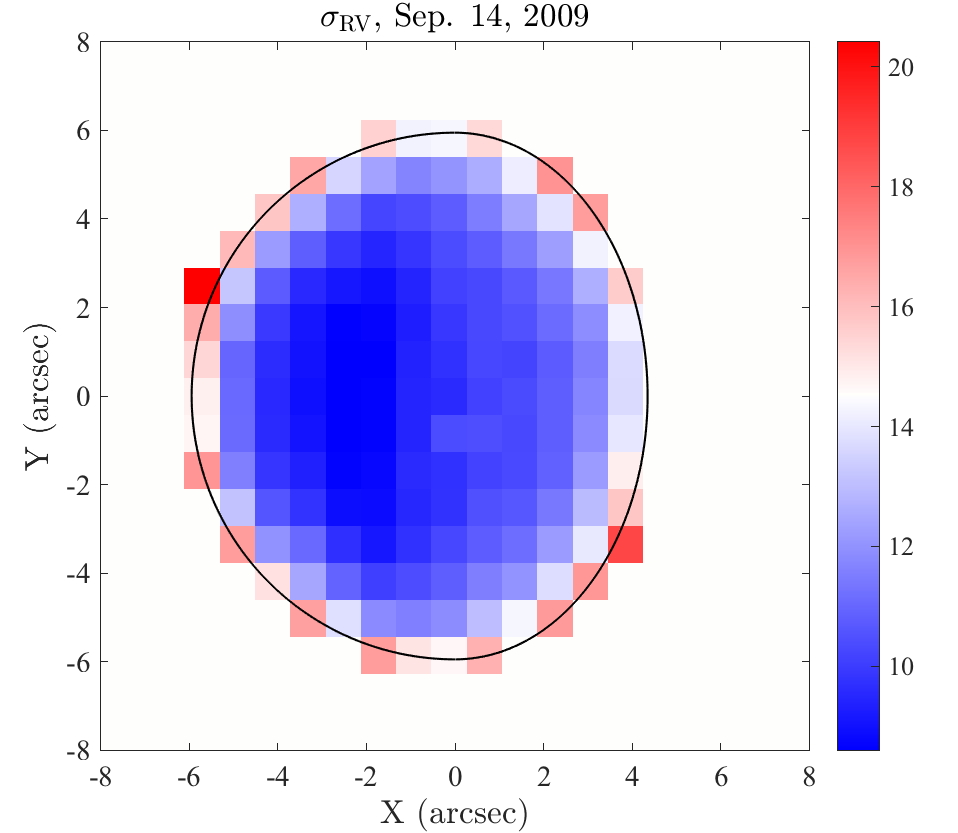}\\
\includegraphics[height=4.1cm]{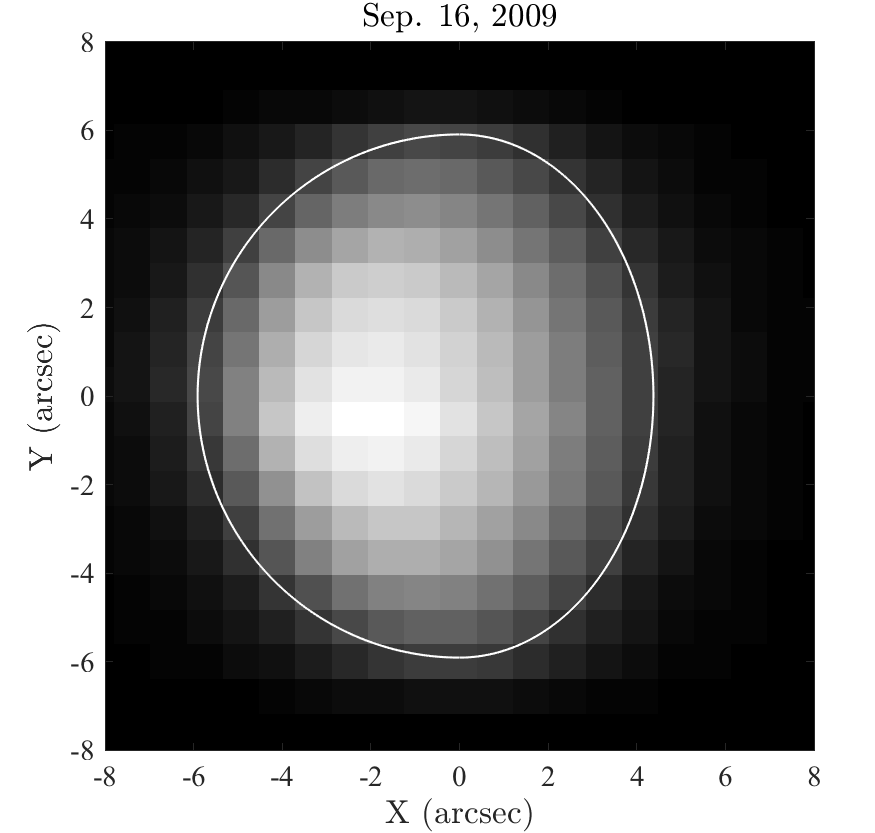}\includegraphics[height=4.1cm]{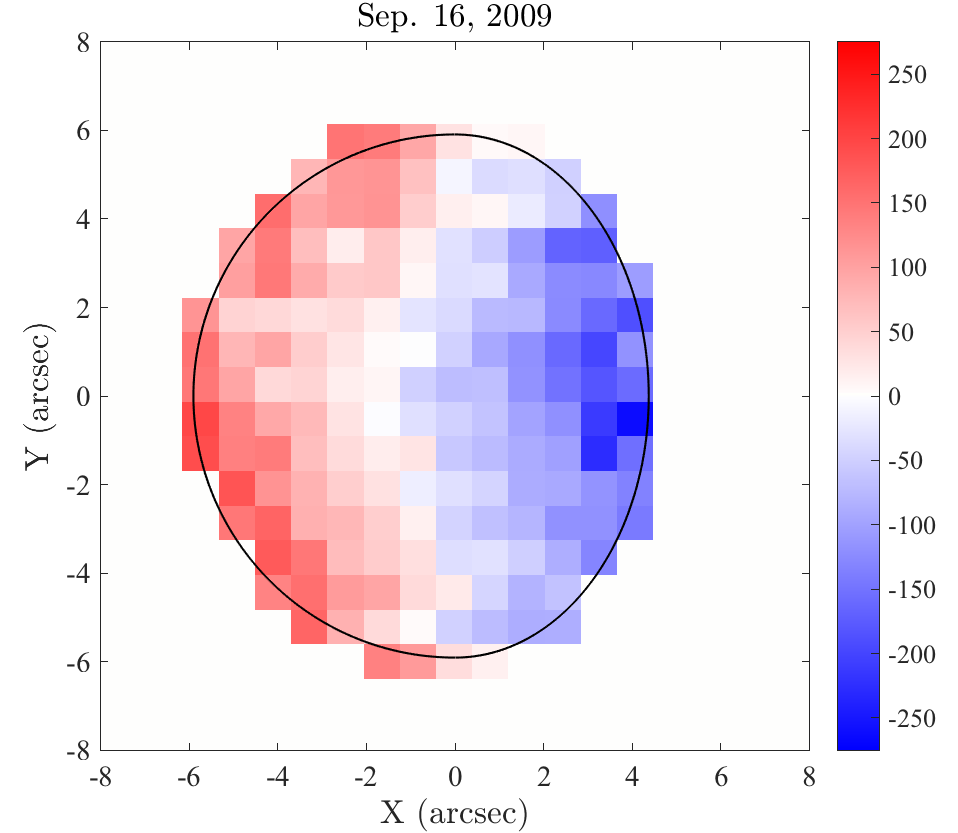}\includegraphics[height=4.1cm]{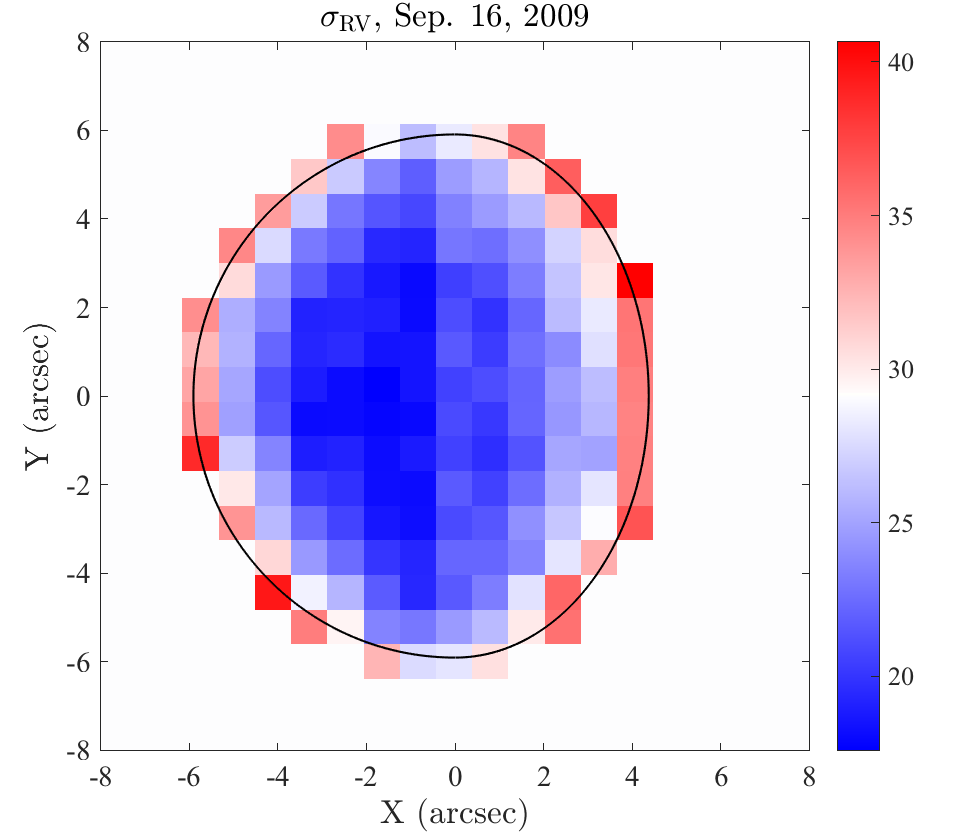}\\
\includegraphics[height=4.1cm]{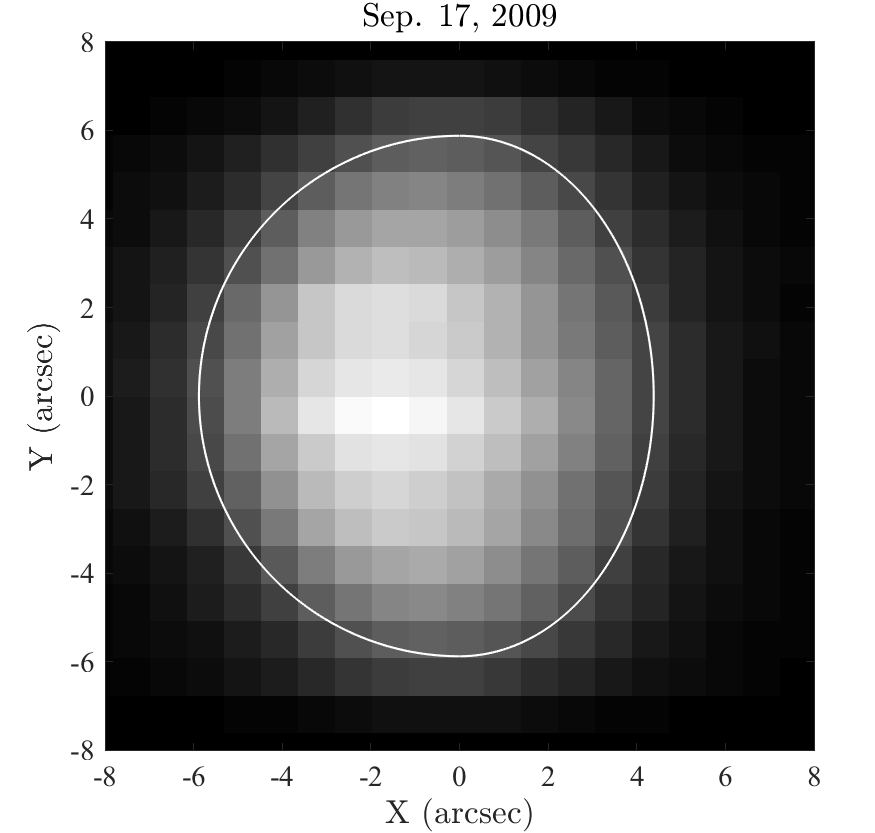}\includegraphics[height=4.1cm]{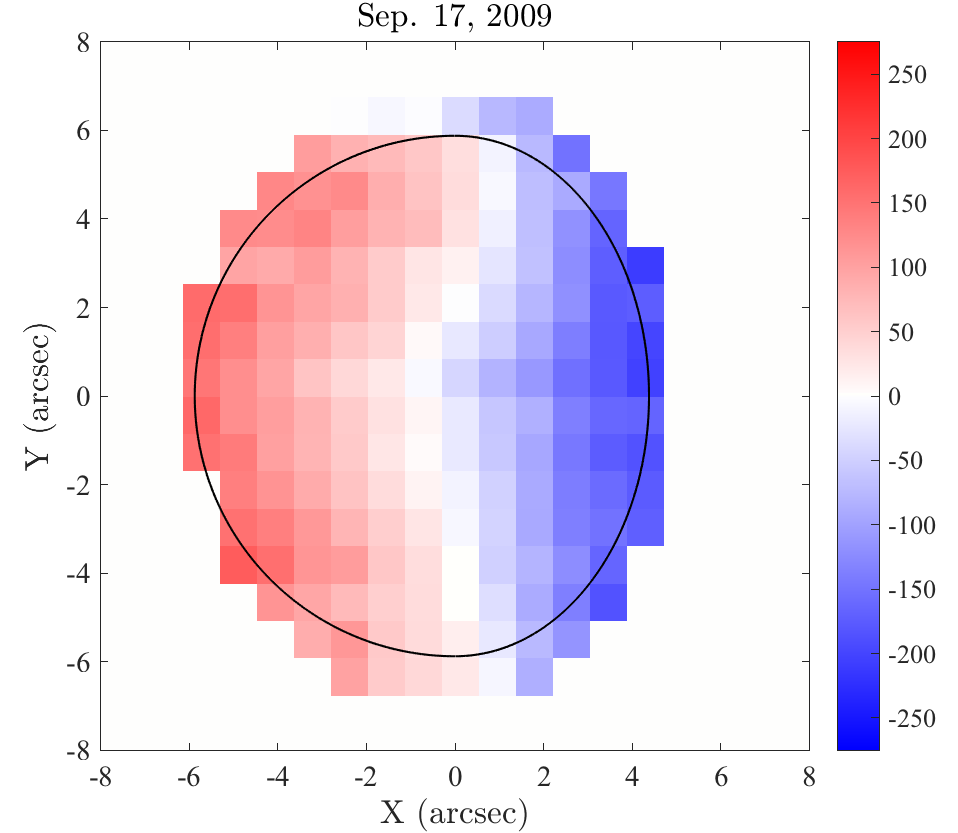}\includegraphics[height=4.1cm]{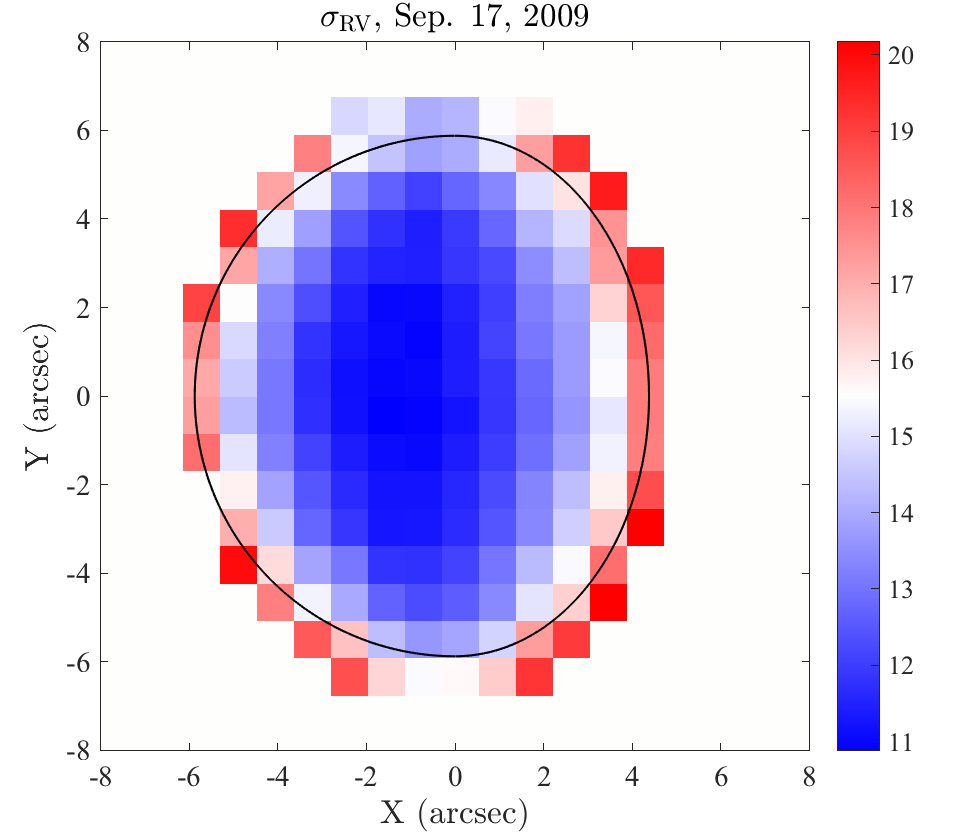}
\caption{From left to right: photometric, Doppler, and Doppler error maps obtained during the 2009 campaign. Top is September 14, middle September 16, and bottom September 17. Middle panels: colors indicate redshift/blueshifts and the color bar indicates the Doppler shift in m s$^{-1}$. Right panels: color indicates the error on Doppler measurements $\sigma\ind{V}$. Spatial coordinates are expressed in arcsec. The white and black contours indicates the theoretical size of the planet on the detector and the maps are oriented such as north is on top. With respect to Fig. \ref{fig_scans_2009}, the images and maps were averaged, then rotated by $90^\circ$ to align the vertical axis with Venus polar axis. Pixel size is $\simeq0.8\times0.8$ arcsec. We note the change in scale for Doppler errors: from 8 to 21 m s$^{-1}$ on Sept. 14, 17 to 41 m s$^{-1}$ on Sept. 16, and 11 to 21 m s$^{-1}$ on Sept. 17. The data corresponding to the 9 maps displayed in this Figure are available in Appendix \ref{sect_app_data}.}
\label{fig_doppler_2009}
\end{figure*}

Fitting a photometric profile that is a function of the pixel FOV and atmospheric seeing (convolution by a Gaussian function) is not straightforward. We developed a dedicated routine based on a Monte Carlo Markov Chain (MCMC) optimization method, including the Metropolis-Hasting algorithm and parallel tempering, and Bayesian inference \citep{Gaulme_2018}. At each step of the iterative process, a 2D photometric profile is generated on a grid of pixels with free FOV, then the image is convolved with a Gaussian function of free full width at half maximum (FWMH). The center of the planet is also a free parameter. The image is then projected along the NS direction and compared with the actual profile. We minimize the square of the difference (least-squares fitting) until convergence is reached. The error bars are retrieved from the posterior density function of each parameter. We note that we make use of a Lambert law to describe the limb darkening because the smooth aspect of Venus in the visible, coupled with a small apparent diameter ($\sim5$ times the seeing value), does not justify models that are undistinguishable from a Lambert sphere in our observing conditions. In Fig. \ref{fig_MCMC_fit_NS}, we present the fitting of the symmetrized photometric profile along the NS direction, done with the MCMC routine for the September 14 data. Seeing is estimated to be $3.21_{-0.78}^{+0.67}$ arcsec and apparent radius\footnote{For computational reasons, we fitted the apparent radius of the planet on the detector instead of the pixel FOV. Obviously, the actual apparent size of the planet in the sky is perfectly known from ephemeris databases.} on the detector $5.54_{-0.22}^{+0.17}$ arcsec if the steps were 0.80 arcsec. It actually means that the steps were of 0.86 arcsec. 

This step is however not sufficient to estimate accurately the seeing and pixel FOV because the atmospheric seeing varies with time and the symmetric photometric profile is an average of the scans. Besides, the fact of having symmetrized the profile could insert a small bias in the apparent dimension of the planet upon the detector. The second step thus consists of attributing a latitude to each position of the slit based on the symmetric NS profile. For each step, the photometric cut along the EW direction is fitted in the same way as the NS profile with an MCMC routine. This process allows us to know the location of the center of the planets at each step of the scan and to monitor the seeing along the day (Fig. \ref{fig_seeing}). On September 14, where we had the best conditions (no clouds), the seeing ranged from about 2 to 3.5 around noon. On September 16 and 17, the seeing conditions were definitely worse with a mean value above 3 arcsec.

Once the relative position along the EW direction are determined, we place each photometric vector in a 2D matrix and reinterpolate (linear interpolation) it on the regular grid along the NS direction, which we got by symmetrizing the profile. The same is done with RV maps, except that the interpolation is performed in a way to not alter the velocity value at the edges. Indeed, by definition the RV field is zero out of Venus (we do not compute it), so that simply interpolating a column on a shifted vertical axis would bias velocities at the edges. Therefore, we extend the RV vector out of Venus with the last value met on the west for the northern space above Venus, and vice versa for the eastern part of the detector. After interpolation, all regions out of Venus are set back to a default zero RV.

\subsection{Biases in the radial-velocity measurements}
\label{sect_biases}
Most of our data were acquired during daytime. Therefore, we can monitor the possible drift of the spectrometer thanks to the solar spectrum that is scattered by the Earth's blue sky. We note that the MTR spectrometer of the THEMIS telescope was not designed to perform high precision RV measurements -- spectro-polarimeteric observations of the Sun instead -- and is not thermally stabilized. Drifts of hundreds of meters per second are expected. 

Figure \ref{fig_drifts_2009} shows the mean RV that is measured on the sky. The variation of RV is much larger than that expected from the Earth's rotation (gray lines). Even though we do not have measurements of the temperature of the spectrometer, it is very likely that the variations we observe are dominated by temperature, which is typical of spectrometers. The amplitude of the observed variations is about 3000 m s$^{-1}$ ($\approx0.06$ \AA) both for observations performed on September 14 and 17, 2009. We can correct from these variations by subtracting the mean RV that is measured on sky, or at least a smooth of it to reduce the measurements uncertainties on the relatively low S/N sky RVs. For the only scan that was done during night time (the first of September 14th), we extrapolated the drift of the sky from a fitting of the six following scans, where the trend is well fitted by a second order polynomial. Again, such a drift was expected and correcting it is straightforward. 

What was not expected and surprised us is an RV drift in between the Earth's sky and Venus. In Fig. \ref{fig_drifts_venus_2009}, we plot the difference of the Earth's sky RVs with the mean RV that is measured on Venus at each position. Since we are scanning the planet along the north-south axis (slit parallel to equator), we do not expect large variations of the average RV along the slit. It should display the meridional circulation on top of relative motions of the planet with respect to the observatory and the Sun. Actually, it arises that whatever the direction of the scan (NS or SN), there is a spurious RV that increases monotonically at each step of the scan. For Fe spectra, the RV drift per scan reaches 2000 m s$^{-1}$ per scan on September 14, while it is of 400 m s$^{-1}$ three days later (Fig. \ref{fig_drifts_venus_2009}). 

In 2008, we scanned the planet along the WE direction and found out this bias. This was particularly terrible as it overlapped and overwhelmed the zonal circulation of Venus. That made the observation taken in 2008 impossible to use scientifically. We thus decided to scan Venus in the perpendicular direction during the 2009 observations. That way, the bias does not compete with the zonal circulation of Venus, and we can combine the RV maps obtained by scanning the planet from SN with those done from NS to hopefully cancel this systematic effect.  

The origin of such a bias is still unclear and we have no definitive explanation. A possible explanation is related to the way we scan the planet. As mentioned earlier, positioning the slit on the planet actually consists of positioning the planet with respect to the slit thanks to a titled mirror (M5). If the M5 mirror is not perfectly placed at the pupil focus of the telescope, the small rotation of the mirror applied to position the planet on the slit can cause some astigmatism. In other words, the optical distortions of the field, which we measure on the solar flat field (see Fig. \ref{fig_solar_flat}), could slightly vary while the mirror is moving. In such a case, an imperfect rectification of geometrical distortion could lead to a differential RV drift in between Venus and the sky.

To correct the RV maps from this bias, we first flatten each individual RV map by fitting a tilted plane (2D linear polynomial) oriented along the NS axis (Fig. \ref{fig_scans_2009}). Indeed, if the RV bias is pretty much stable from one scan to the next, it actually seems to diminish slightly with time. Therefore by subtracting a tilted plane to the RV map is a way to include the amplitude variation of the bias. Subtracting a tilted plane affects any uniform pole-to-pole meridional RV field, but not equator-to-pole meridional fields. When we then combine the RV maps, we assume the residuals to cancel each other. This assumption is the main weakness of the whole process as we cannot be fully sure that the bias completely cancels out when averaging scans done in both directions. We discuss that aspect in Sect. \ref{sect_prospects} regarding the strong variations of zonal wind speed as a function of local time that we observe.

In Fig. \ref{fig_scans_2009}, we show the individual flattened RV maps and they appear to be very consistent from day to day and spectral line to line, although the data taken on September 14 are the only data that benefited from good observing conditions. From the photometric images, it clearly appears that the data taken on September 17 suffered from worse atmospheric seeing and transparency (clouds). We note RV discontinuity during severe photometric drops caused by clouds, which is especially visible in the second half of the data taken on September 17.  Conditions on September 16 were poor and allowed only for two scans (NS and SN). The error bars on Doppler velocity are about twice as bad as for the other two nights. For each day, all flattened RV maps were then averaged to get a final RV map (Fig. \ref{fig_doppler_2009}). For obvious observing condition differences, we only consider the data taken on September 14 to model the atmospheric circulation. However, it is worth noticing that despite significantly worse conditions and strong biases in some latitude ranges, the overall aspects of the September 16 and 17 RV maps are very consistent with the September 14 map.

\begin{figure}[t!]
\includegraphics[width=8cm]{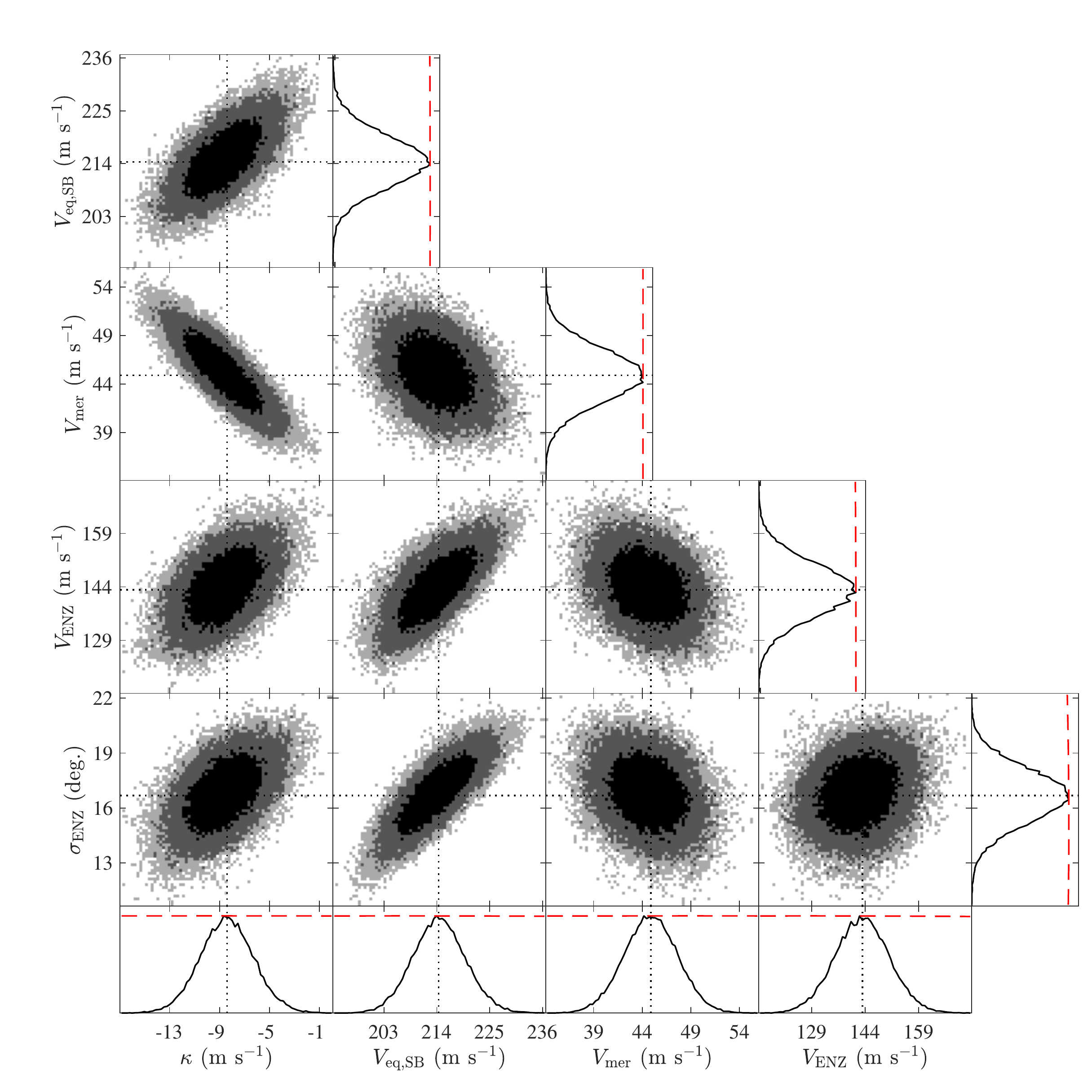}
\includegraphics[width=8cm]{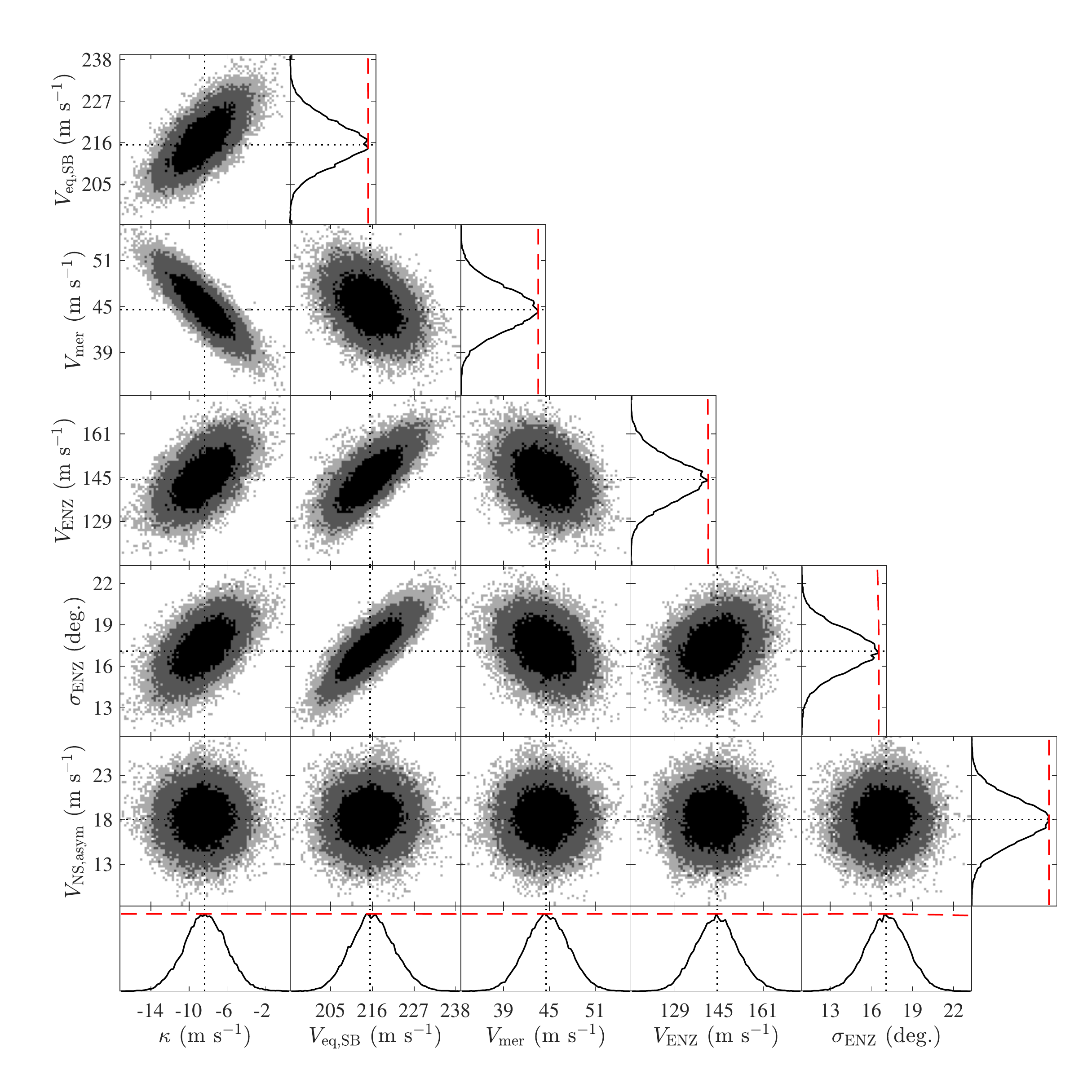}
\caption{Fitting the global circulation parameters with the MCMC routine. Top panel: models 4; bottom: model 7, both obtained from 250,000 iterations. Color scales and subpanels are the same as in Fig. \ref{fig_MCMC_fit_NS}.}
\label{fig_mcmc_modele_circ}
\end{figure}

\begin{figure*}[t!]
\center
\includegraphics[height=3.8cm]{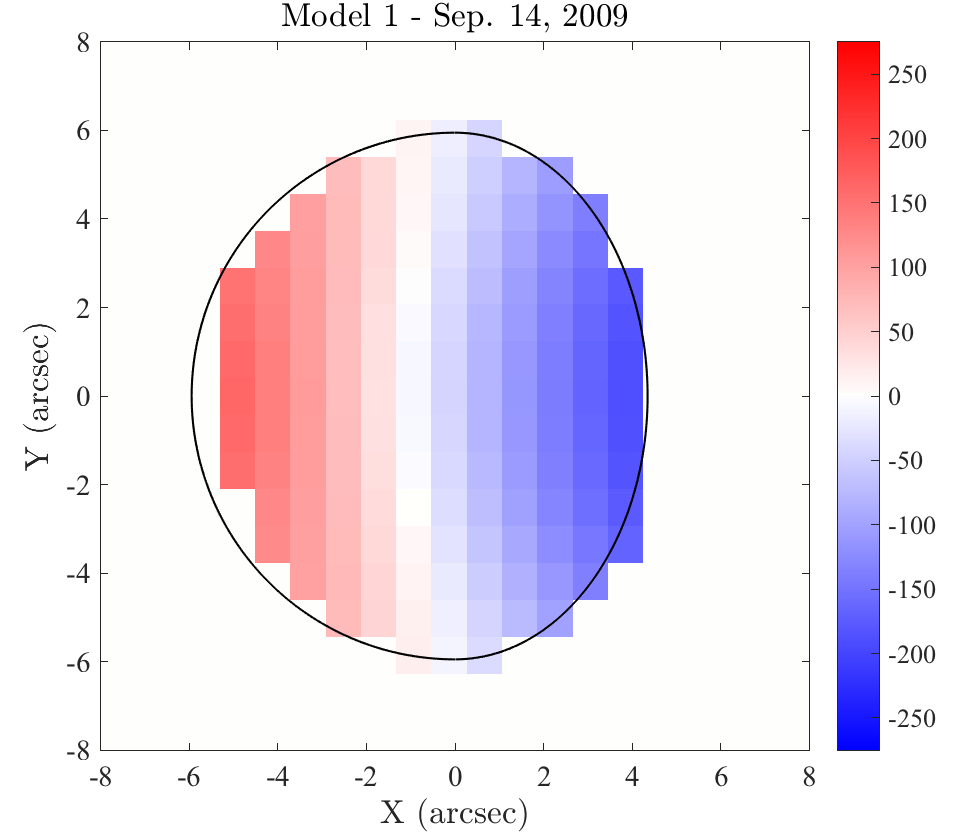}\includegraphics[height=3.8cm]{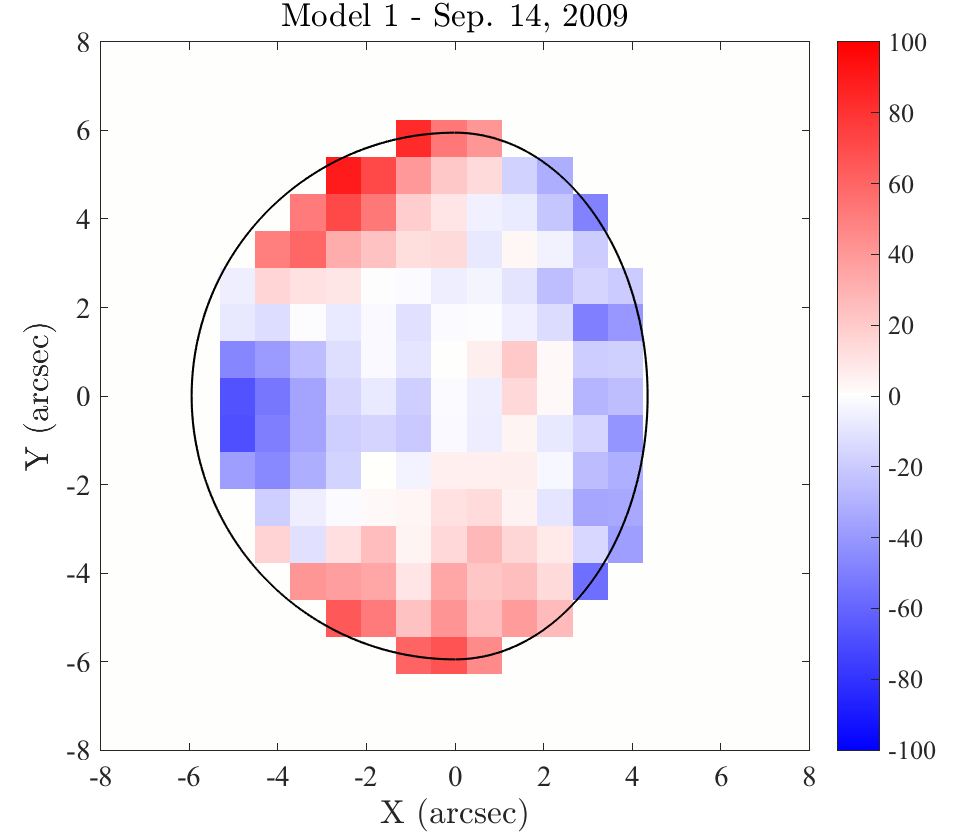}\includegraphics[height=3.8cm]{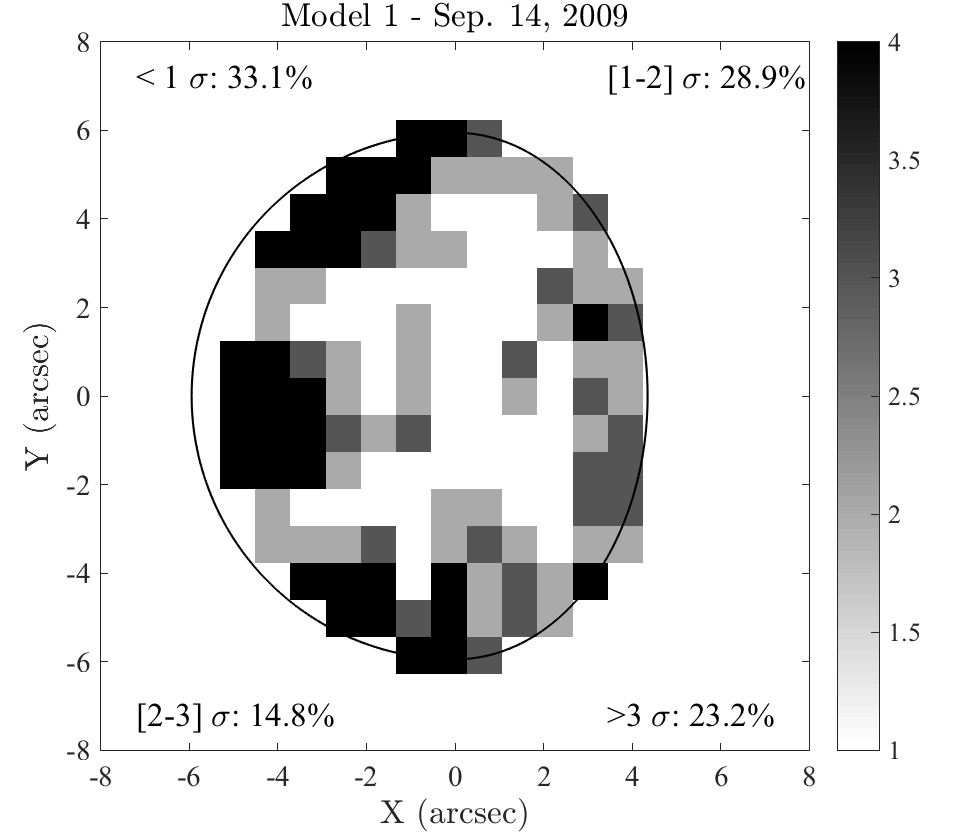}\includegraphics[height=3.8cm]{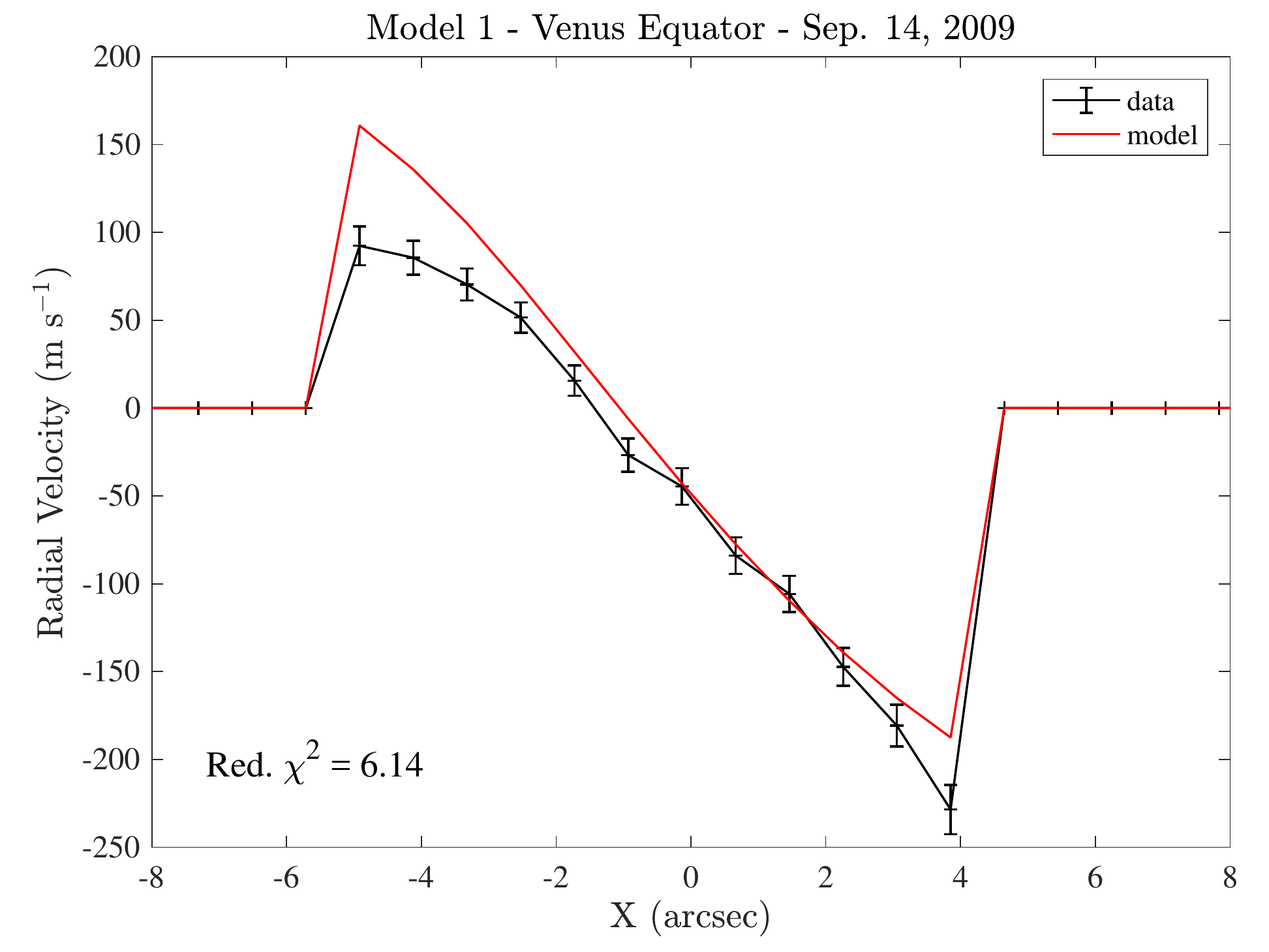}\\
\includegraphics[height=3.8cm]{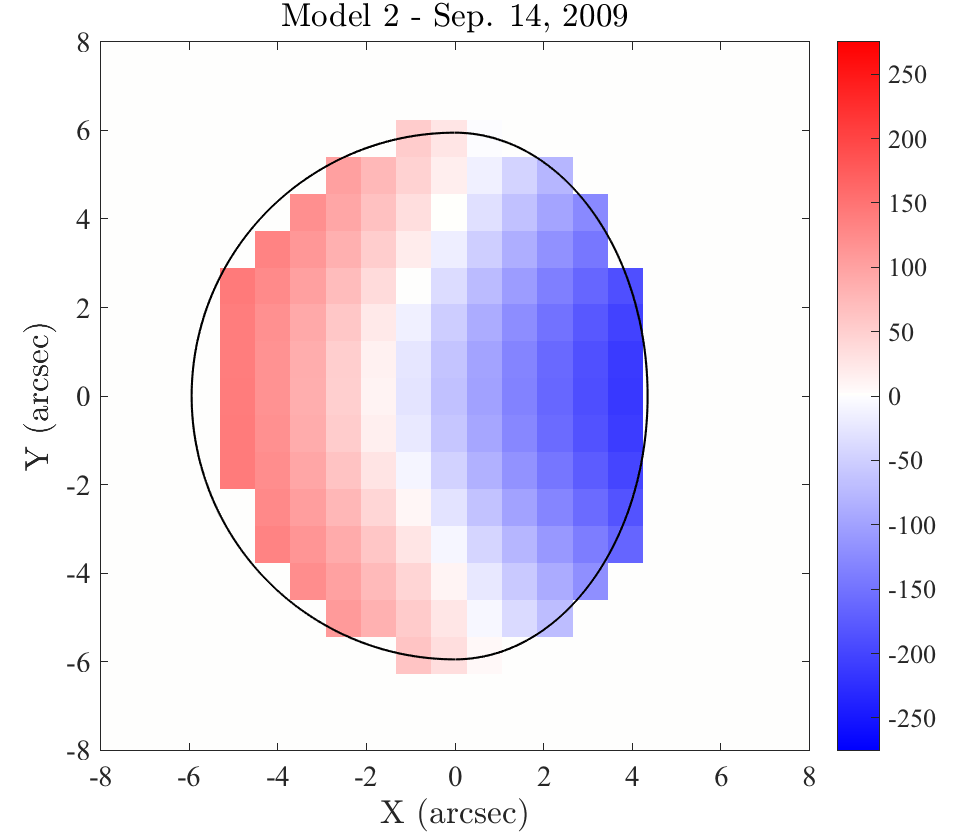}\includegraphics[height=3.8cm]{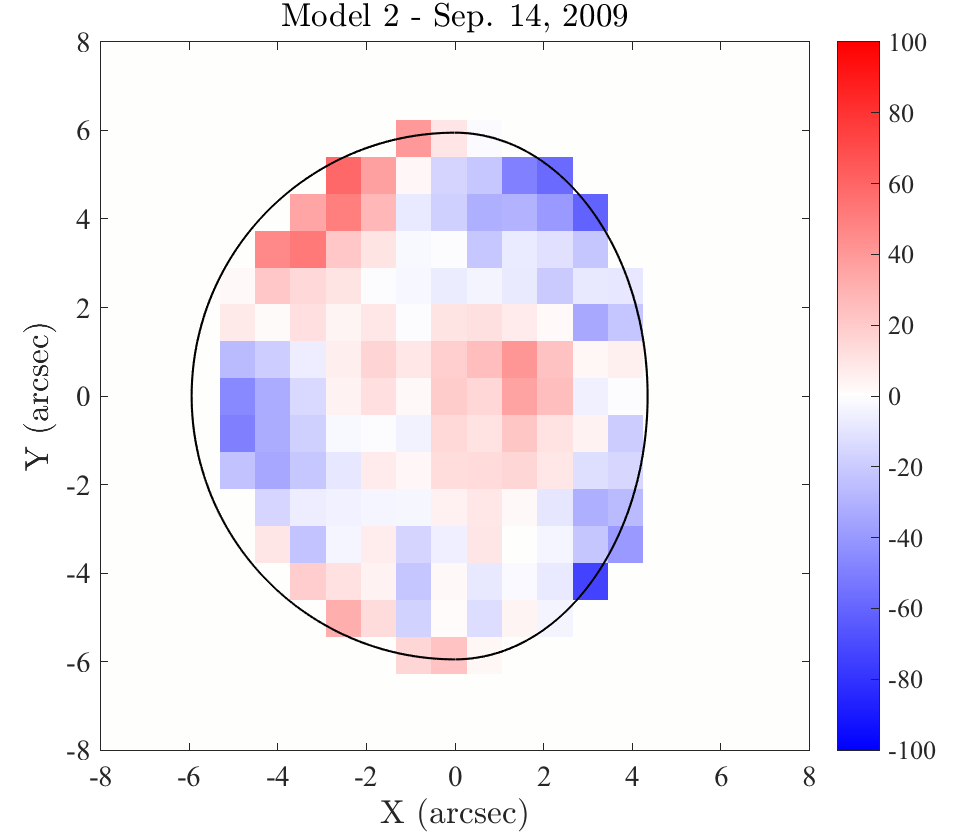}\includegraphics[height=3.8cm]{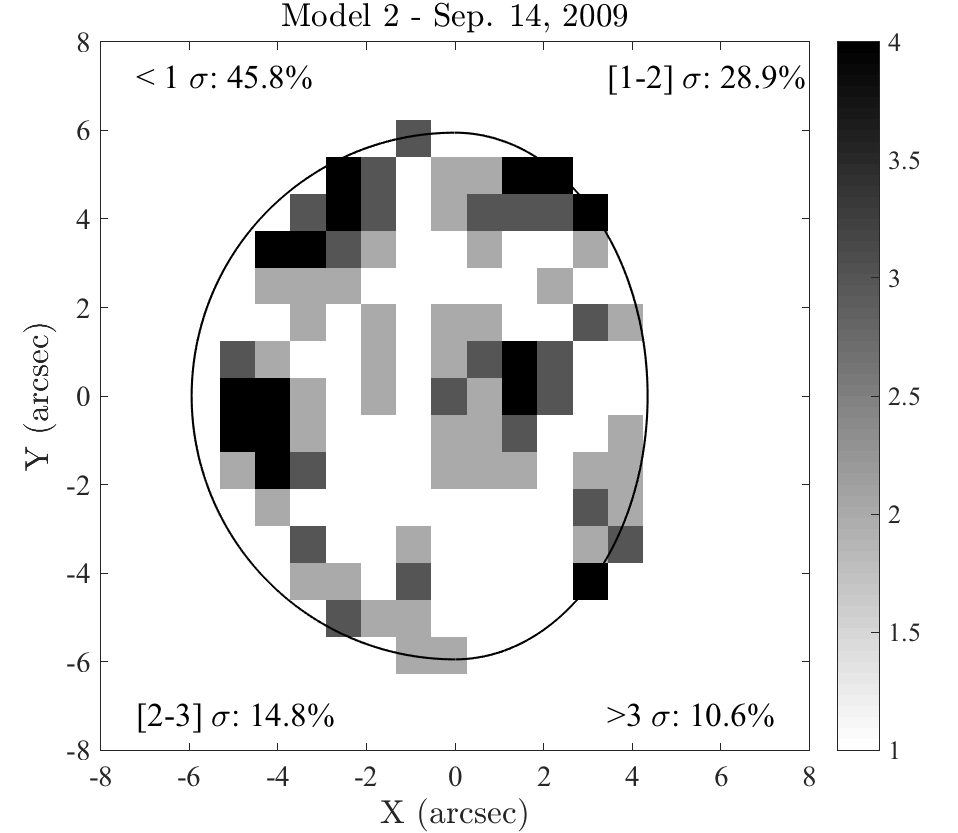}\includegraphics[height=3.8cm]{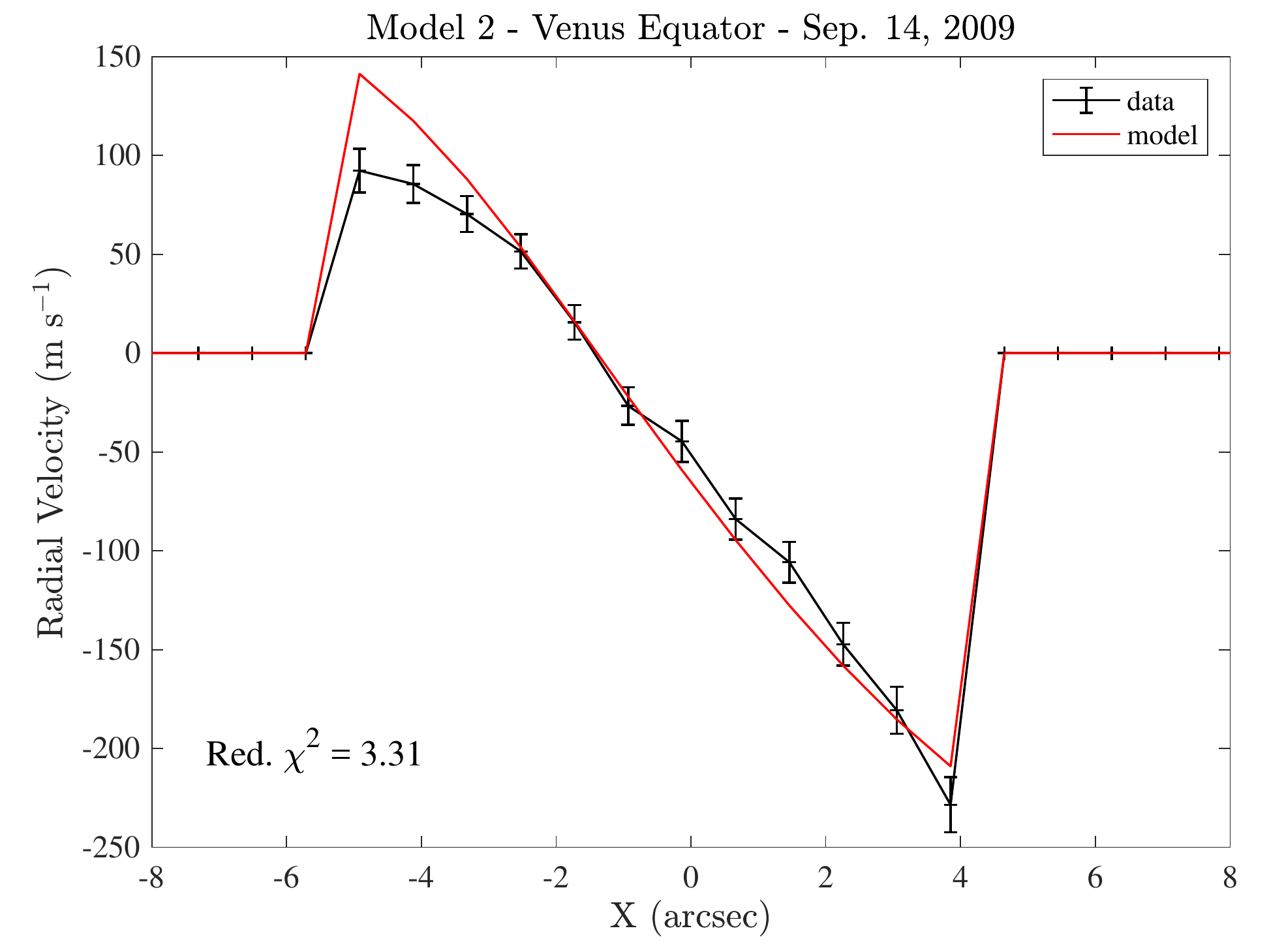}\\
\includegraphics[height=3.8cm]{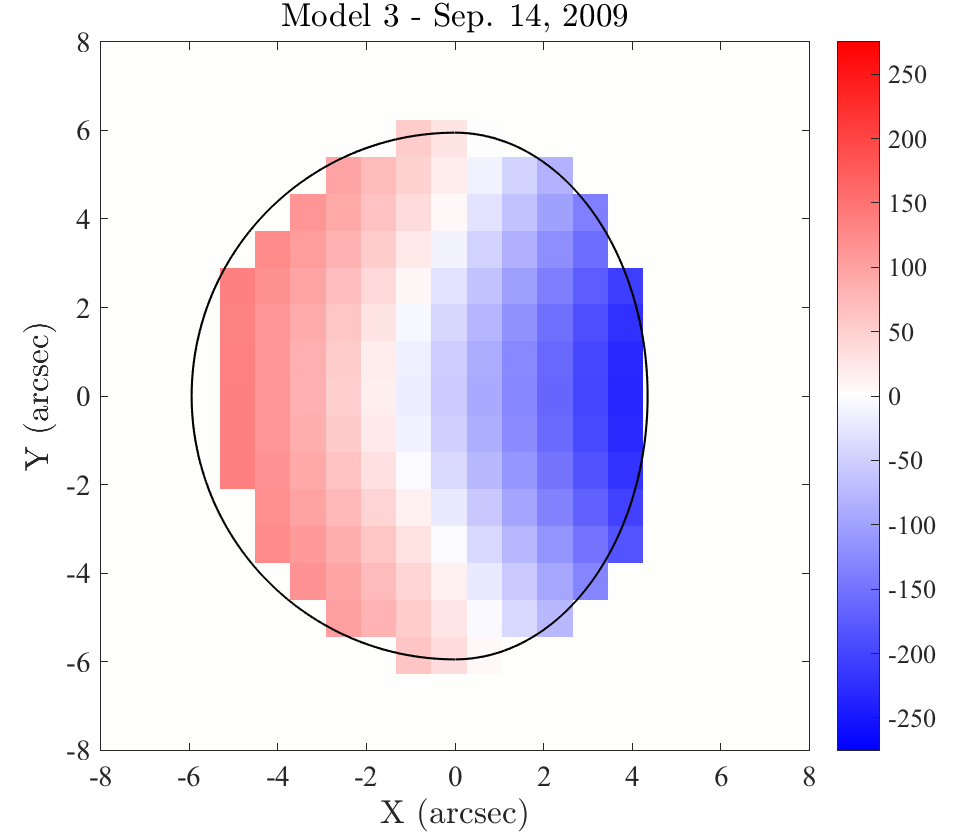}\includegraphics[height=3.8cm]{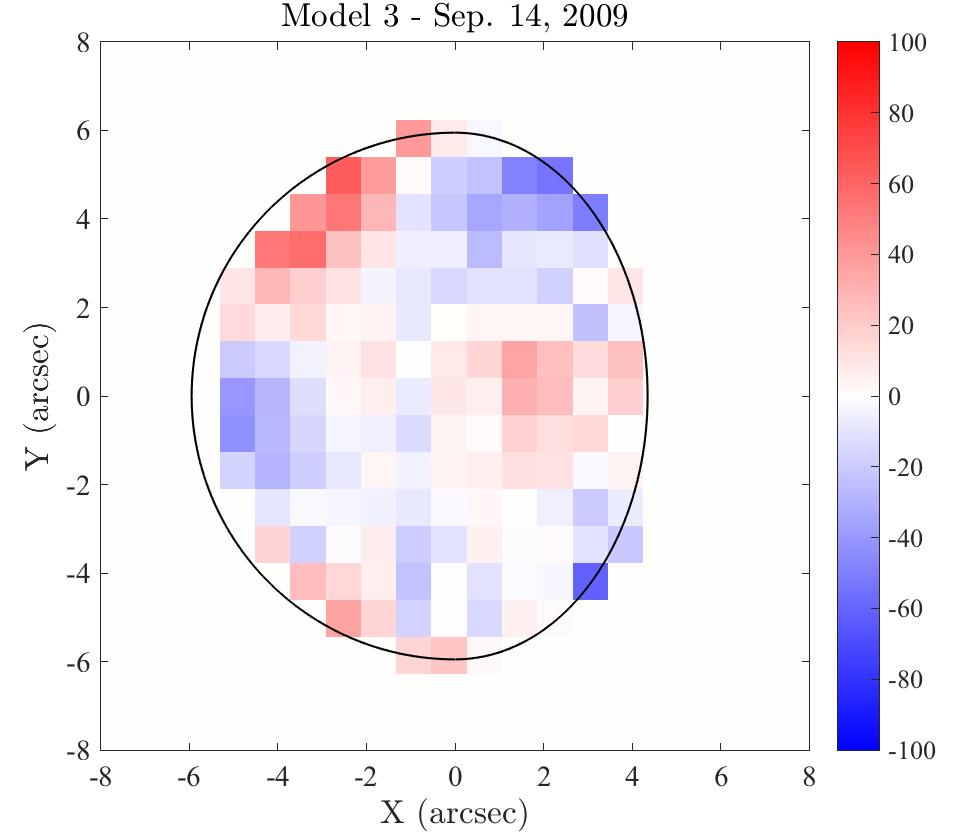}\includegraphics[height=3.8cm]{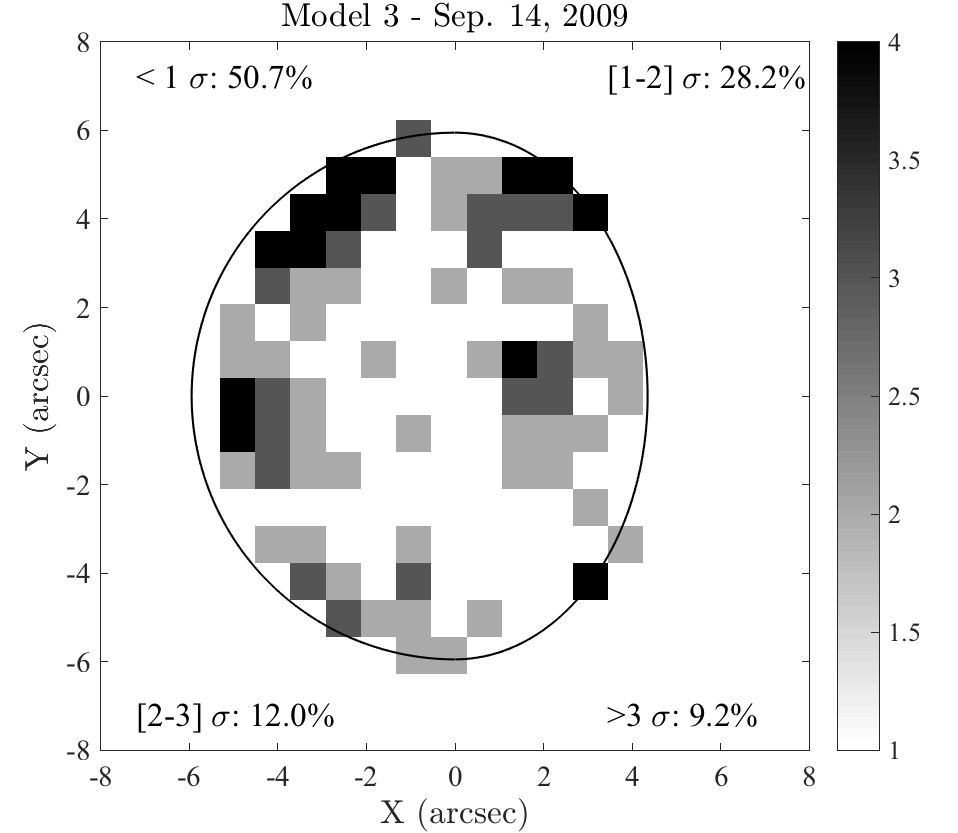}\includegraphics[height=3.8cm]{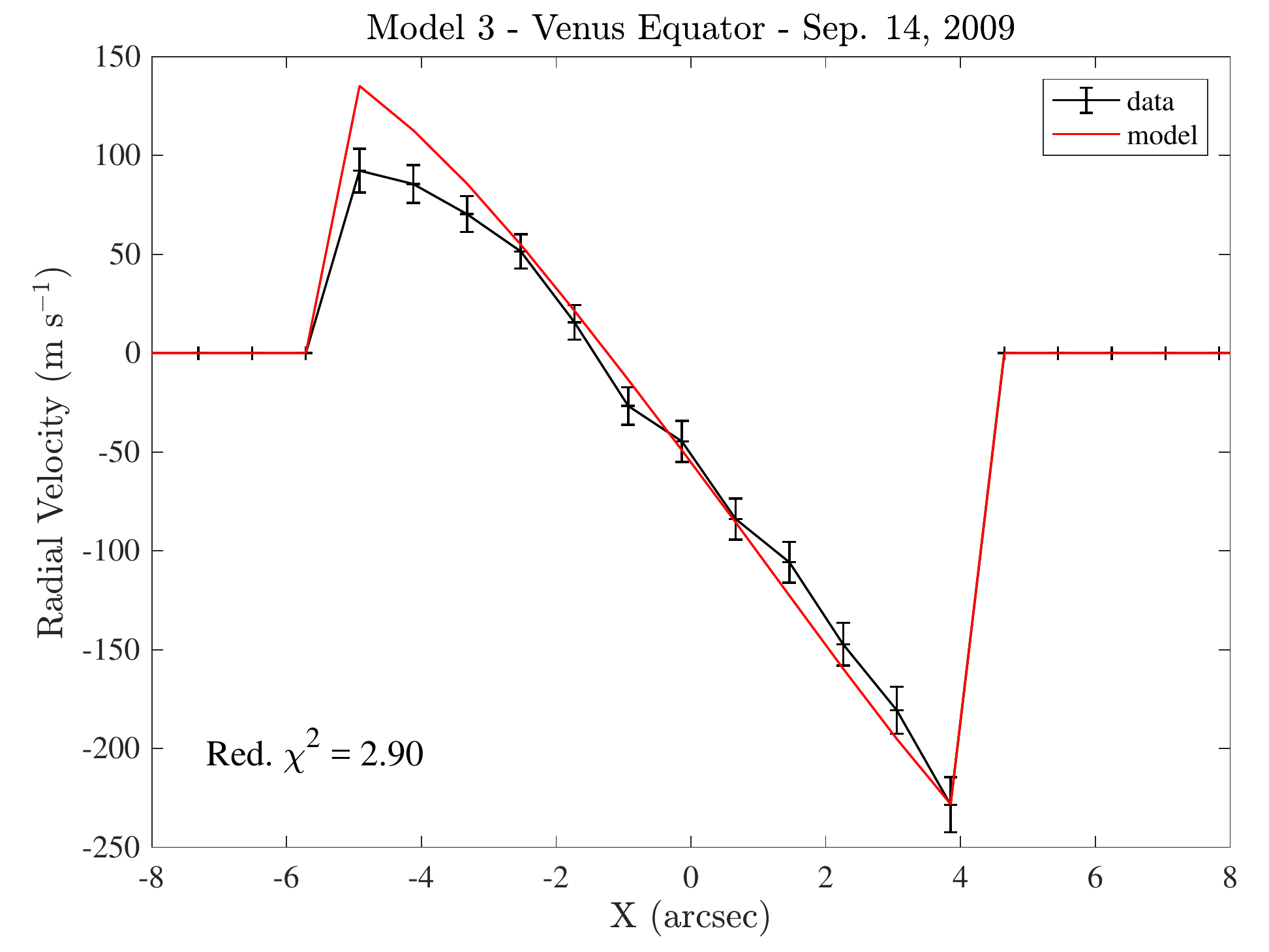}\\
\includegraphics[height=3.8cm]{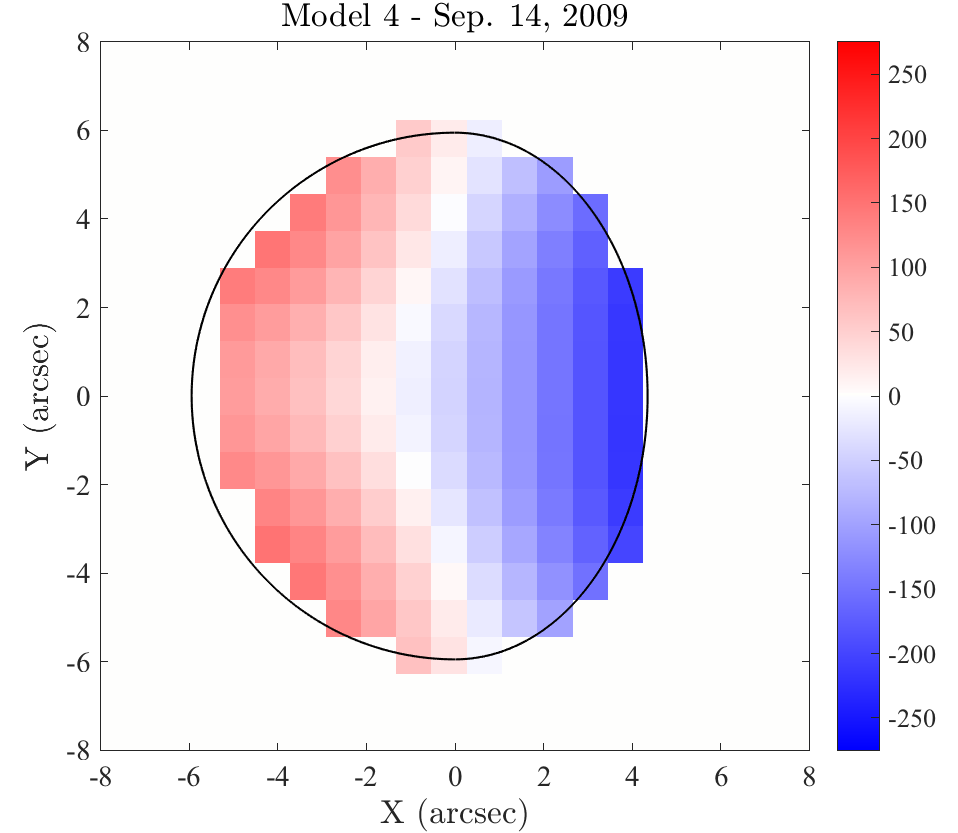}\includegraphics[height=3.8cm]{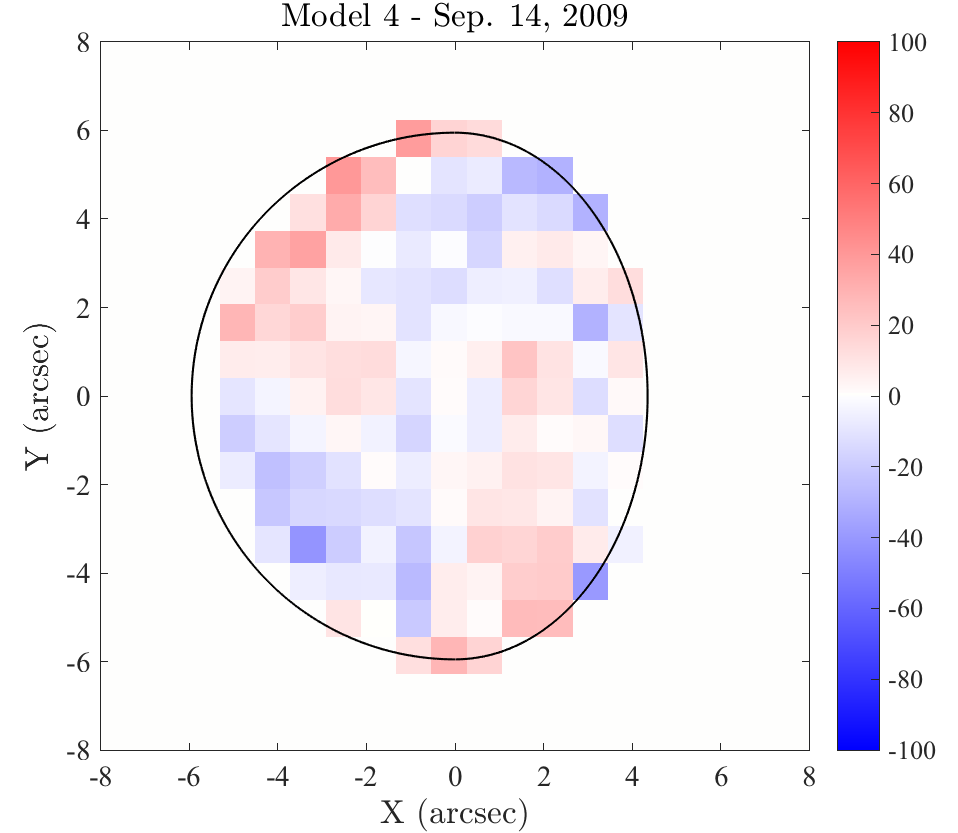}\includegraphics[height=3.8cm]{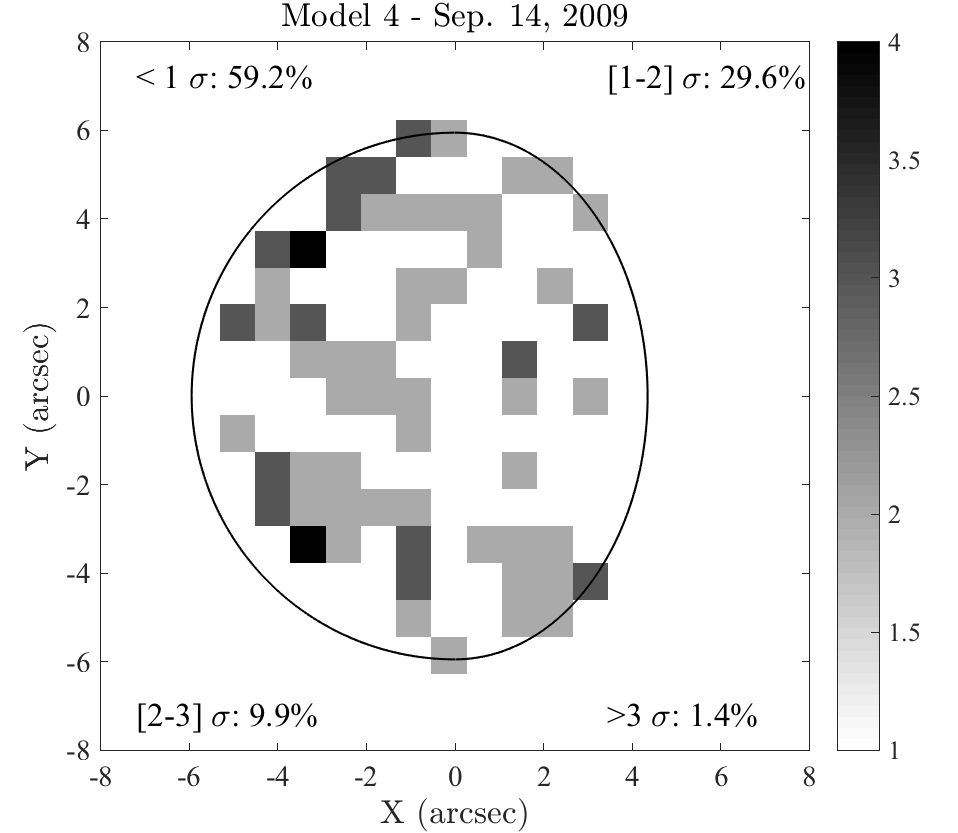}\includegraphics[height=3.8cm]{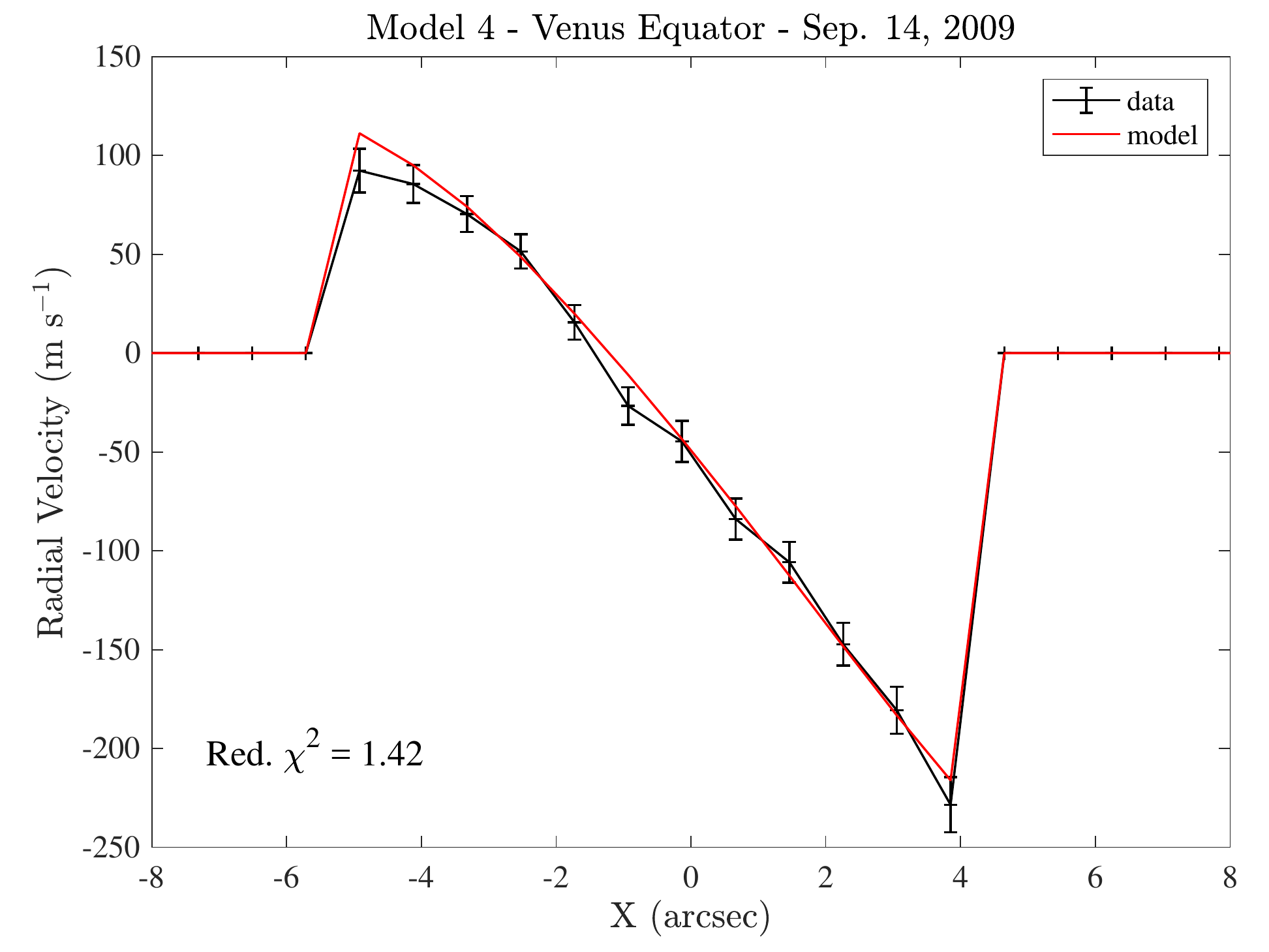}
\caption{Model vs. data for models 1 to 4 for data taken on September 14, 2009. Left panels: best-fit model. Second column of panels: observation minus model, i.e., residual maps. Third column of panels: residual maps in terms of measurement errors. White color indicates agreement within $1 \sigma$, light gray in between 1 and $2 \sigma$, dark gray in between 2 and 3 $\sigma$, and black above 3$\sigma$. In the four corners of each error map are indicated the fractions of points in each group ($[0-1]\,\sigma$, $[1-2]\,\sigma$, $[2-3]\,\sigma$, or $>3\,\sigma$). Right panels: best fit models overplotted with data along the equator. Error bars are 1$\sigma$ errors.}
\label{fig_models_1_4}
\end{figure*}

\begin{figure}[t]
\includegraphics[width=8.cm]{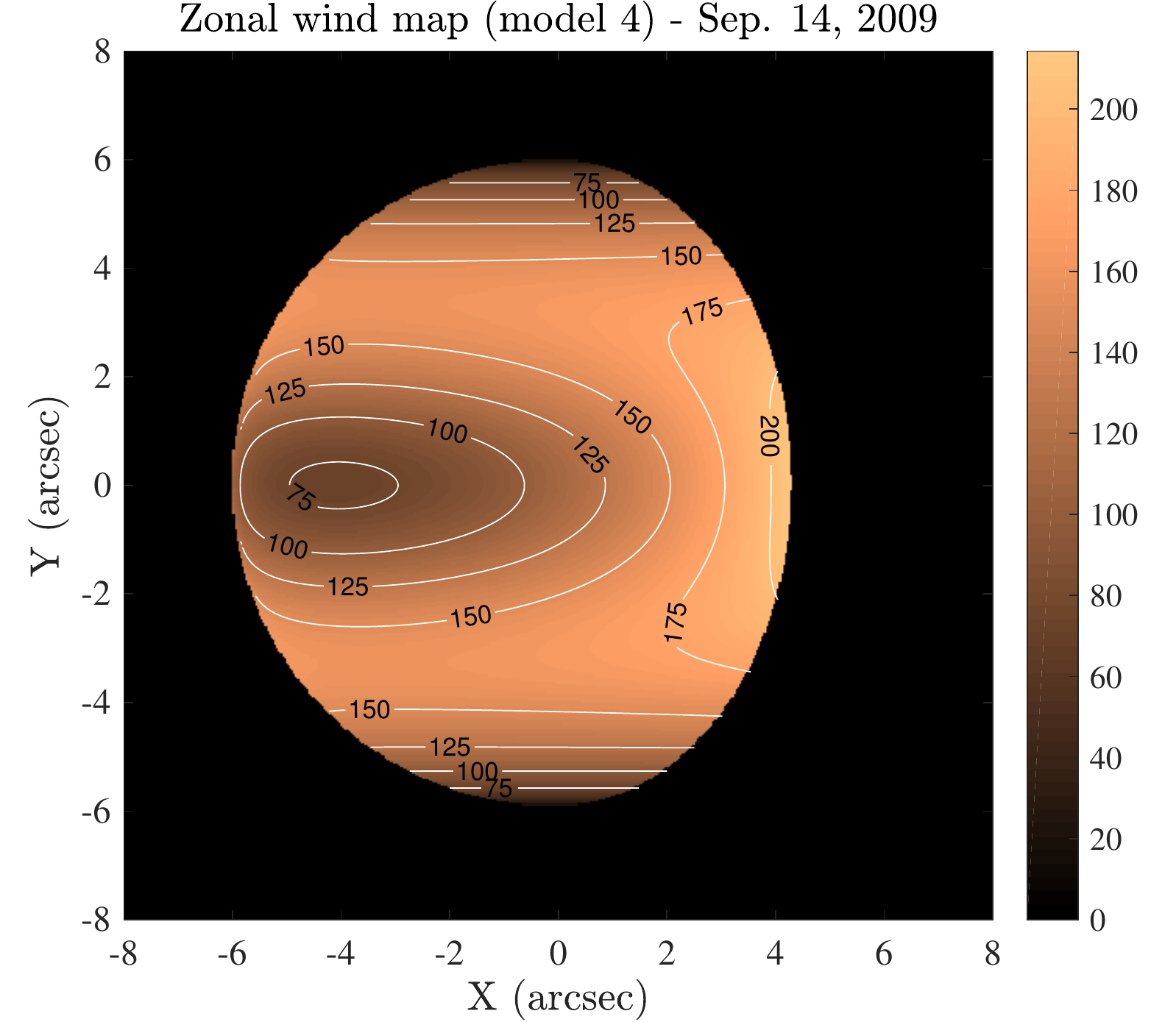}
\includegraphics[width=8.cm]{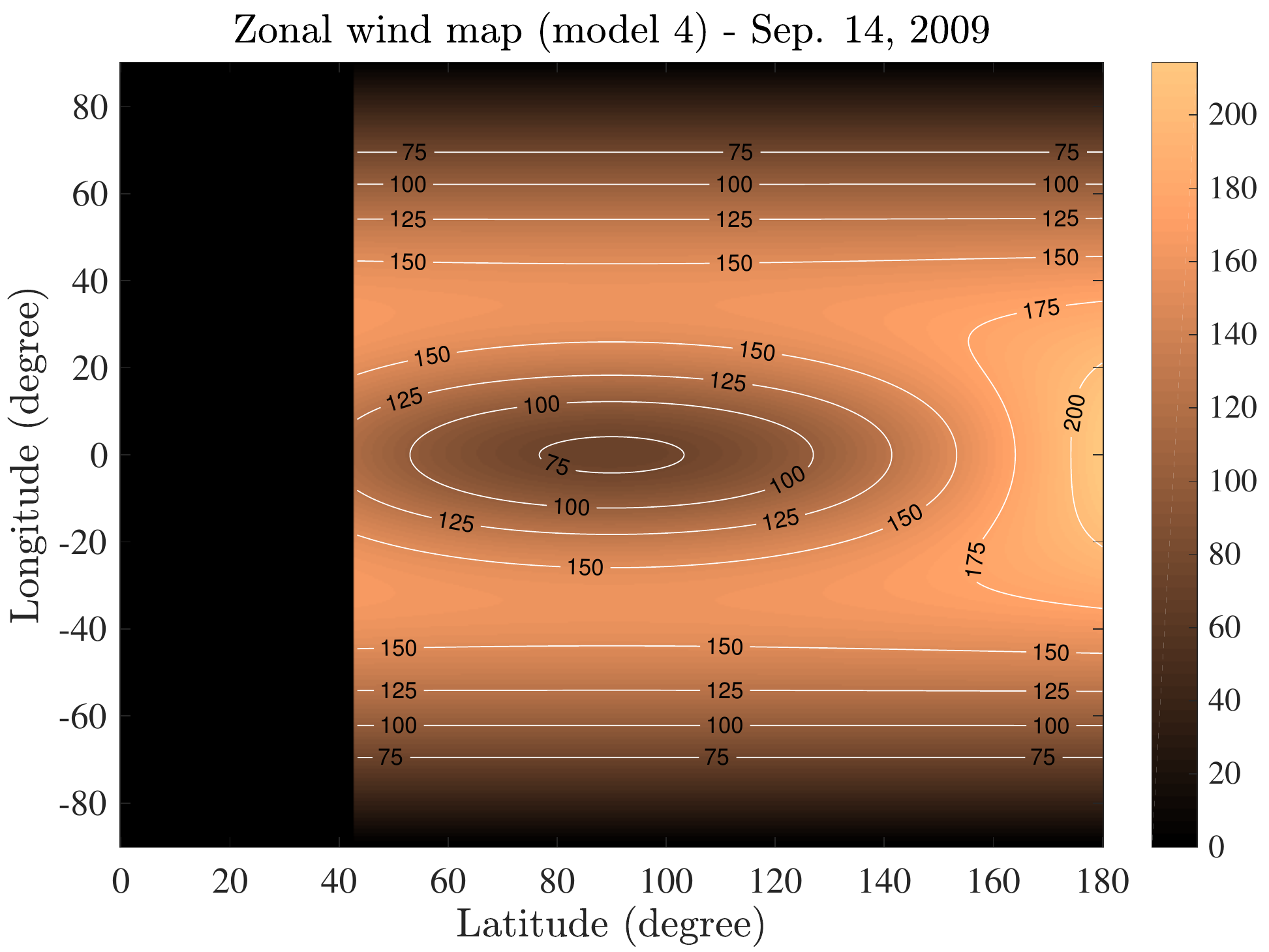}
\caption{Zonal wind map that corresponds to Model 4.}
\label{fig_model4}
\end{figure}

\begin{figure*}[t]
\center
\includegraphics[height=3.8cm]{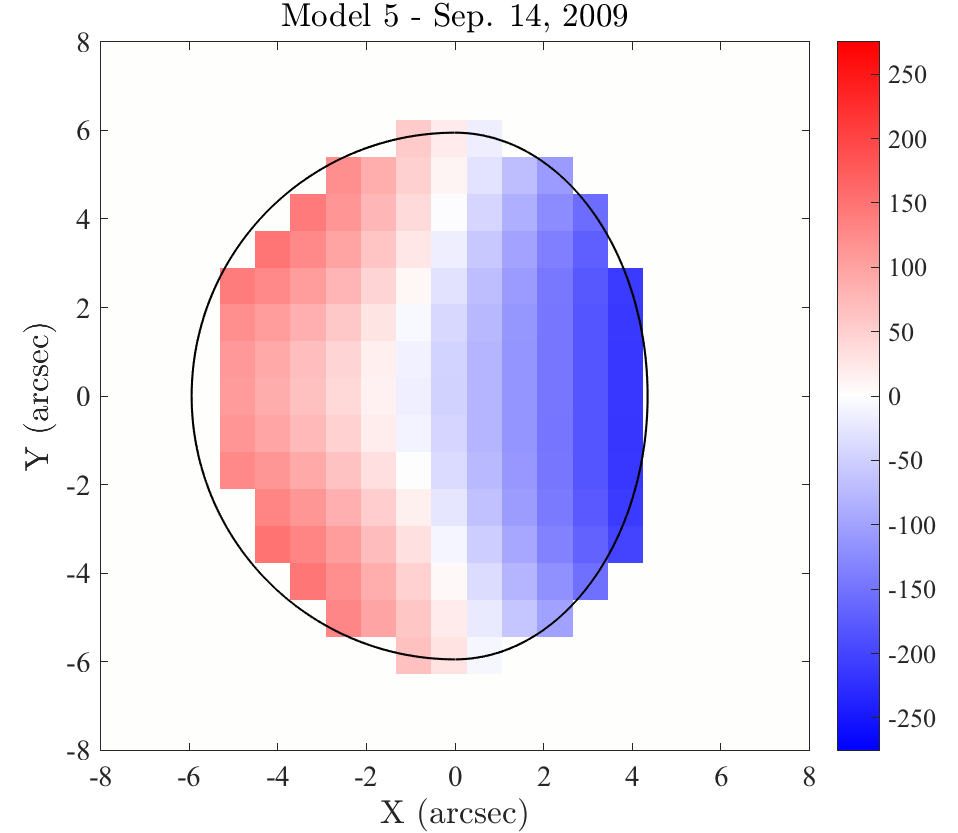}\includegraphics[height=3.8cm]{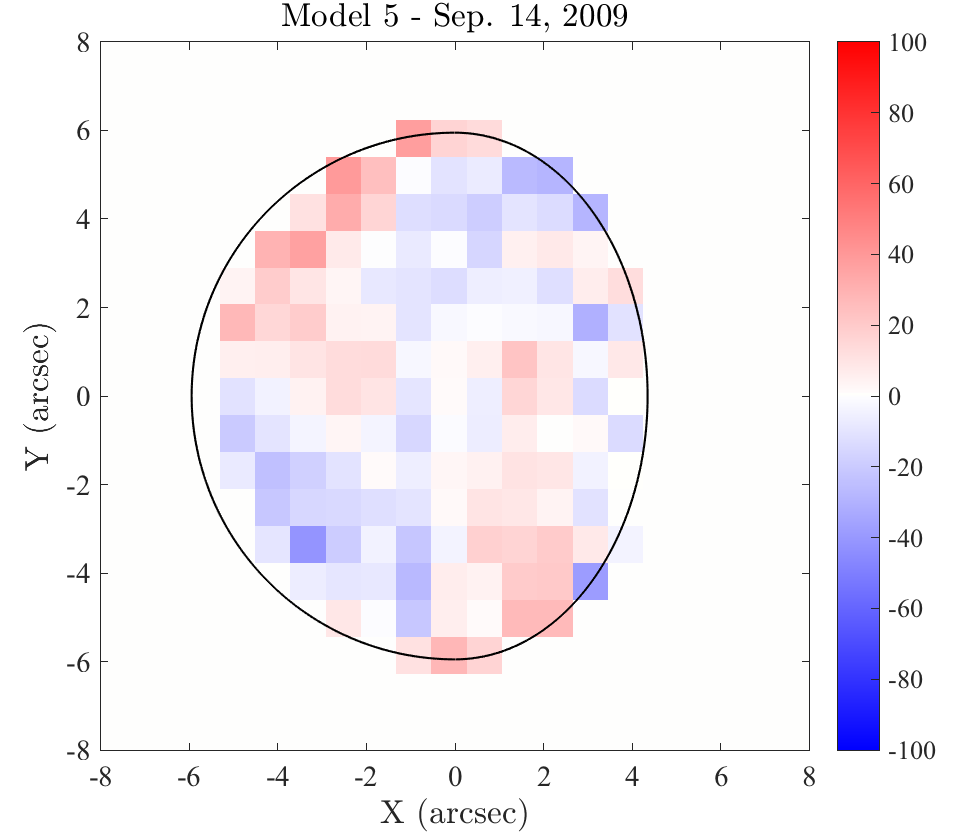}\includegraphics[height=3.8cm]{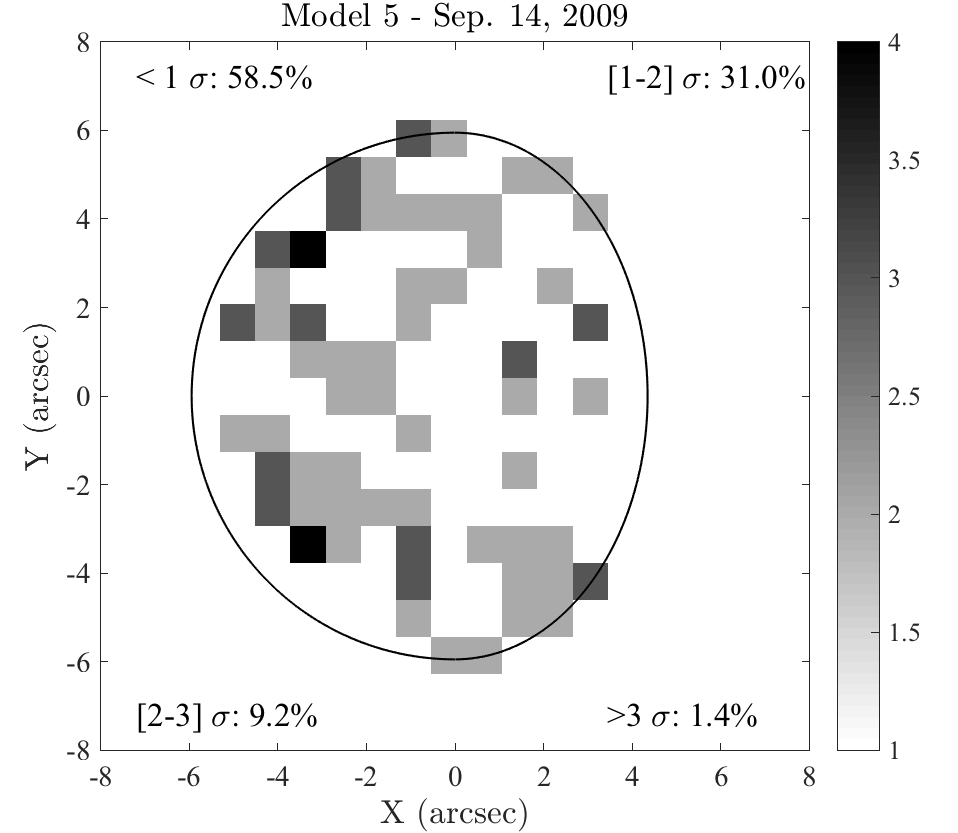}\includegraphics[height=3.8cm]{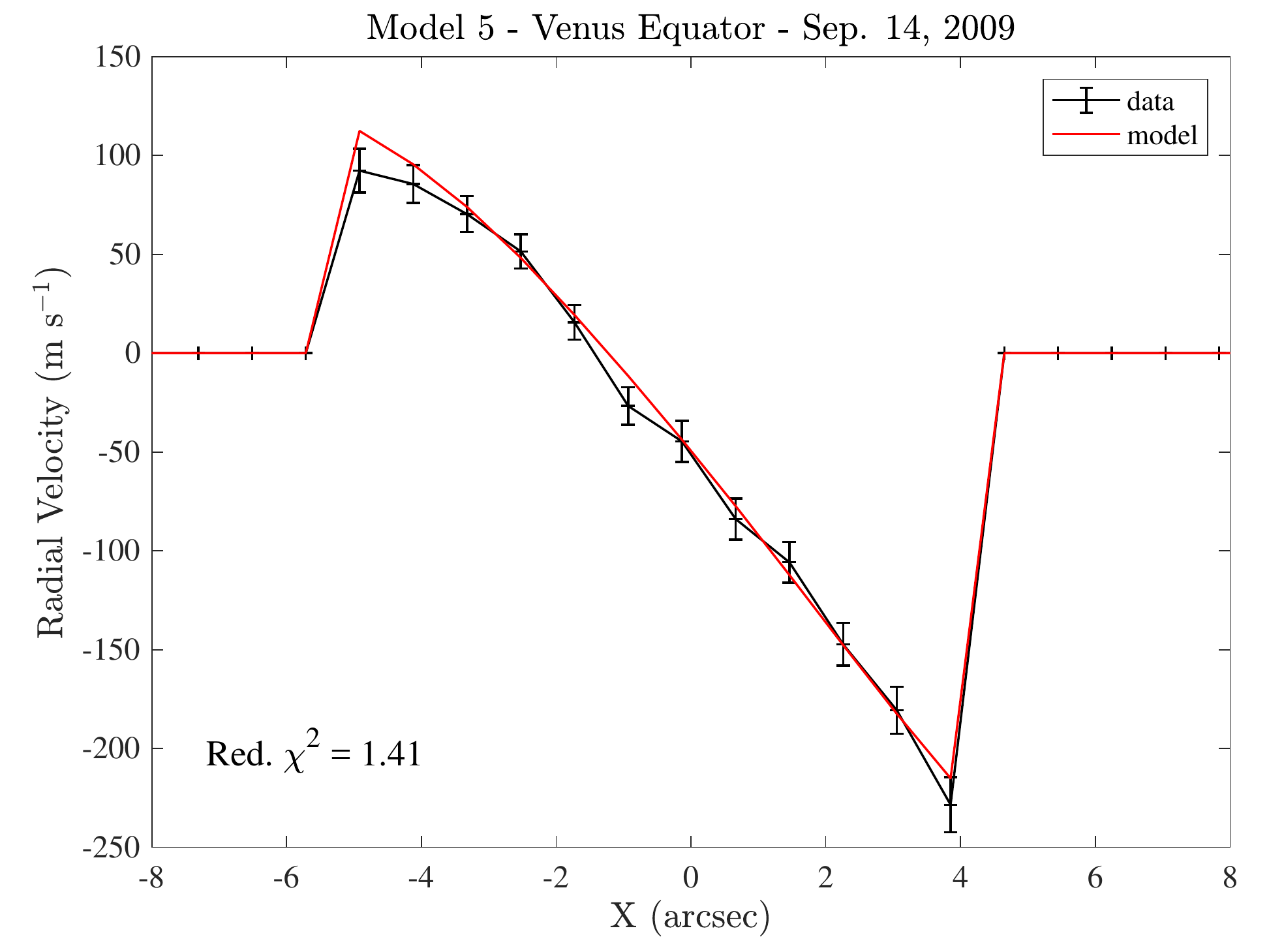}\\
\includegraphics[height=3.8cm]{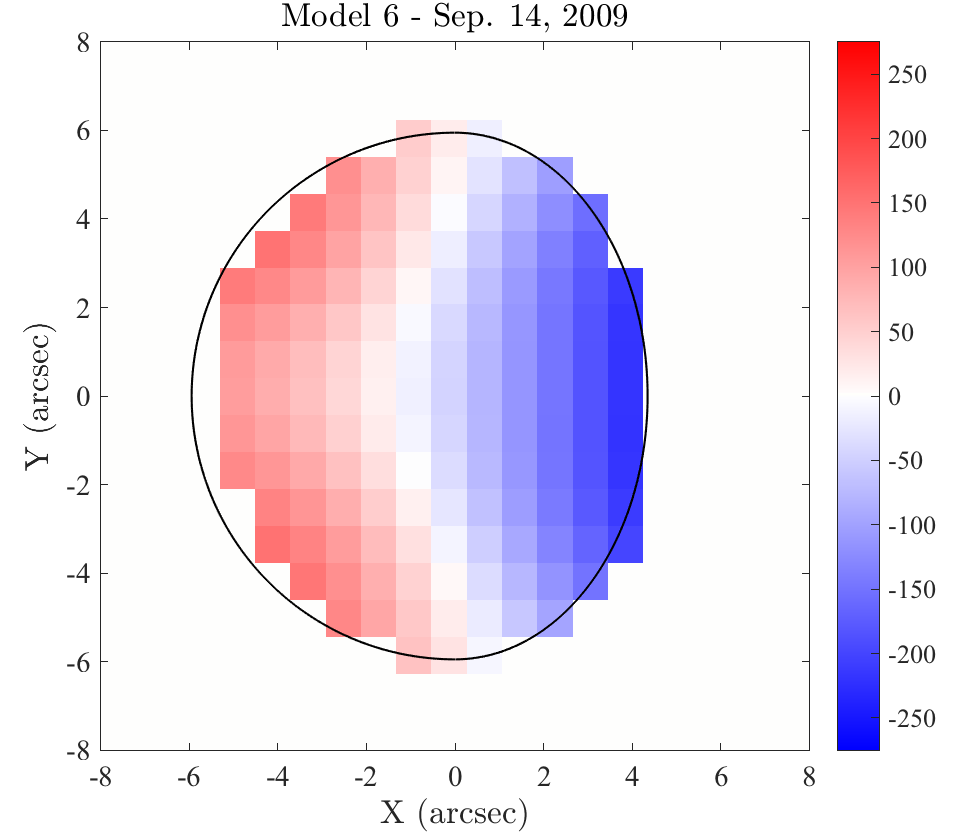}\includegraphics[height=3.8cm]{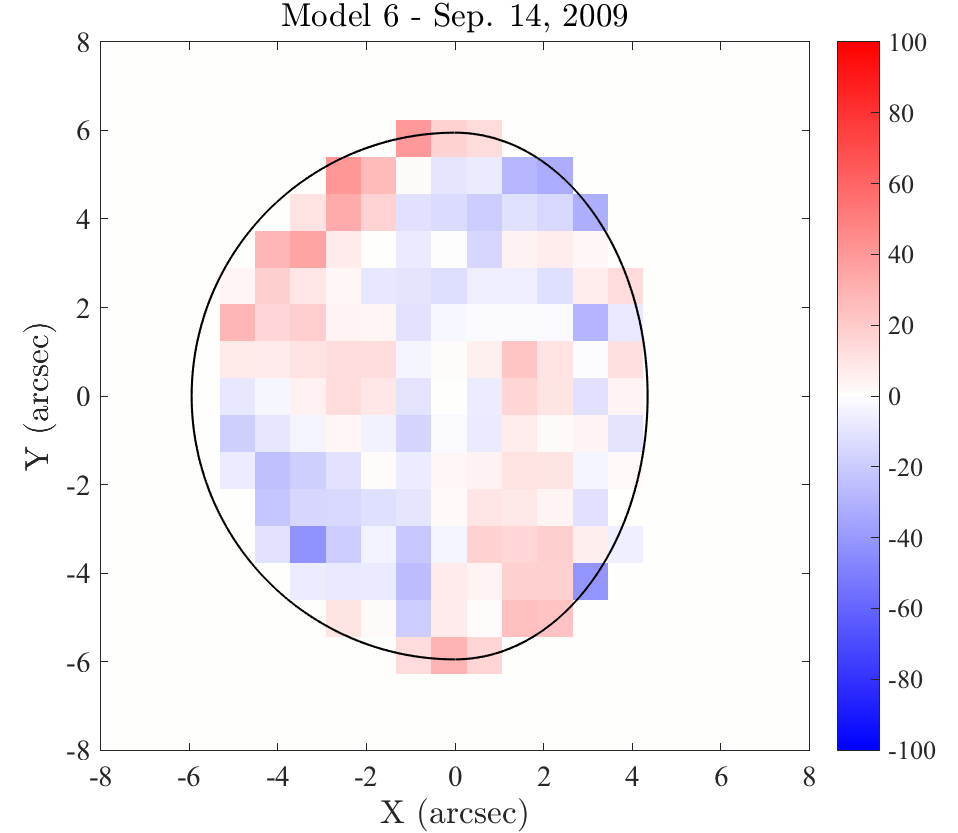}\includegraphics[height=3.8cm]{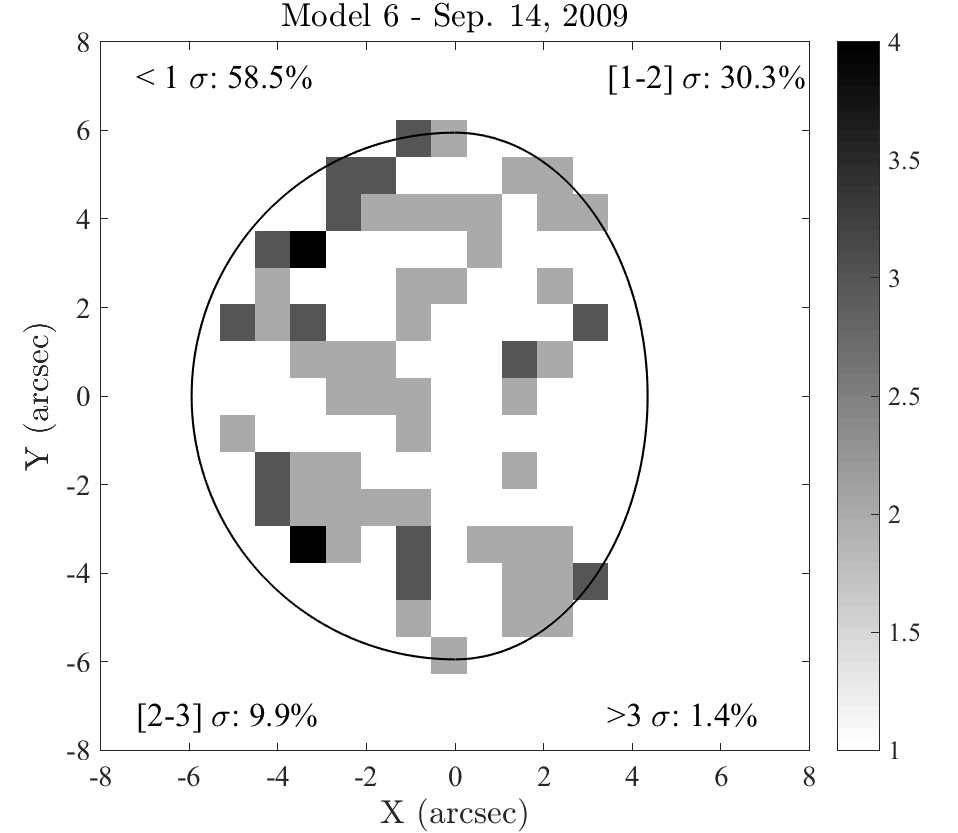}\includegraphics[height=3.8cm]{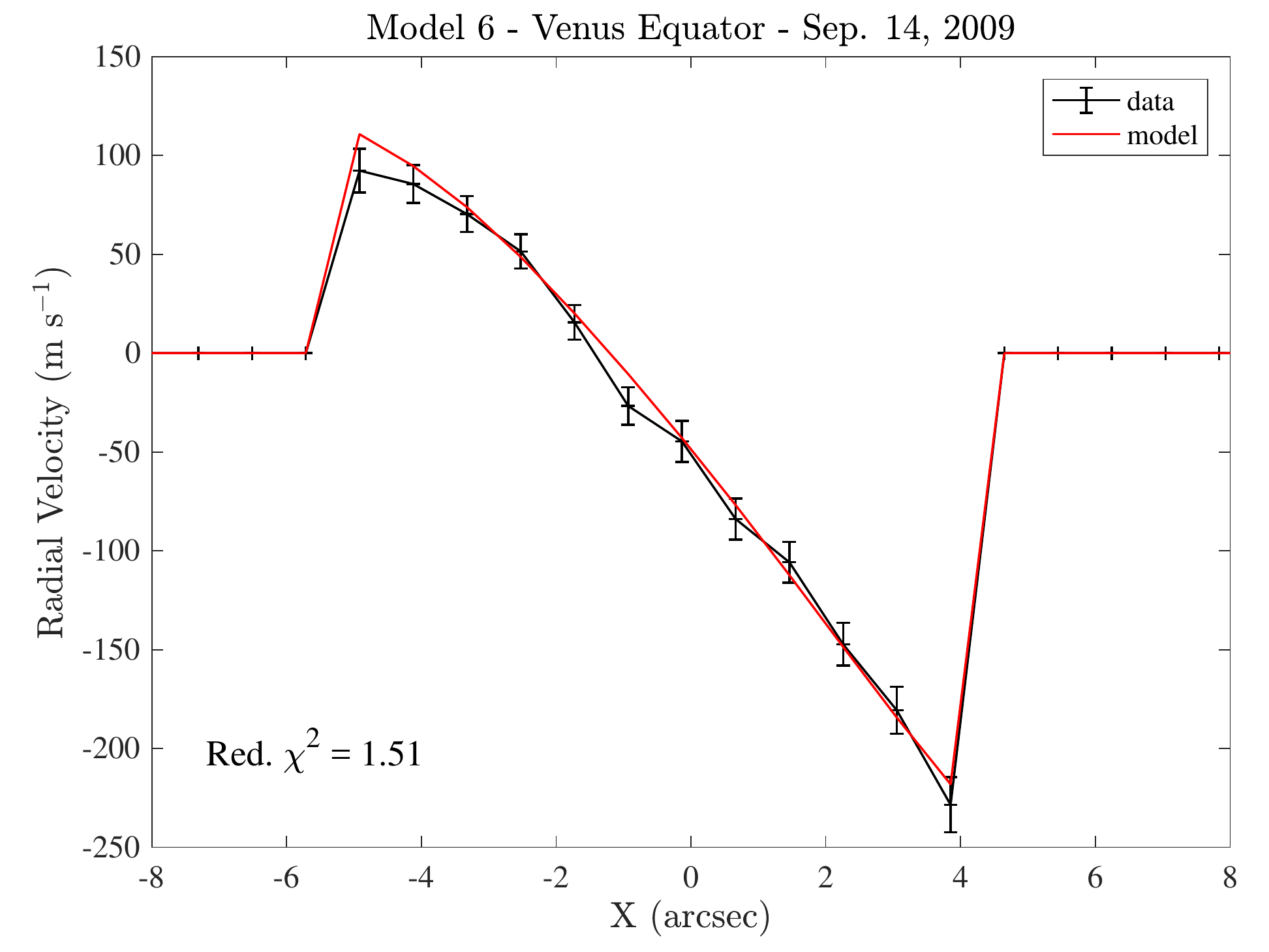}\\
\includegraphics[height=3.8cm]{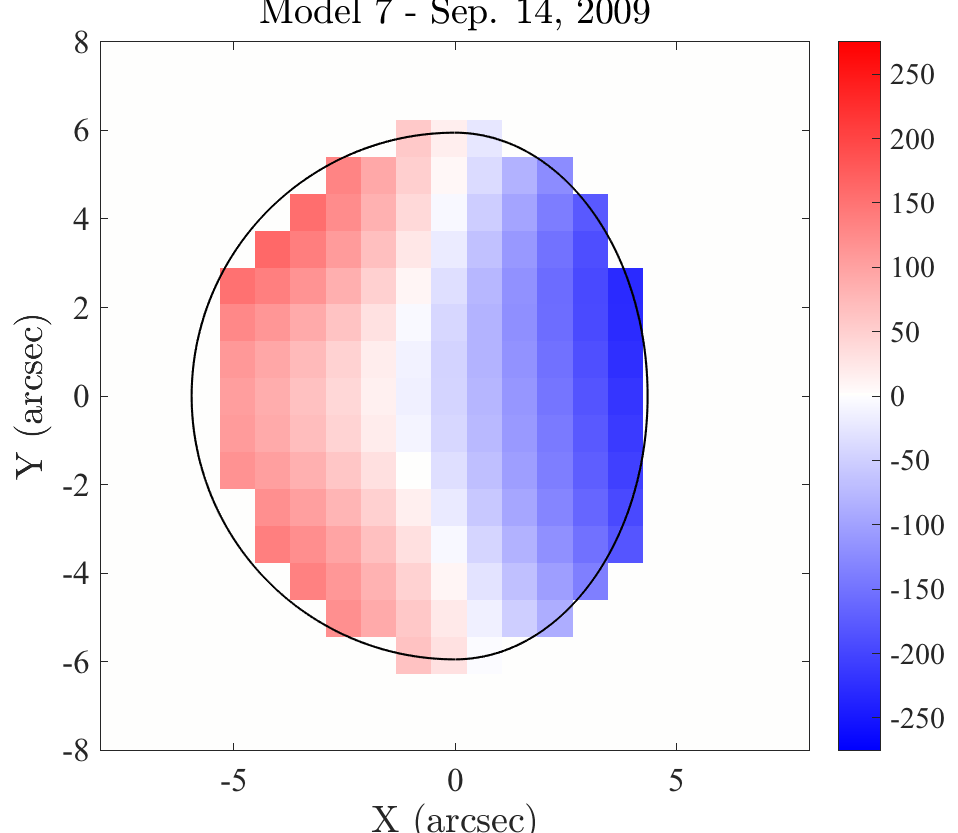}\includegraphics[height=3.8cm]{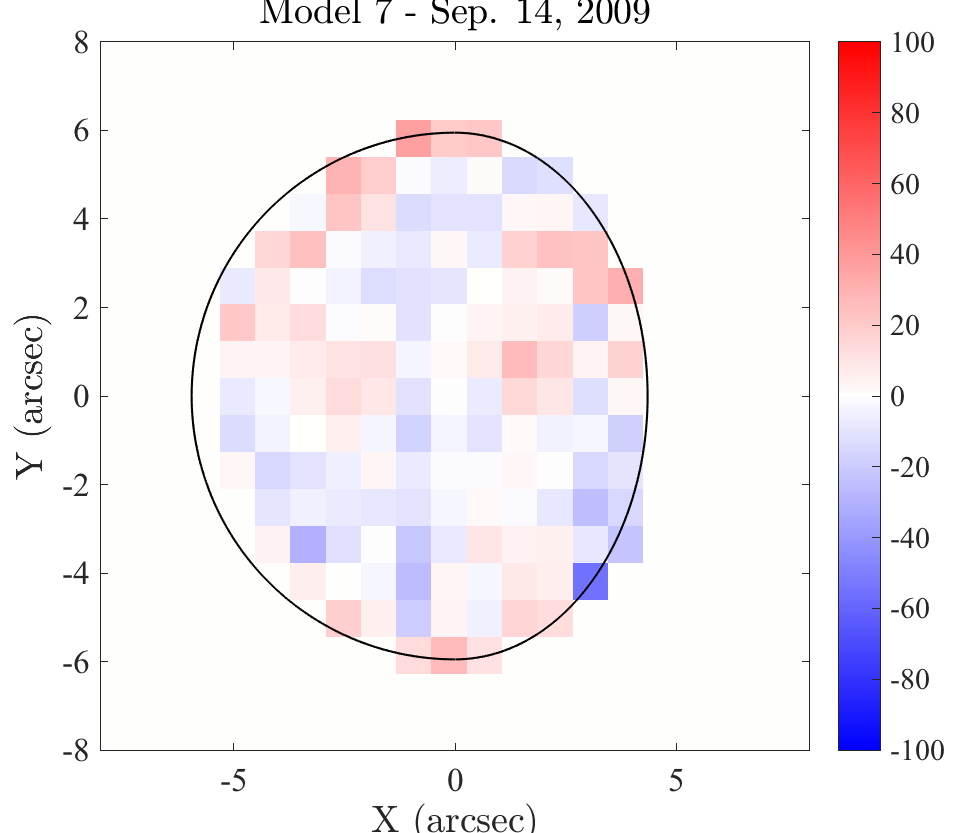}\includegraphics[height=3.8cm]{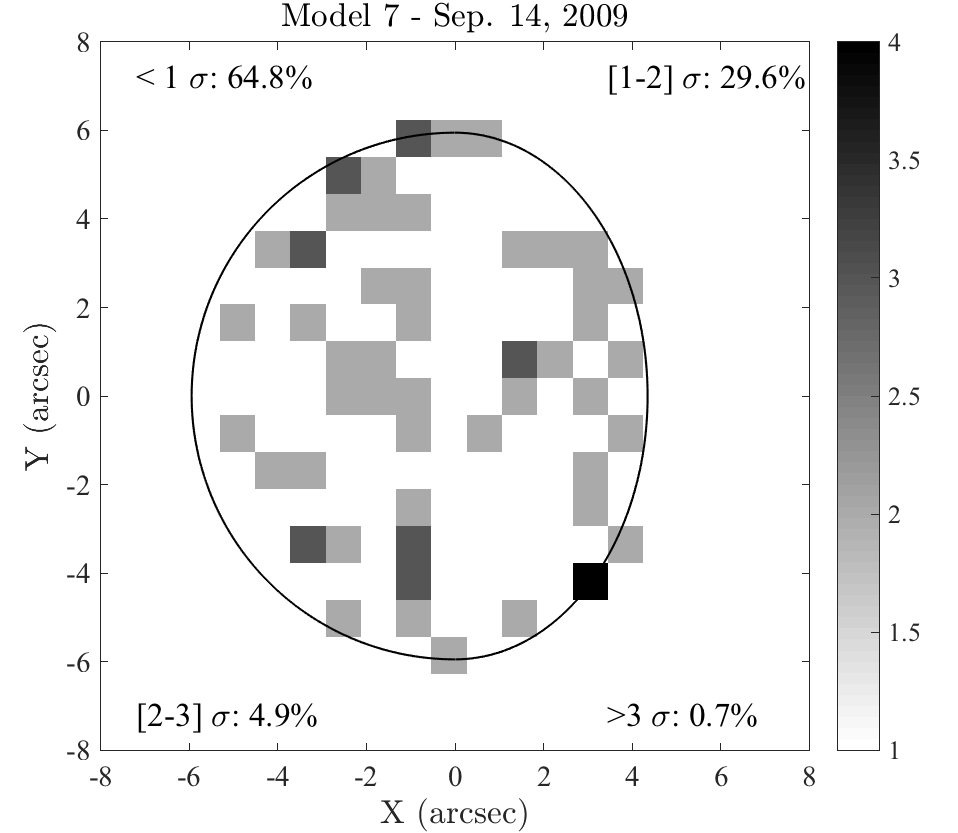}\includegraphics[height=3.8cm]{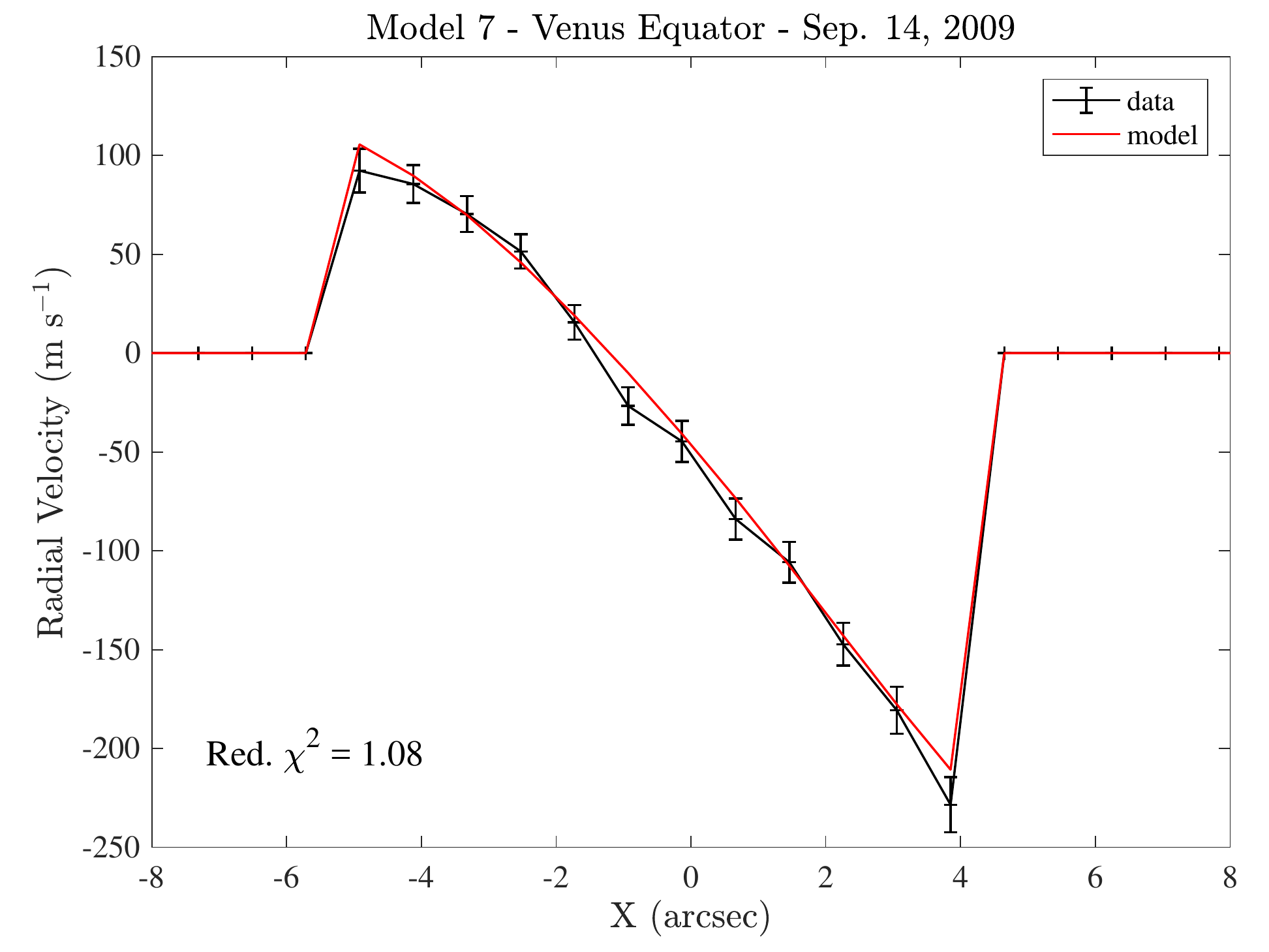}
\caption{Same as Fig. \ref {fig_models_1_4} for models 5 to 7.}
\label{fig_models_5_7}
\end{figure*}

\section{Models of atmospheric circulation}
\label{sect_inversion}

In this section, we aim interprete the RV map obtained on September 14, 2009 (Fig. \ref{fig_doppler_2009}, upper panel). We first note that this map was computed by stacking data over about eight hours, which represents a little less than 10\,\% of Venus' rotation at that altitude. We can therefore consider this map as an ``instantaneous'' Doppler snapshot of Venus. As far as we know, it is the first instantaneous optically resolved RV map of any planet of the solar system in the visible domain. The RV image of Jupiter obtained by \citet{Goncalves_2019} with the JOVIAL/JIVE instrument are averaged over several complete planetary rotations. 

Because we expect, or at least we cannot exclude, the presence of both zonal and meridional winds and the variations of winds with local time, we must extract the atmospheric circulation pattern with forward modeling. It consists of fitting the RV map with a simulated RV map that is composed of a sum of different terms (zonal, meridional, etc), which we alter according to the atmospheric seeing. The model is optimized with an MCMC routine that we developed to interpret these data. Such an approach, with respect to classical least-squares fitting, is a straightforward and reliable way to compute error bars on the estimated parameters. Two examples of outputs of the MCMC code are shown in Fig. \ref{fig_mcmc_modele_circ}. These examples correspond with two models that we comment on the next paragraphs.

We consider several types of types of models, from a simple solid-body rotator to circulation patterns involving variations of wind speed with local time and hemispheric asymmetries. We show that if a simple solid-body rotator is not sufficient to properly fit our observed map, models that are too sophisticated cannot be discriminated as the S/N level of the RV map does not allow us to look for small-scale structures. We note that the atmospheric seeing is about 3 arcsec in average on the map\ and the apparent diameter 11.9 arcsec. 

The first impression we get by looking at the RV map is that it looks similar to the simulated solid-body rotator from Fig. \ref{fig_RV_simu}  (top panel), which is expressed as
\begin{equation}
\Delta V\ind{model\,1} =  V\ind{z,0} \cos{\lambda} + \kappa,
\label{eq_model_1}
\end{equation} 
where $V\ind{z,0}$ is the zonal wind amplitude at the equator, $\lambda$ is the latitude (from 0 to $\pi/2$ in each hemisphere), and $\kappa$ is an offset. We note that the offset term $\kappa$ is meant to take into account the fact that our measurements are not absolute. We introduce the subscript ``0'' in $V\ind{z,0}$ to be consistent with other models (3 to 7), where more than one term ie employed to describe the zonal circulation. However, fitting the map with a pure solid-body zonal circulation pattern plus a global RV offset (Model~1, see Fig. \ref{fig_models_1_4} and Table \ref{tab_modeles}) appears to be rather unsatisfactory given the structure of the residuals (Fig. \ref{fig_models_1_4}, top row, middle panel), and the large value of the reduced $\chi^2$ at 6.14. To help evaluate the fitting quality, we also plot both data and model across the planet's equator in the third column of Fig. \ref{fig_models_1_4}. The disagreement with a solid rotator is obvious. Furthermore, the residual map shows a symmetrical feature in both hemispheres at mid-high latitudes, which recalls the Doppler signature of an equator-to-pole circulation peaking at mid-latitudes (Fig. \ref{fig_RV_simu}, bottom panel). 

We then tested a second model composed of a solid-body zonal and an equator-to-pole meridional circulation pattern, i.e.,
\begin{equation}
\Delta V\ind{model\,2} =  V\ind{z,0} \cos{\lambda} + V\ind{m,0}\, |\sin{2\lambda}| + \kappa,
\label{eq_model_2}
\end{equation} 
where $V\ind{m,0}$ is the speed of the meridional circulation at $\lambda = 45^\circ$.
The results of model~2 are shown on the second row of Fig. \ref{fig_models_1_4}. The residual map and reduced $\chi^2$ (3.31) indicate a better agreement of the model with respect to the data, but a redshift bulb located in the equatorial region indicates that there are more terms to include in the model. Model~3 includes a variation of the zonal circulation as a function of longitude with either a maximum or a minimum located at the subsolar longitude (local noon meridian), i.e.,
\begin{equation}
\Delta V\ind{model\,3} =  \left(V\ind{z,0} - V\ind{z,1} \sin{\phi}\right) \cos{\lambda} + V\ind{m,0}\, |\sin{2\lambda}| + \kappa,
\label{eq_model_3}
\end{equation} 
where $\phi$ is the longitude, ranging from 0 to $\pi$ from the evening to the morning terminators, and $V\ind{z,1}$ is the amplitude of the noon-centered departure to a uniform solid-body rotation. The fit does not significantly improve with respect to the previous fit (red. $\chi^2 = 2.90$). To make the model agree better, we add the option that the longitudinal variation is concentrated at low latitudes, i.e., within $\pm20/30^\circ$ around the equator. Model 4 is built upon this assumption, by including a departure to zonal wind that is modulated by a Gaussian function centered around the equator
\begin{equation}
\begin{split}
\Delta V\ind{model\,4} =\ &  V\ind{z,0} \cos{\lambda} - V\ind{z,1} \sin{\phi} \, e^{-\lambda^2/(2 \sigma\ind{z,1})^2}  \\ 
& +  V\ind{m,0}\, |\sin{2\lambda}| + \kappa,
\label{eq_model_4}
\end{split}
\end{equation}
where $V\ind{z,0}$ is the solid-body component at equator (i.e., the equatorial velocity at dawn for this specific model), $V\ind{z,1}$ is the amplitude of the ``equatorial noon-centered'' departure from a solid-body zonal circulation, and $\sigma\ind{z,1}$ is the standard deviation of the Gaussian function. Residuals are still present but are much less significant; i.e., $\approx90\,\%$ of the points within $2 \sigma$ with respect to the model.  This is illustrated by a much lower reduced $\chi^2$ (1.42) and the cut along the equator, which shows a satisfactory agreement between the data and model.  In Fig. \ref{fig_model4} we represent the map of the zonal wind\footnote{We do not make a similar plot for meridional circulation as it is a simple equator to pole regime (sine curve from equator to pole), which is uniform as a function of the longitude.} deduced from the model 4, where it appears that zonal winds are larger than 200 m s$^{-1}$ in the morning -- and evening, by extrapolating -- and get slower in the subsolar area, down to 70 m s$^{-1}$. These values are not considered as accurate values of the wind speed at noon or in the morning, but as the result of a global fit of the atmospheric circulation of Venus in which global trends are retrieved but not local wind speed. The main fact here is that zonal winds are faster by a factor 2 to 3 in the morning and evening  with respect to the subsolar region. In the following, we refer to model 4 as a circulation pattern in which global zonal winds display a hot spot structure -- recalling some hot Jupiter's nomenclature -- in which winds are slower in the region where solar heating is maximum. 

That being said, the difficulty is to not overinterpret data that likely suffer from biases and were obtained in nonoptimal conditions. Is it possible to go further without interpreting noise? We make other three attempts to refine the atmospheric circulation model. The first (model 5) lets the hot spot free to shift along the equator, instead of being centered around noon, i.e.,
\begin{equation}
\begin{split}
\Delta V\ind{model\,5} =\ &  V\ind{z,0} \cos{\lambda} - (V\ind{z,1,sin} \sin{\phi} + V\ind{z,1,cos} \cos{\phi}) \, e^{-\lambda^2/(2 \sigma\ind{z,1})^2}  \\ 
& +  V\ind{m,0}\, |\sin{2\lambda}| + \kappa,
\label{eq_model_6}
\end{split}
\end{equation}
where $V\ind{z,0}$ is the solid-body component at equator, $V\ind{z,1,sin}$ and $V\ind{z,1,cos}$ are the amplitude of the sin and cos component of the departure to pure solid-body zonal wind. The result is that the spot shifts of $\delta\phi\sim7^\circ$ away for the subsolar meridian (noon), which is not significant given the noise\footnote{$\delta\phi = \tan^{-1}\left(\frac{V\ind{z,1,cos}}{V\ind{z,1,sin}}\right)$}. Model 6 is identical to model 4  (hot spot is centered around noon) with the exception that both zonal and meridional flows may be periodic functions of the longitude, i.e.,
\begin{equation}
\begin{split}
\Delta V\ind{model\,6} =\ &  V\ind{z,0} \cos{\lambda} - V\ind{z,1} \sin{\phi} \, e^{-\lambda^2/(2 \sigma\ind{z,1})^2}  \\ 
& +  V\ind{m,0}\, |\sin{2\lambda}| + V\ind{m,1}\sin{\phi}  + \kappa, \label{eq_model_6}
\end{split}
\end{equation}
where $V\ind{m,1}$ is the amplitude of the noon-meridional departure to uniform Hadley-cell meridional circulation. Unfortunately, in both cases the residual maps and reduced $\chi^2$ (1.41 and 1.51, respectively) do not show any significant difference with respect to model 4; this tends to indicate that model 4 reached the best possible model of this RV map and that going beyond is not reliable. 

\begin{figure}[t]
\includegraphics[width=8.cm]{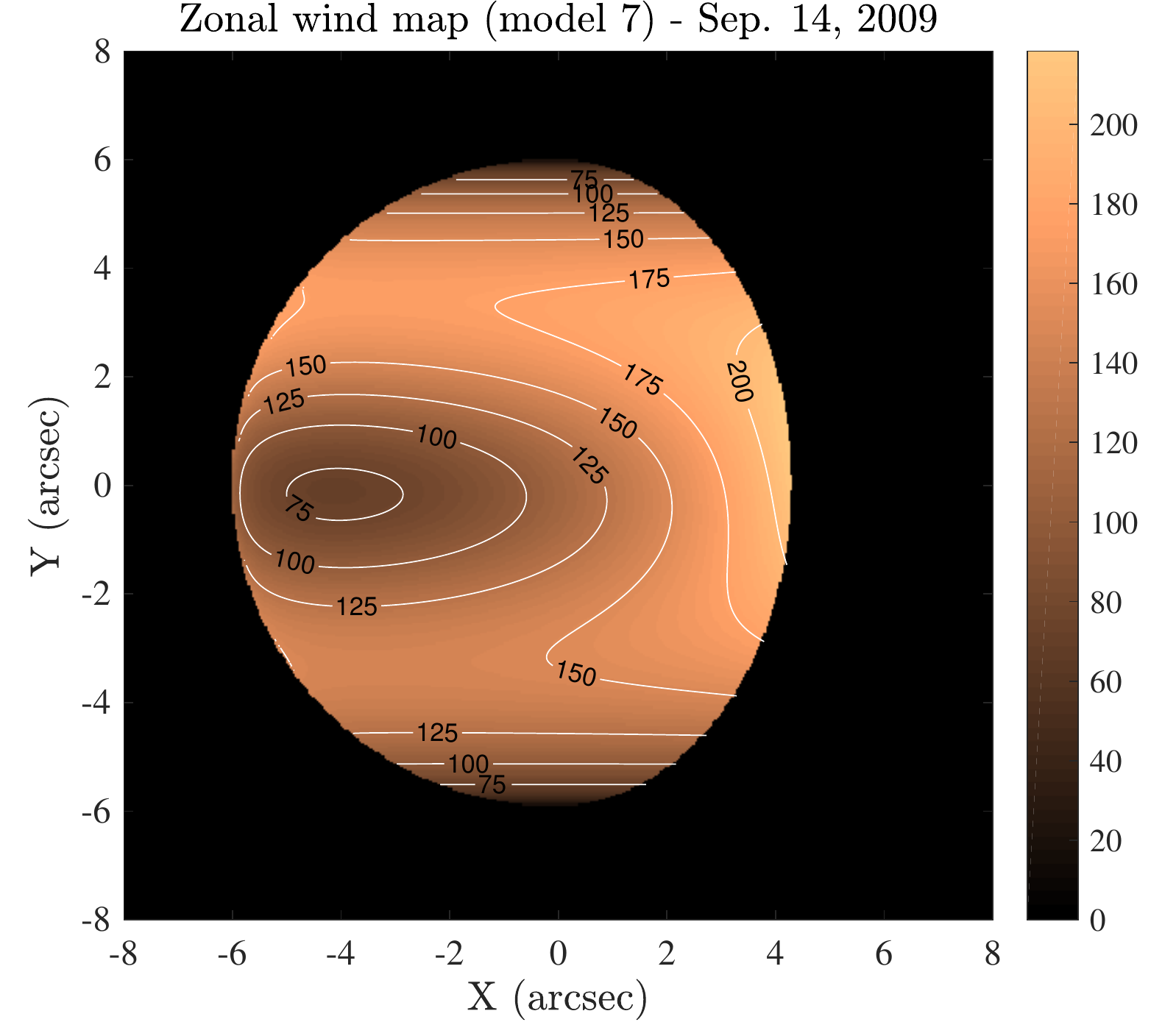}
\includegraphics[width=8.cm]{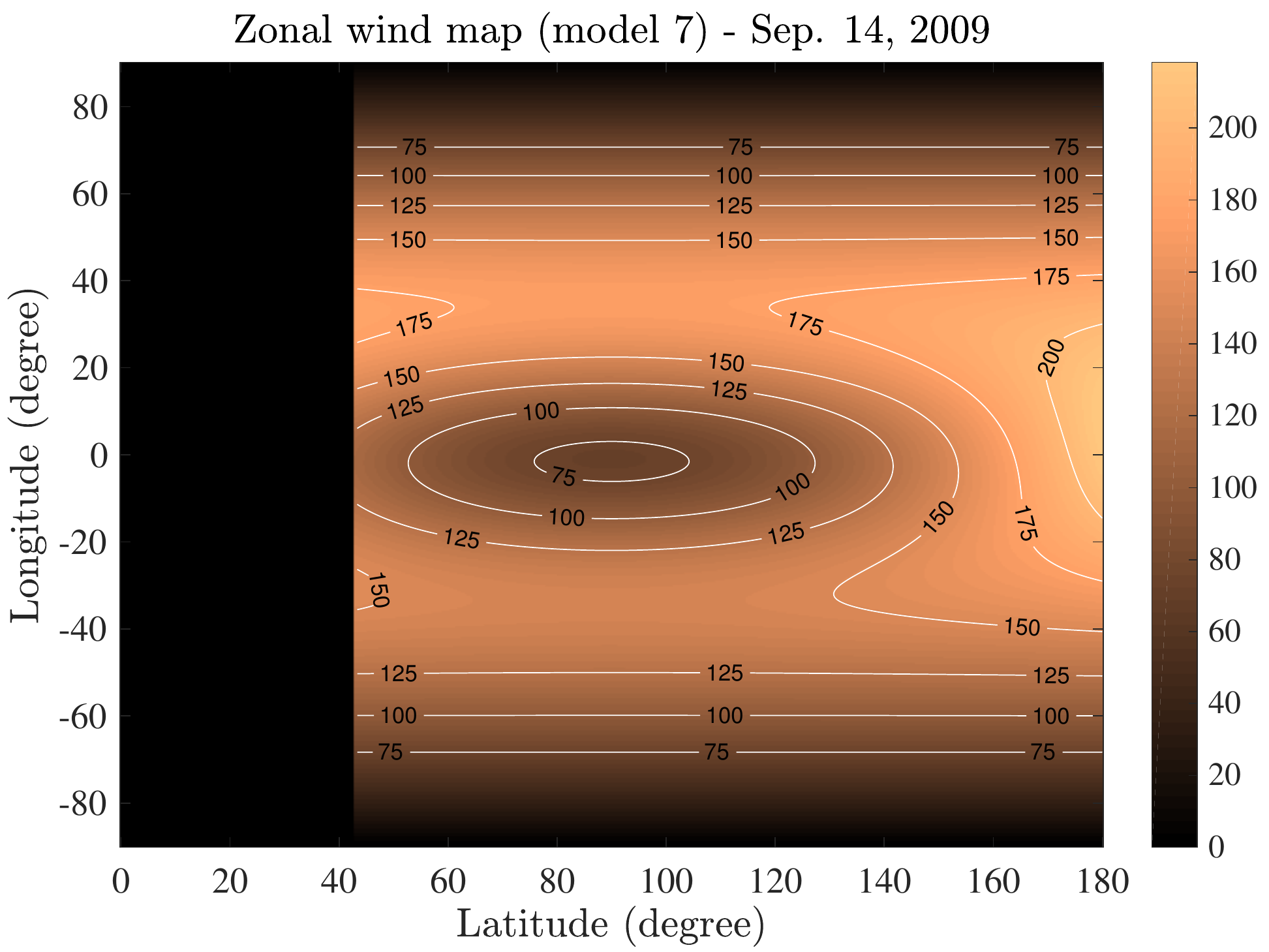}
\caption{Zonal wind map that corresponds to Model 7.}
\label{fig_model7}
\end{figure}

An option to interpret these data a little further consists of identifying large-scale patterns among the residuals of model 4. We may note, by eye, that to the west of the bisector meridian, the residuals are mostly red in the northern hemisphere and blue in the southern, and vice versa to the east of the meridian. 
Interpreting the ``quadrupolar'' appearance of the residuals out of the equatorial region could be caused by an asymmetric zonal wind circulation, faster in the southern hemisphere. Model 7 includes the possibility of a faster zonal circulation in either hemisphere as follows: 
\begin{equation}
\begin{split}
\Delta V\ind{model\,7} =\ &  V\ind{z,0} \cos{\lambda} - V\ind{z,1} \sin{\phi} \, e^{-\lambda^2/(2 \sigma\ind{z,1})^2}  \\ 
& -V\ind{z,2}\cos{\lambda}\sin{(2\lambda)}  +  V\ind{m,0}\, |\sin{2\lambda}| + \kappa,
\label{eq_model_7}
\end{split}
\end{equation}
where the term $V\ind{z,2}\cos{\lambda}\sin{(2\lambda)}$ is a departure to symmetrical solid-body rotation. The term $\sin{(2\lambda)}$ involves that the NS zonal wind asymmetry is smooth (0 at equator and maximum at 45$^\circ$). Contrarily to models 5 and 6, model 7 improves the match between data and fit (red. $\chi^2 = 1.08$, Fig. \ref{fig_model7}, Table \ref{tab_modeles}). According to the results shown in Table \ref{tab_modeles}, the zonal wind would be slightly larger in the northern hemisphere by about $18 \times\cos(\pi/4) \approx 13$ m s$^{-1}$ at mid latitudes.

\section{Discussion and prospects}
\label{sect_prospects}
This paper presents the first complete Doppler snapshots of a planet in the visible domain. The map obtained on September 14, 2009 is the result of integrating eight hours of data, which represents about 10\% of the rotation period at the cloud-top altitude\footnote{The maps obtained on September 16 and 17, 2009 are the result of integrating 2 and 3.5 hours of data, respectively, which represent about 2.5 and 4.5\,\% of the rotation period at the cloud-top altitude.}. Despite poor observing conditions -- even on September 14 the average seeing was 3 arcsec -- we note a strong coherence of the RV maps (Fig. \ref{fig_doppler_2009}). Beyond biases on\ RVs, Doppler shifts vary on the same range from side to side (color scale are the same in Fig. \ref{fig_doppler_2009}), indicating a clear retrograde rotation.

From a technical point of view, this paper makes use of an innovative method to measure and analyze the RVs of planetary atmosphere, following the recipes introduced by \citet{Gaulme_2018} and applied in parallel to this work to observations of Jupiter \citep{Goncalves_2019}. The main improvements consist of using the correct expression of the Young effect, and including the atmospheric seeing both to locate (a posteriori) the spectrometer on the planet and to model RVs. Besides, we perform forward modeling to extract the wind components, by simultaneously fitting different components of the atmospheric circulation. The development of a dedicated MCMC routine ensures a good exploration of the parameter space and a proper estimate of error bars.

The first main result confirms what was expected from both cloud tracking and recent spectroscopic observations: solid body rotation alone is not sufficient to model observations and equator-to-pole meridional circulation is needed. However, we find the speed of zonal and meridional winds to be larger than previously measured. It is hard to compare the zonal wind values because we identify a strong longitudinal and latitudinal variation, however, no observations have indicated winds as large as 200 m s$^{-1}$ at the morning terminator in the equatorial region, so far. As regards meridional winds, our model is identical to that used to interpret previous observations \citep[e.g.,][]{Machado_2017} and we find an amplitude about twice as large (about 45 instead of 22 m s$^{-1}$ at mid latitudes).

\begin{figure}[t]
\includegraphics[width=8.cm]{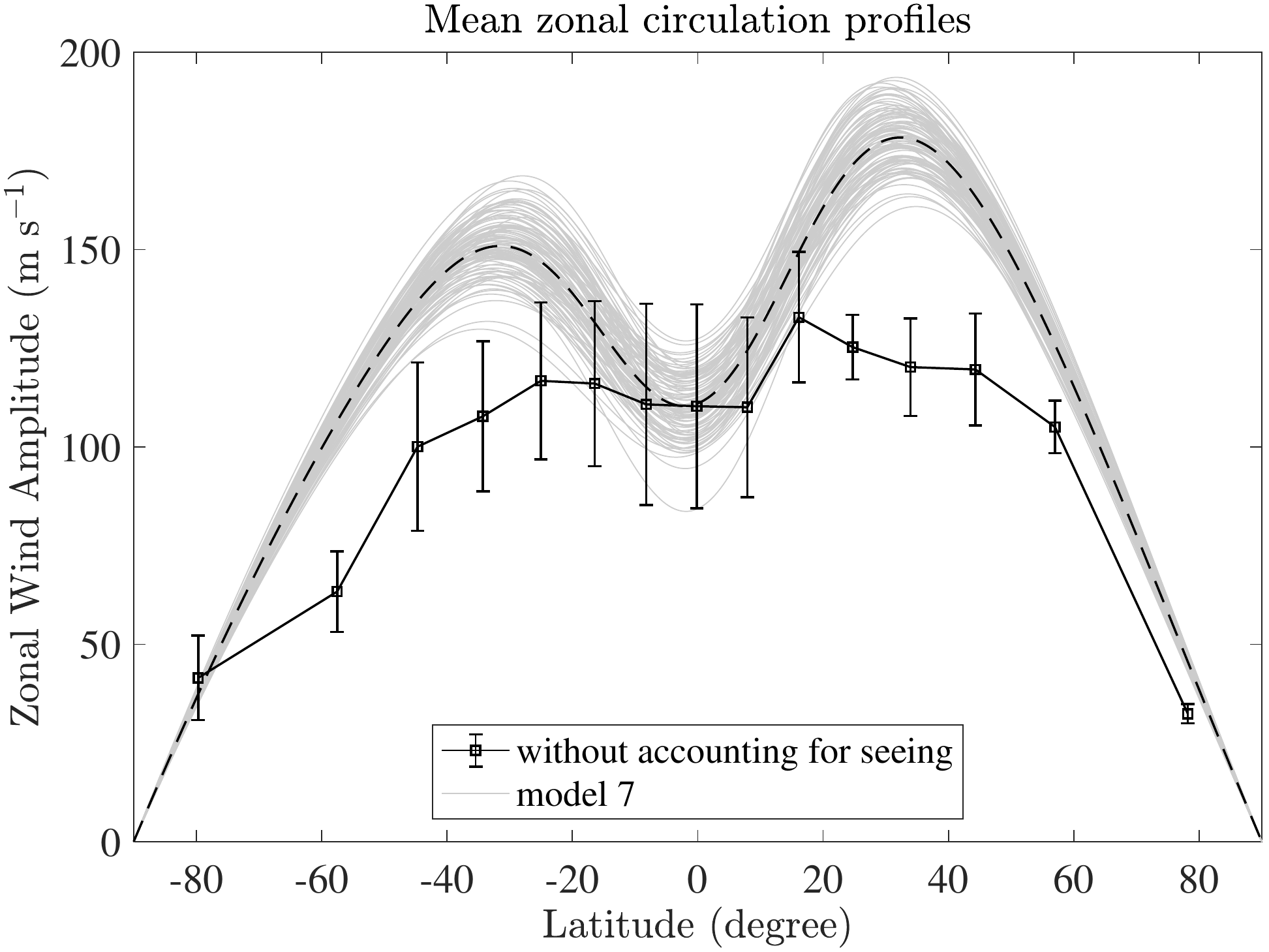}
\caption{Plain black line: fitting of the RV map obtained from the observations done on September 14, 2009 without taking account of the atmospheric seeing on RV profiles. Gray lines: mean zonal wind on the visible part of the dayside of Venus as computed from model 7 based on the observations done on September 14, 2009. In other words, it is the mean of the map shown in Fig \ref{fig_model7}, bottom panel, with 100 executions to take into account the errors on the parameters of the model. The black dashed line indicates the average mean profile of model 7.}
\label{fig_M_shape}
\end{figure}

The speeds of the winds that we report are significantly higher than those that have been reported so far from both cloud tracking and spectroscopic measurements. We first note that no study has ever produced a complete RV map of Venus in the visible, which makes a direct comparison hard to perform. However, we now explore how compatible our results are with respect to those from the literature, in particular the most recent results by \citet{Machado_2012,Machado_2014,Machado_2017}. For this, we fit the RV map obtained on September 14, 2009 with a simple zonal plus meridional wind without taking account the biases on RVs caused by the atmospheric seeing. The model consists of an equator-to-pole meridional circulation ($\propto |\sin2\lambda|$) and a zonal wind whose amplitude is fitted for each band of latitude. The zonal wind profile as a function latitude obtained that way is shown in Fig. \ref{fig_M_shape}. We observe a (noisy) M-shaped profile where  zonal winds peak at mid-latitudes with an amplitude of about $120$ m s$^{-1}$ and a local minimum around the equator at about 100 m s$^{-1}$, which is very similar to what is reported in the literature. This suggests that if atmospheric seeing and instrumental PSF were included in the previous high-resolution spectroscopic works, a larger zonal wind would have been measured. We are aware that reported seeing and pointing accuracy by \citet{Machado_2014,Machado_2017} are below one arcsec each. However, the diameter of the spectrometer's fiber is 1.6 arcsec, which implies that combined instrumental plus atmospheric PSF is larger than 2 arcsec and that their RV measurements are likely to be biased too, especially toward the planetary edges. 
Interestingly, we note that wind measurements of Venus by \citet{Machado_2012, Machado_2014, Machado_2017} provide very consistent results, with average  zonal winds estimated in between 117 and 123 m s$^{-1}$, whereas \citet{Widemann_2007,Gabsi_2008} reported much lower values ($\approx 75$ m s$^{-1}$). The earlier results were obtained either with the CFHT or the VLT telescopes, which are located on sites where the seeing is very good (maximum seeing reported is 1.4 arcsec in \citealt{Machado_2012}). To the contrary, both papers \citet{Widemann_2007} and \citet{Gabsi_2008} reported rather poor observing conditions at the Observatoire de Haute Provence, which could explain in part why they obtain such low wind estimates. 

The second main result is the hot spot structure of the atmospheric circulation, at least for the zonal component, which had never been suggested that clearly so far. Cloud tracking measurements and Doppler spectroscopic measurements indicate possible longitudinal variations of the wind as functions of local time with faster circulation toward the terminator \citep[e.g.,][]{Khatuntsev_2013,Hueso_2015,Machado_2017}. In particular, \citet{Khatuntsev_2013} found diurnal variations of the zonal winds but at the $\sim10$ m s$^{-1}$ level only for local times 6 h to 18 h. Since then, based on VIRTIS VEx observations, \citet{Hueso_2015} reported zonal winds variations as a function of longitude, from to 90 m s$^{-1}$ slightly before noon (local time) up to 130 m s$^{-1}$ at 17 h for latitudes of about $-30^\circ$. \citet{Machado_2017} also reported zonal wind variations at $[20, 40]^\circ$ latitudes, ranging from 125 at $\approx10$ h to 167 m s$^{-1}$ at $\sim7$ h. We also note that their local values of zonal winds are a little larger in the  northern hemisphere, as we find in this work. However, no such hot spot pattern has been identified. In contrast, from clouds tracking measurements performed on images from the VEx VMC, \citet{Bertaux_2016} did not find zonal wind speed variation as a function of local time, but as a function of geographical features with values ranging from 101 to 83 m s$^{-1}$ at latitudes in between 5 and $15^\circ$S. 

We note that a hot spot regime of zonal winds is compatible with an M-shaped mean zonal wind profile as a function of latitude. If we average the zonal winds at a given latitude, we obtain that type of profile, with no wind at poles, a maximum amplitude at mid-latitudes and a local minimum in the equatorial region. Figure \ref{fig_M_shape} also shows the zonal profile obtained by averaging the zonal wind along longitudes on the visible dayside of Venus: it shows that the observed M-shape can be the result of a hot spot structure.
Such a regime recalls global circulation models that are proposed to explain light curves of stars hosting hot Jupiters. In particular, \citet{Showman_Guillot_2002} studied the atmospheric circulation of hot Jupiters that are locked to their host stars. Venus can be seen to be somewhat similar as its atmosphere is very dense (90 bars on the ground) and its rotation period (243 days) is very close to its orbital period (225 days). A significant difference from the case studied by \citet{Showman_Guillot_2002} regards the global circulation regime, which is assumed to be in geostrophic regime for hot Jupiters, and which is considered to be cyclostrophic in the case of planets with little planetary rotation such as Venus or Titan. Detailed numerical simulations are needed to investigate this question further, which is beyond the scope of our paper.

Our two main results, i.e., both the zonal and meridonal large wind speeds as well as the hot spot pattern of the zonal flow,  are significantly different with respect to what has been measured so far. They will likely remain controversial until further new independent observations, either by another team or with another instrument. Indeed, the presence of uncontrolled and not fully understood biases (see Sect \ref{sect_biases}) makes the conclusion partially questionable.
This works points out the difficulty of constructing RV maps from individual spectra, even with a long-slit spectrometer that covers the whole planetary diameter at once. The constantly changing observation conditions and pointing stability of the telescope, as well as the wavelength stability of the spectrometer, are major challenges to conduct clean measurements. 

There are two options for go further. The first consists of using the same type of instrumentation (long-slit spectrometry) with adaptive optics (AO) to reduce the seeing issue, which has a huge impact on map reconstruction. At the same time, even with AO, these kinds of observations are dependent on thermal drifts of the spectrometer and on clouds during observations. The THEMIS telescope plans to get an AO system soon, however it will be designed for working on the Sun, which is likely not adapted for observing planetary targets. In the short term, the most promising alternative is the JOVIAL/JIVE project  (principal investigator is second author Schmider), which consists of three identical Doppler imaging spectrometers that will be installed in France, New Mexico, and Japan \citep{Goncalves_2016,Goncalves_2019}. The original purpose of the project is to get RV maps of Jupiter and Saturn at short cadence (30 s) to conduct seismic observations of these planets similar to techniques of helioseismology. This second objective of this instrument is to measure the atmospheric dynamics of the dense atmospheres of the solar system, i.e., mostly Venus, Jupiter, and Saturn, given that Uranus, Neptune, and Titan are faint and barely resolved from the ground. The JOVIAL/JIVE instrument is able to provide a complete RV map in only one exposure and does not have any of the issues we encountered in this work in building maps from 1D segments. A paper dedicated to the first Doppler measurements of the wind of Jupiter was recently published by \citet{Goncalves_2019}. Preliminary observations of Venus were led in spring 2018, and more observations could be done with AO with the C2PU 1 m telescope at the Calern observatory (France) in 2019 or 2020. We will first look at whether the hot spot structure is real with the help of high spatial resolution. The second exciting opportunity will be to make use of the three instruments mentioned above in order to observe Venus continuously for several days to study time variations of the atmospheric circulation of the planet.



\section*{Acknowledgments} P. Gaulme was supported in part by the German space agency (Deutsches Zentrum f\"ur Luft- und Raumfahrt) under PLATO data grant 50OO1501. 
I. Gon\c{c}alves's PhD was granted by Observatoire de la C\^ote d'Azur and the JIVE in NM project (NASA EPSCoR grant \#NNX14AN67A). THEMIS is a UPS of the CNRS funded by INSU. The authors are grateful to the referee, E. Lellouch, for his intensive review of the paper, which contributed to improving its quality. 
P. Gaulme wishes to apologize to P. Gava for leaving her home on her thirtieth birthday while observing Venus at Teide observatory in 2008.

\begin{table*}[h!]
\small
 \caption{Global circulation models. The value $V\ind{z,0}$ is the amplitude of zonal winds along the equator, $V\ind{z,1}$ is the amplitude of the solid-body (SB) zonal component for models that have longitudinal variations of the zonal circulation (models 4 to 7), $V\ind{m,0}$ is the amplitude of meridional winds at latitude $\lambda =45^\circ$, and $\kappa$ is a global RV offset. Model 3: $V\ind{z,1}$ (``noon zonal'' component) is the amplitude of departure to solid body rotation which is minimum at noon. Models 4, 6,  and 7: $V\ind{z,1}$ is the amplitude of the equatorial noon-centered zonal component. Model 5:  $V\ind{z,1,sin}$ and $V\ind{z,1,cos}$ are the amplitudes of the equatorial decentered zonal components for the sine and cosine, respectively. Model 6: $V\ind{z,1}$ is the amplitude of the noon-centered meridional component. All terms are in m s$^{-1}$ except $\sigma\ind{z,1}$, which is in degrees.
}
\center
\begin{tabular}{l  | l l | l l | c}
\hline
\multicolumn{5}{c}{\sf{Model 1}} & Red. $\chi^2$\\
\hline
\multirow{1}{*}{Pure zonal  (solid-body)}  & $V\ind{z,0}\ \cos{\lambda}$ & Zonal amplitude & $V\ind{z,0}$      & $157.6_{-2.3}^{+0.3}$ & \multirow{2}{*}{6.14}\\
& $+ \kappa$  &  Offset & $\kappa$    & $5.5_{-1.0} ^{+0.7}$  &  \\ 
\hline
\multicolumn{5}{c}{\sf{Model 2}}\\
\hline
\multirow{1}{*}{Zonal (solid-body)}            & $V\ind{z,0}\ \cos{\lambda}$ & Zonal amplitude & $V\ind{z,0}$      & $156.9_{-1.4}^{+1.2}$ &   \multirow{3}{*}{3.31}\\
\& Meridional       & $+V\ind{m,0}\ |\sin{\lambda}|$ &  Meridional  &$V\ind{m,0}$    & $57.9_{-2.7}^{+2.4}$ &   \\ 
&$+\kappa$  &  Offset & $\kappa$    & $-25.3_{-1.8}^{+1.4}$  & \\ 
\hline
\multicolumn{5}{c}{\sf{Model 3}}\\
\hline
 \multirow{1}{*}{Zonal (solid-body),}         & $V\ind{z,0}\ \cos{\lambda}$ & Zonal amplitude & $V\ind{z,0}$      & $211.8_{-6.4}^{+7.8}$ &  \multirow{4}{*}{2.90} \\
\multirow{1}{*}{Noon zonal variation}     & $ - V\ind{z,1}\ \sin{\phi}$ &  Departure to zonal  &$V\ind{z,1}$    & $79.9_{-8.4}^{+12.0}$ & \\ 
\& Meridional  & $+V\ind{m,0}\ |\sin{\lambda}|$ &  Meridional  &$V\ind{m,0}$    & $47.9_{-2.7}^{+2.8}$ & \\ 
 &$+\kappa$  &  Offset & $\kappa$    & $-10.4_{-2.6}^{2.3}$ & \\ 
\hline
\multicolumn{5}{c}{\sf{Model 4}}\\
\hline
 \multirow{1}{*}{Zonal (solid-body),}                       & $V\ind{z,0} \cos{\lambda}$ & SB zonal amplitude       & $V\ind{z,0}$      & $214.4 _{-5.8}^{+5.1}$ &  \multirow{5}{*}{1.42}\\
\multirow{2}{*}{Equatorial noon zonal variation}     &\multirow{2}{*}{$-V\ind{z,1}\ \sin{\phi} \, e^{-\lambda^2/(2 \sigma\ind{z,1})^2} $}          &  Departure to zonal   &$V\ind{z,1}$    & $143.2_{-8.3}^{+6.8}$ &  \\ 
    &&  Sigma of the Gaussian  &$\sigma\ind{z,1}$ & $16.7_{-1.7}^{+1.3}$   & \\ 
\& Meridional  & $+V\ind{m,0}\ |\sin{\lambda}|$ &  Meridional  & $V\ind{m,0}$  & $44.9_{-3.0}^{+2.5}$ & \\ 
 &$+\kappa$  &  Offset & $\kappa$    & $-8.4_{-2.3}^{+1.9}$   & \\ 
\hline
\multicolumn{5}{c}{\sf{Model 5}}\\
\hline
 \multirow{1}{*}{Zonal (solid-body),}                       & $V\ind{z,0} \cos{\lambda}$ & SB zonal amplitude       & $V\ind{z,0}$      & $235.4_{-12.9}^{+16.3}$    &      \multirow{6}{*}{1.41}\\
\multirow{3}{*}{Equatorial decentered zonal variation}   & \multirow{2}{*}{$-V\ind{z,1,sin}\ \sin{\phi} \, e^{-\lambda^2/(2 \sigma\ind{z,1})^2} $}  &  Departure to zonal   &$V\ind{z,1,sin}$    & $157.4_{-13.2}^{+15.5}$  & \\ 
                                                                                       & \multirow{2}{*}{$-V\ind{z,1,cos}\ \cos{\phi} \, e^{-\lambda^2/(2 \sigma\ind{z,1})^2} $}  &  Departure to zonal   &$V\ind{z,1,cos}$    & $20.2_{-2.4}^{+2.7}$ &  \\ 
    &&  Sigma of the Gaussian  &$\sigma\ind{z,1}$   & $20.2_{-2.4}^{+2.7}$ \\ 
\& Meridional  & $+V\ind{m,0}\ |\sin{\lambda}|$ &  Meridional  & $V\ind{m,0}$  & $51.8_{-342}^{+3.5}$ & \\ 
 &$+\kappa$  &  Offset & $\kappa$    & $-14.2_{-3.9}^{+2.6}$  &  \\ 
\hline
\multicolumn{5}{c}{\sf{Model 6}}\\
\hline
 \multirow{1}{*}{Zonal (solid-body),}                       & $V\ind{z,0} \cos{\lambda}$ & SB zonal amplitude       & $V\ind{z,0}$      & $218.6_{-11.5}^{+6.7}$   &       \multirow{6}{*}{1.51}\\
\multirow{2}{*}{Equatorial noon zonal variation,}   & \multirow{2}{*}{$-V\ind{z,1}\ \sin{\phi} \, e^{-\lambda^2/(2 \sigma\ind{z,1})^2} $}  &  Departure to zonal   &$V\ind{z,1}$    & $149.9_{-17.2}^{+9.2}$  & \\ 
                                                                            &&  Sigma of the Gaussian  &$\sigma\ind{z,1}$  & $16.8_{-1.8}^{+1.2}$  & \\ 
\& Meridional including & $+V\ind{m,0}\ |\sin{\lambda}|$ &  Meridional  & $V\ind{m,0}$  & $49.2_{-19.6}^{+22.9}$ & \\ 
longitudinal variation (noon centered)   & \multirow{1}{*}{$+V\ind{m,1}\ \sin{\phi} $}  &  Departure to meridional   &$V\ind{m,1}$    &$-1.6_{-35.1}^{+16.3} $  & \\ 
 &$+\kappa$  &  Offset & $\kappa$    & $-7.6_{-3.0}^{+1.9}$  &  \\ 
\hline
\multicolumn{5}{c}{\sf{Model 7}}\\
\hline
 \multirow{1}{*}{Zonal (solid-body),}                       & $V\ind{z,0} \cos{\lambda}$ & SB zonal amplitude       & $V\ind{z,0}$      & $215.4_{-5.5}^{+6.0}$  &       \multirow{6}{*}{1.08}\\
\multirow{2}{*}{Equatorial noon zonal variation}     & \multirow{2}{*}{$-V\ind{z,1}\ \sin{\phi} \, e^{-\lambda^2/(2 \sigma\ind{ENZ})^2} $}          &  Departure to zonal   &$V\ind{z,1}$    & $144.4_{-8.5}^{+6.9}$ & \\ 
    &&  Sigma of the Gaussian  &$\sigma\ind{z,1}$  &   $17.1_{-1.6}^{+1.4} $  & \\ 
\multirow{1}{*}{N-S zonal asymmetry}     & \multirow{1}{*}{$-V\ind{z,2}\ \cos{\lambda}\ \sin{(2\lambda)} $}          &  Amplitude NS asymmetry   &$V\ind{z,2}$    & $18.0_{-2.7}^{+2.1}$   & \\ 
\& Meridional  & $+V\ind{m,0}\ |\sin{\lambda}|$ &  Meridional  & $V\ind{m,0}$  & $44.6_{-2.9}^{+2.7}$  & \\ 
 &$+\kappa$  &  Offset & $\kappa$    &  $-8.4_{-2.3}^{+2.1}$  &  \\ 
\hline
\end{tabular}
\label{tab_modeles}
\end{table*}


\bibliographystyle{aa} 
 \bibliography{bibi_venus.bib}

\begin{appendix}
\section{Data}
\label{sect_app_data}

\hspace{13cm}
\begin{sidewaystable}
\vspace{-9.5cm}
\captionsetup{width=21cm}
\tiny
\caption{\tiny Two-dimensional maps of Venus from data taken on September 14, 2009. The top section is the photometry, the middle is line-of-sight velocity (Doppler signal), the bottom is the 1$\sigma$ error on velocity. Velocity data are provided for the data points that are reliable, i.e., for all the point visible on Fig. \ref{fig_doppler_2009} (middle panel), whereas photometry is provided for out-of-mask regions in case a user would like to fit the complete photometric profile. The step along the $x$-axis (in between columns) is 0.80 arcsec and the step along the $y$-axis (line to line) is 0.83 arcsec. Photometric flux is in analog digital unit, and velocities are in m s$^{-1}$.}
\label{sep14_table}
\centering
\scalebox{0.85}{
\begin{tabular}{c c c c c c c c c c c c c c c c c c c c}
\hline 
\multicolumn{20}{c}{Photometric map} \\ 
\hline 
8.2 &   14.0 &   21.9 &   34.1 &   53.2 &   79.5 &  117.5 &  165.5 &  212.8 &  252.0 &  272.1 &  248.0 &  189.5 &  127.5 &   78.1 &   45.7 &   26.1 &   15.0 &    9.5 &5.6 \\ 
14.1 &   22.3 &   35.3 &   59.3 &   99.0 &  162.1 &  259.1 &  393.5 &  536.7 &  626.4 &  626.8 &  551.1 &  434.5 &  290.0 &  160.8 &   84.4 &   45.5 &   25.3 &   15.3 &10.0 \\ 
23.8 &   39.5 &   67.0 &  113.6 &  195.5 &  339.0 &  572.1 &  846.5 & 1051.3 & 1145.5 & 1124.0 &  996.4 &  795.1 &  554.5 &  328.9 &  174.0 &   82.9 &   41.7 &   23.7 &14.8 \\ 
40.5 &   65.1 &  106.7 &  186.3 &  342.5 &  653.2 & 1045.9 & 1442.2 & 1656.9 & 1675.3 & 1563.3 & 1356.4 & 1112.2 &  827.9 &  537.2 &  294.9 &  146.0 &   73.3 &   41.9 &27.1 \\ 
52.6 &   89.1 &  156.6 &  291.3 &  571.6 & 1058.3 & 1593.8 & 1959.0 & 2075.9 & 2008.6 & 1859.8 & 1647.5 & 1365.6 & 1067.5 &  761.5 &  459.4 &  230.7 &  104.4 &   54.5 &32.8 \\ 
62.6 &  106.5 &  193.7 &  384.4 &  872.0 & 1535.1 & 2126.6 & 2385.0 & 2443.1 & 2374.3 & 2195.1 & 1939.8 & 1628.8 & 1292.0 &  936.3 &  575.3 &  286.5 &  124.1 &   62.0 &36.3 \\ 
80.8 &  143.5 &  263.2 &  529.0 & 1178.4 & 1945.2 & 2452.2 & 2670.1 & 2708.6 & 2602.9 & 2385.0 & 2101.5 & 1773.1 & 1414.2 & 1039.3 &  664.0 &  347.6 &  153.5 &   73.9 &41.7 \\ 
94.8 &  170.3 &  303.0 &  636.3 & 1373.6 & 2160.5 & 2638.5 & 2794.6 & 2808.9 & 2698.3 & 2485.5 & 2207.6 & 1876.8 & 1511.9 & 1119.1 &  733.5 &  402.5 &  182.8 &   85.8 &47.1 \\ 
98.4 &  177.7 &  325.0 &  713.8 & 1392.9 & 2215.9 & 2673.6 & 2816.1 & 2807.5 & 2687.4 & 2455.6 & 2188.4 & 1858.3 & 1511.9 & 1142.7 &  776.5 &  445.0 &  212.9 &   99.7 &52.5 \\ 
97.7 &  172.4 &  315.3 &  657.0 & 1335.4 & 2135.0 & 2640.9 & 2825.9 & 2808.3 & 2685.5 & 2469.0 & 2181.0 & 1859.3 & 1502.2 & 1139.8 &  768.7 &  449.3 &  209.4 &   96.8 &52.5 \\ 
92.1 &  161.3 &  291.3 &  551.1 & 1150.9 & 1895.6 & 2489.2 & 2701.4 & 2778.9 & 2670.3 & 2458.4 & 2175.0 & 1841.2 & 1470.9 & 1089.9 &  706.7 &  382.4 &  183.9 &   90.3 &49.9 \\ 
84.4 &  146.0 &  254.7 &  453.1 &  909.3 & 1537.1 & 2080.7 & 2408.1 & 2466.4 & 2414.7 & 2233.4 & 1989.7 & 1672.2 & 1316.2 &  945.1 &  579.9 &  315.2 &  145.5 &   72.9 &41.6 \\ 
65.1 &  109.0 &  182.5 &  306.7 &  594.8 & 1069.1 & 1553.5 & 1940.5 & 2081.1 & 2036.7 & 1880.8 & 1633.2 & 1357.9 & 1069.7 &  749.0 &  437.3 &  225.2 &  106.5 &   55.7 &32.3 \\ 
46.1 &   74.2 &  122.4 &  204.0 &  353.3 &  635.7 & 1065.2 & 1369.1 & 1577.1 & 1631.2 & 1532.5 & 1333.1 & 1099.2 &  825.4 &  529.7 &  295.4 &  150.9 &   74.8 &   39.5 &24.3 \\ 
31.4 &   49.9 &   79.4 &  126.3 &  201.7 &  328.5 &  570.6 &  891.1 & 1077.5 & 1151.6 & 1119.4 &  963.7 &  777.7 &  526.0 &  315.5 &  177.1 &   91.2 &   47.8 &   27.7 &16.7 \\ 
18.8 &   28.3 &   42.5 &   65.8 &  101.0 &  157.9 &  255.8 &  397.3 &  558.9 &  626.5 &  631.2 &  549.5 &  412.6 &  281.6 &  177.3 &  100.1 &   50.2 &   26.6 &   15.5 &9.1 \\ 
5.4 &    9.0 &   13.7 &   21.1 &   33.3 &   50.6 &   81.8 &  133.5 &  195.3 &  234.9 &  221.6 &  181.0 &  126.2 &   77.1 &   42.2 &   22.6 &   12.5 &    7.5 &    4.5 &2.0 \\ 
\hline 
\multicolumn{20}{c}{Line-of-sight velocity map} \\ 
\hline 
0.0 &    0.0 &    0.0 &    0.0 &    0.0 &    0.0 &    0.0 &    0.0 &    0.0 &    0.0 &    0.0 &    0.0 &    0.0 &    0.0 &    0.0 &    0.0 &    0.0 &    0.0 &    0.0 &0.0 \\ 
0.0 &    0.0 &    0.0 &    0.0 &    0.0 &    0.0 &    0.0 &    0.0 &    0.0 &   76.8 &   56.9 &    8.2 &    0.0 &    0.0 &    0.0 &    0.0 &    0.0 &    0.0 &    0.0 &0.0 \\ 
0.0 &    0.0 &    0.0 &    0.0 &    0.0 &    0.0 &    0.0 &  137.4 &   95.5 &   37.9 &   26.0 &  -19.4 &  -34.6 &  -73.3 &    0.0 &    0.0 &    0.0 &    0.0 &    0.0 &0.0 \\ 
0.0 &    0.0 &    0.0 &    0.0 &    0.0 &    0.0 &  139.8 &  110.5 &   77.6 &   20.8 &   12.9 &  -31.5 &  -58.0 &  -96.5 & -191.4 &    0.0 &    0.0 &    0.0 &    0.0 &0.0 \\ 
0.0 &    0.0 &    0.0 &    0.0 &    0.0 &  140.5 &   89.6 &   84.9 &   66.0 &   10.3 &  -13.7 &  -34.0 &  -76.8 & -112.6 & -160.4 & -204.7 &    0.0 &    0.0 &    0.0 &0.0 \\ 
0.0 &    0.0 &    0.0 &    0.0 &    0.0 &  109.2 &   96.3 &   70.5 &   39.7 &    4.7 &  -23.4 &  -55.2 &  -95.6 & -138.5 & -188.8 & -210.0 &    0.0 &    0.0 &    0.0 &0.0 \\ 
0.0 &    0.0 &    0.0 &    0.0 &  118.4 &   86.8 &   72.8 &   54.1 &   34.7 &   -7.1 &  -33.8 &  -68.5 & -100.1 & -137.8 & -186.8 & -214.2 &    0.0 &    0.0 &    0.0 &0.0 \\ 
0.0 &    0.0 &    0.0 &    0.0 &   92.3 &   85.6 &   70.3 &   51.5 &   15.6 &  -26.7 &  -44.6 &  -83.9 & -105.8 & -147.3 & -180.7 & -228.5 &    0.0 &    0.0 &    0.0 &0.0 \\ 
0.0 &    0.0 &    0.0 &    0.0 &   94.9 &   83.6 &   70.8 &   54.0 &   23.0 &  -25.3 &  -45.1 &  -84.7 &  -96.9 & -138.3 & -195.3 & -213.8 &    0.0 &    0.0 &    0.0 &0.0 \\ 
0.0 &    0.0 &    0.0 &    0.0 &  113.3 &   96.9 &   79.5 &   57.2 &   29.6 &  -16.2 &  -43.3 &  -72.2 &  -89.9 & -137.4 & -184.4 & -205.7 &    0.0 &    0.0 &    0.0 &0.0 \\ 
0.0 &    0.0 &    0.0 &    0.0 &  146.8 &  119.5 &  103.1 &   62.2 &   31.5 &  -15.5 &  -41.9 &  -75.8 & -113.1 & -149.4 & -211.9 & -224.3 &    0.0 &    0.0 &    0.0 &0.0 \\ 
0.0 &    0.0 &    0.0 &    0.0 &  144.2 &  145.1 &  114.4 &   80.6 &   35.9 &   -2.0 &  -42.7 &  -74.6 & -112.4 & -155.9 & -172.4 & -197.5 &    0.0 &    0.0 &    0.0 &0.0 \\ 
0.0 &    0.0 &    0.0 &    0.0 &    0.0 &  177.1 &  162.1 &  104.6 &   62.2 &   16.0 &  -17.1 &  -72.7 &  -92.2 & -127.7 & -167.3 &    0.0 &    0.0 &    0.0 &    0.0 &0.0 \\ 
0.0 &    0.0 &    0.0 &    0.0 &    0.0 &    0.0 &  152.8 &  143.7 &   92.7 &   26.2 &  -16.0 &  -62.1 &  -93.9 & -135.5 & -186.3 &    0.0 &    0.0 &    0.0 &    0.0 &0.0 \\ 
0.0 &    0.0 &    0.0 &    0.0 &    0.0 &    0.0 &    0.0 &  158.9 &  110.6 &   48.9 &    0.4 &  -35.9 &  -94.4 & -134.5 &    0.0 &    0.0 &    0.0 &    0.0 &    0.0 &0.0 \\ 
0.0 &    0.0 &    0.0 &    0.0 &    0.0 &    0.0 &    0.0 &    0.0 &    0.0 &   93.0 &   36.1 &   -2.8 &    0.0 &    0.0 &    0.0 &    0.0 &    0.0 &    0.0 &    0.0 &0.0 \\ 
0.0 &    0.0 &    0.0 &    0.0 &    0.0 &    0.0 &    0.0 &    0.0 &    0.0 &    0.0 &    0.0 &    0.0 &    0.0 &    0.0 &    0.0 &    0.0 &    0.0 &    0.0 &    0.0 &0.0 \\ 
\hline 
\multicolumn{20}{c}{Velocity error map} \\ 
\hline 
0.0 &    0.0 &    0.0 &    0.0 &    0.0 &    0.0 &    0.0 &    0.0 &    0.0 &    0.0 &    0.0 &    0.0 &    0.0 &    0.0 &    0.0 &    0.0 &    0.0 &    0.0 &    0.0 &0.0 \\ 
0.0 &    0.0 &    0.0 &    0.0 &    0.0 &    0.0 &    0.0 &    0.0 &    0.0 &   15.1 &   14.7 &   16.3 &    0.0 &    0.0 &    0.0 &    0.0 &    0.0 &    0.0 &    0.0 &0.0 \\ 
0.0 &    0.0 &    0.0 &    0.0 &    0.0 &    0.0 &    0.0 &   13.8 &   11.9 &   11.6 &   11.9 &   13.0 &   14.3 &   16.8 &    0.0 &    0.0 &    0.0 &    0.0 &    0.0 &0.0 \\ 
0.0 &    0.0 &    0.0 &    0.0 &    0.0 &    0.0 &   12.5 &   10.9 &   10.1 &   10.4 &   10.8 &   11.5 &   12.0 &   13.8 &   16.9 &    0.0 &    0.0 &    0.0 &    0.0 &0.0 \\ 
0.0 &    0.0 &    0.0 &    0.0 &    0.0 &   12.0 &   11.1 &    9.7 &    9.1 &    9.8 &   10.3 &   10.8 &   11.2 &   12.2 &   14.0 &   18.8 &    0.0 &    0.0 &    0.0 &0.0 \\ 
0.0 &    0.0 &    0.0 &    0.0 &    0.0 &   10.6 &    9.8 &    8.9 &    8.9 &    9.5 &    9.8 &   10.5 &   10.6 &   11.4 &   13.0 &   15.8 &    0.0 &    0.0 &    0.0 &0.0 \\ 
0.0 &    0.0 &    0.0 &    0.0 &   11.6 &    9.9 &    9.4 &    8.7 &    8.8 &    9.6 &    9.8 &   10.2 &   10.3 &   10.9 &   12.2 &   14.9 &    0.0 &    0.0 &    0.0 &0.0 \\ 
0.0 &    0.0 &    0.0 &    0.0 &   11.1 &    9.6 &    9.1 &    8.6 &    8.7 &    9.5 &   10.4 &   10.4 &   10.3 &   10.8 &   11.9 &   14.0 &    0.0 &    0.0 &    0.0 &0.0 \\ 
0.0 &    0.0 &    0.0 &    0.0 &   11.1 &    9.6 &    9.0 &    8.6 &    8.7 &    9.5 &    9.6 &   10.1 &   10.3 &   10.8 &   11.7 &   13.7 &    0.0 &    0.0 &    0.0 &0.0 \\ 
0.0 &    0.0 &    0.0 &    0.0 &   11.0 &    9.7 &    9.0 &    8.6 &    8.6 &    9.4 &    9.8 &   10.3 &   10.2 &   10.7 &   11.5 &   13.7 &    0.0 &    0.0 &    0.0 &0.0 \\ 
0.0 &    0.0 &    0.0 &    0.0 &   11.9 &   10.0 &    9.1 &    8.7 &    8.8 &    9.3 &    9.9 &   10.3 &   10.5 &   11.1 &   11.9 &   14.2 &    0.0 &    0.0 &    0.0 &0.0 \\ 
0.0 &    0.0 &    0.0 &    0.0 &   13.2 &   10.7 &    9.6 &    9.1 &    9.0 &    9.5 &   10.1 &   10.3 &   10.7 &   11.4 &   12.7 &   15.7 &    0.0 &    0.0 &    0.0 &0.0 \\ 
0.0 &    0.0 &    0.0 &    0.0 &    0.0 &   12.2 &   10.8 &    9.9 &    9.5 &    9.9 &   10.4 &   10.8 &   11.4 &   12.3 &   14.2 &    0.0 &    0.0 &    0.0 &    0.0 &0.0 \\ 
0.0 &    0.0 &    0.0 &    0.0 &    0.0 &    0.0 &   12.7 &   11.2 &   10.2 &   10.4 &   10.8 &   11.5 &   12.5 &   13.9 &   16.8 &    0.0 &    0.0 &    0.0 &    0.0 &0.0 \\ 
0.0 &    0.0 &    0.0 &    0.0 &    0.0 &    0.0 &    0.0 &   13.6 &   12.4 &   11.7 &   12.1 &   12.7 &   14.1 &   17.0 &    0.0 &    0.0 &    0.0 &    0.0 &    0.0 &0.0 \\ 
0.0 &    0.0 &    0.0 &    0.0 &    0.0 &    0.0 &    0.0 &    0.0 &    0.0 &   14.3 &   14.3 &   15.4 &    0.0 &    0.0 &    0.0 &    0.0 &    0.0 &    0.0 &    0.0 &0.0 \\ 
0.0 &    0.0 &    0.0 &    0.0 &    0.0 &    0.0 &    0.0 &    0.0 &    0.0 &    0.0 &    0.0 &    0.0 &    0.0 &    0.0 &    0.0 &    0.0 &    0.0 &    0.0 &    0.0 &0.0 \\ 
\hline
\end{tabular}
}
\end{sidewaystable}

\begin{sidewaystable*}
\tiny
\caption{Two-dimensional maps of Venus from data taken on September 16, 2009. Pixel size is 0.82 arcsec in X, and 0.78 arcsec in Y. }\label{sep16_table}
\centering
\scalebox{0.9}{
\begin{tabular}{c c c c c c c c c c c c c c c c c c c c}
\hline 
\multicolumn{20}{c}{Photometric map} \\ 
\hline 
23.0 &   36.0 &   55.8 &   87.0 &  137.2 &  214.6 &  328.3 &  467.5 &  602.3 &  689.6 &  701.4 &  634.9 &  512.6 &  369.3 &  241.5 &  146.1 &   85.9 &   51.6 &   32.0 &19.7 \\ 
33.9 &   53.6 &   85.7 &  138.8 &  231.1 &  379.9 &  595.3 &  830.0 & 1010.1 & 1090.8 & 1065.7 &  949.1 &  770.9 &  569.3 &  379.8 &  228.4 &  130.4 &   76.1 &   45.7 &28.8 \\ 
41.7 &   67.5 &  111.2 &  189.1 &  332.6 &  579.4 &  929.1 & 1251.1 & 1447.0 & 1501.5 & 1433.1 & 1266.4 & 1033.2 &  770.4 &  517.0 &  306.3 &  168.0 &   92.8 &   53.8 &32.2 \\ 
59.2 &   96.9 &  164.8 &  291.6 &  539.5 &  956.3 & 1437.0 & 1776.9 & 1933.6 & 1938.5 & 1818.7 & 1606.0 & 1328.1 & 1017.8 &  712.1 &  436.5 &  239.6 &  127.9 &   70.9 &41.4 \\ 
74.6 &  123.1 &  212.1 &  386.2 &  726.7 & 1245.9 & 1756.1 & 2074.0 & 2196.6 & 2174.7 & 2034.5 & 1802.9 & 1511.3 & 1185.5 &  859.6 &  554.2 &  318.5 &  171.6 &   94.4 &54.7 \\ 
91.4 &  153.1 &  269.0 &  508.7 &  979.5 & 1608.8 & 2129.0 & 2404.4 & 2483.8 & 2426.3 & 2255.1 & 1998.9 & 1684.7 & 1336.0 &  984.9 &  649.3 &  378.3 &  202.1 &  108.3 &61.2 \\ 
104.7 &  176.2 &  318.3 &  624.0 & 1237.7 & 1953.8 & 2438.2 & 2647.9 & 2683.8 & 2591.3 & 2392.0 & 2109.3 & 1777.3 & 1413.0 & 1047.8 &  689.4 &  396.1 &  206.5 &  109.9 &62.5 \\ 
111.8 &  194.0 &  367.7 &  754.6 & 1459.1 & 2196.0 & 2651.0 & 2829.2 & 2835.1 & 2715.6 & 2488.9 & 2189.4 & 1839.9 & 1466.6 & 1087.9 &  716.5 &  411.7 &  207.7 &  107.0 &59.7 \\ 
143.7 &  245.2 &  443.9 &  830.0 & 1448.5 & 2075.5 & 2485.9 & 2663.3 & 2681.0 & 2577.3 & 2372.4 & 2093.4 & 1766.2 & 1416.6 & 1064.3 &  723.2 &  439.8 &  244.7 &  135.5 &79.1 \\ 
132.3 &  224.8 &  406.4 &  758.5 & 1325.5 & 1932.4 & 2365.6 & 2567.1 & 2606.8 & 2518.5 & 2328.8 & 2062.6 & 1747.1 & 1404.5 & 1059.5 &  722.2 &  445.0 &  249.7 &  138.7 &79.7 \\ 
124.6 &  208.1 &  367.9 &  670.9 & 1177.1 & 1754.0 & 2193.3 & 2410.2 & 2465.2 & 2393.2 & 2217.1 & 1964.0 & 1656.7 & 1326.1 &  994.5 &  672.5 &  410.4 &  233.3 &  131.2 &75.8 \\ 
102.3 &  170.3 &  296.5 &  539.2 &  964.9 & 1512.5 & 1974.5 & 2224.1 & 2298.4 & 2241.2 & 2078.2 & 1835.5 & 1538.2 & 1215.7 &  893.3 &  585.5 &  345.3 &  189.1 &  104.5 &60.5 \\ 
89.8 &  144.7 &  240.5 &  415.3 &  724.5 & 1163.3 & 1582.7 & 1847.9 & 1956.1 & 1933.1 & 1802.5 & 1589.0 & 1322.3 & 1030.3 &  742.2 &  477.2 &  281.2 &  158.3 &   91.4 &54.8 \\ 
69.1 &  107.6 &  171.6 &  282.4 &  476.8 &  780.6 & 1134.1 & 1406.7 & 1548.8 & 1565.1 & 1472.3 & 1295.5 & 1064.2 &  807.6 &  560.9 &  348.1 &  201.8 &  115.7 &   68.8 &41.6 \\ 
50.4 &   75.7 &  117.4 &  182.6 &  298.3 &  493.5 &  758.9 & 1009.0 & 1172.0 & 1220.4 & 1159.6 & 1015.3 &  817.1 &  600.0 &  398.9 &  238.2 &  136.7 &   79.9 &   48.3 &29.8 \\ 
35.6 &   53.9 &   81.8 &  121.3 &  184.7 &  285.6 &  430.0 &  592.6 &  729.8 &  799.0 &  783.3 &  693.1 &  552.5 &  398.2 &  260.6 &  156.2 &   91.3 &   54.2 &   33.2 &20.5 \\ 
13.4 &   20.0 &   29.3 &   42.4 &   62.1 &   91.3 &  130.7 &  175.5 &  214.5 &  233.7 &  225.6 &  193.0 &  148.0 &  103.8 &   68.3 &   42.7 &   26.5 &   16.5 &   10.5 &6.6 \\ 
\hline 
\multicolumn{20}{c}{Line-of-sight velocity map} \\ 
\hline 
0.0 &    0.0 &    0.0 &    0.0 &    0.0 &    0.0 &    0.0 &    0.0 &    0.0 &  106.9 &   34.6 &   14.4 &    0.0 &    0.0 &    0.0 &    0.0 &    0.0 &    0.0 &    0.0 &0.0 \\ 
0.0 &    0.0 &    0.0 &    0.0 &    0.0 &    0.0 &    0.0 &   82.6 &   38.2 &    3.8 &  -48.3 &  -70.7 &  -86.3 &  -85.3 &    0.0 &    0.0 &    0.0 &    0.0 &    0.0 &0.0 \\ 
0.0 &    0.0 &    0.0 &    0.0 &    0.0 &    0.0 &  154.5 &  105.3 &   94.5 &   38.2 &   20.7 &  -43.1 &  -80.1 &  -64.1 &    0.0 &    0.0 &    0.0 &    0.0 &    0.0 &0.0 \\ 
0.0 &    0.0 &    0.0 &    0.0 &    0.0 &    0.0 &  145.7 &   70.6 &   53.2 &   32.6 &  -33.5 &  -30.9 &  -49.6 &  -85.8 & -129.6 &    0.0 &    0.0 &    0.0 &    0.0 &0.0 \\ 
0.0 &    0.0 &    0.0 &    0.0 &    0.0 &  166.0 &   83.1 &   75.9 &   51.1 &   14.7 &  -45.8 &  -66.0 &  -77.5 & -117.0 & -116.2 & -142.6 &    0.0 &    0.0 &    0.0 &0.0 \\ 
0.0 &    0.0 &    0.0 &    0.0 &    0.0 &  114.3 &   80.7 &   51.3 &   30.4 &  -16.8 &  -31.8 &  -45.0 &  -87.5 &  -90.7 & -114.1 & -132.3 &    0.0 &    0.0 &    0.0 &0.0 \\ 
0.0 &    0.0 &    0.0 &    0.0 &  135.2 &  140.8 &   68.1 &   37.4 &   17.4 &   27.6 &  -57.4 &  -71.9 &  -89.1 &  -99.9 & -226.4 & -153.7 &    0.0 &    0.0 &    0.0 &0.0 \\ 
0.0 &    0.0 &    0.0 &    0.0 &  132.4 &   91.6 &   73.2 &   28.3 &   -2.0 &  -31.4 &  -48.0 &  -62.3 &  -97.3 & -117.7 & -211.5 & -260.0 &    0.0 &    0.0 &    0.0 &0.0 \\ 
0.0 &    0.0 &    0.0 &    0.0 &   95.6 &   39.4 &   44.1 &   16.6 &    9.2 &  -49.2 &  -69.3 &  -67.2 & -116.0 & -150.4 & -182.3 & -158.2 &    0.0 &    0.0 &    0.0 &0.0 \\ 
0.0 &    0.0 &    0.0 &    0.0 &   76.6 &   94.6 &   52.6 &   26.8 &    3.2 &   -0.3 &  -47.6 &  -91.9 & -118.0 & -160.8 & -198.7 & -115.7 &    0.0 &    0.0 &    0.0 &0.0 \\ 
0.0 &    0.0 &    0.0 &    0.0 &   45.7 &   40.2 &   31.0 &   36.8 &   14.1 &  -26.9 &  -38.3 &  -72.4 &  -74.4 & -124.3 & -160.9 & -188.5 &    0.0 &    0.0 &    0.0 &0.0 \\ 
0.0 &    0.0 &    0.0 &    0.0 &    0.0 &  144.5 &   89.2 &   55.5 &   59.0 &    8.2 &  -31.3 &  -28.3 &  -90.3 & -123.3 & -126.6 & -104.0 &    0.0 &    0.0 &    0.0 &0.0 \\ 
0.0 &    0.0 &    0.0 &    0.0 &    0.0 &  142.7 &   68.3 &   17.6 &   58.1 &   16.8 &  -31.1 &  -51.6 & -105.1 & -166.6 & -170.8 &    0.0 &    0.0 &    0.0 &    0.0 &0.0 \\ 
0.0 &    0.0 &    0.0 &    0.0 &    0.0 &    0.0 &   95.3 &  108.6 &  114.6 &   52.4 &   15.5 &    7.6 &  -19.8 &  -47.5 & -119.1 &    0.0 &    0.0 &    0.0 &    0.0 &0.0 \\ 
0.0 &    0.0 &    0.0 &    0.0 &    0.0 &    0.0 &    0.0 &  109.6 &  113.0 &   65.5 &   -8.7 &  -36.8 &  -32.5 &  -48.9 &    0.0 &    0.0 &    0.0 &    0.0 &    0.0 &0.0 \\ 
0.0 &    0.0 &    0.0 &    0.0 &    0.0 &    0.0 &    0.0 &    0.0 &  141.5 &   93.1 &   29.8 &    4.4 &    7.6 &    0.0 &    0.0 &    0.0 &    0.0 &    0.0 &    0.0 &0.0 \\ 
0.0 &    0.0 &    0.0 &    0.0 &    0.0 &    0.0 &    0.0 &    0.0 &    0.0 &    0.0 &    0.0 &    0.0 &    0.0 &    0.0 &    0.0 &    0.0 &    0.0 &    0.0 &    0.0 &0.0 \\ 
\hline 
\multicolumn{20}{c}{Velocity error map} \\ 
\hline 
0.0 &    0.0 &    0.0 &    0.0 &    0.0 &    0.0 &    0.0 &    0.0 &    0.0 &   27.6 &   28.0 &   30.5 &    0.0 &    0.0 &    0.0 &    0.0 &    0.0 &    0.0 &    0.0 &0.0 \\ 
0.0 &    0.0 &    0.0 &    0.0 &    0.0 &    0.0 &    0.0 &   29.6 &   23.6 &   23.0 &   24.6 &   26.1 &   30.0 &   35.5 &    0.0 &    0.0 &    0.0 &    0.0 &    0.0 &0.0 \\ 
0.0 &    0.0 &    0.0 &    0.0 &    0.0 &    0.0 &   28.6 &   25.9 &   21.8 &   19.4 &   21.7 &   23.3 &   27.9 &   36.0 &    0.0 &    0.0 &    0.0 &    0.0 &    0.0 &0.0 \\ 
0.0 &    0.0 &    0.0 &    0.0 &    0.0 &    0.0 &   24.6 &   22.5 &   20.0 &   19.3 &   22.2 &   22.3 &   23.7 &   28.0 &   32.7 &    0.0 &    0.0 &    0.0 &    0.0 &0.0 \\ 
0.0 &    0.0 &    0.0 &    0.0 &    0.0 &   26.0 &   22.4 &   20.7 &   18.7 &   18.3 &   21.0 &   21.5 &   24.2 &   26.6 &   29.1 &   36.8 &    0.0 &    0.0 &    0.0 &0.0 \\ 
0.0 &    0.0 &    0.0 &    0.0 &    0.0 &   25.1 &   20.4 &   19.8 &   18.2 &   18.2 &   21.7 &   20.6 &   22.6 &   25.7 &   28.0 &   34.8 &    0.0 &    0.0 &    0.0 &0.0 \\ 
0.0 &    0.0 &    0.0 &    0.0 &   26.9 &   23.7 &   19.0 &   19.2 &   18.2 &   18.8 &   20.6 &   19.7 &   21.4 &   25.2 &   25.0 &   34.7 &    0.0 &    0.0 &    0.0 &0.0 \\ 
0.0 &    0.0 &    0.0 &    0.0 &   24.9 &   21.5 &   18.1 &   18.2 &   17.9 &   18.0 &   21.0 &   20.2 &   22.2 &   24.5 &   26.0 &   34.7 &    0.0 &    0.0 &    0.0 &0.0 \\ 
0.0 &    0.0 &    0.0 &    0.0 &   25.2 &   21.1 &   18.9 &   18.2 &   17.6 &   18.6 &   20.6 &   21.1 &   22.2 &   24.8 &   26.2 &   34.8 &    0.0 &    0.0 &    0.0 &0.0 \\ 
0.0 &    0.0 &    0.0 &    0.0 &   25.7 &   22.2 &   19.3 &   19.6 &   18.5 &   18.5 &   21.7 &   20.3 &   22.6 &   24.0 &   27.8 &   35.2 &    0.0 &    0.0 &    0.0 &0.0 \\ 
0.0 &    0.0 &    0.0 &    0.0 &   25.6 &   23.6 &   19.2 &   19.3 &   19.1 &   18.1 &   21.1 &   19.9 &   22.3 &   26.1 &   28.3 &   35.3 &    0.0 &    0.0 &    0.0 &0.0 \\ 
0.0 &    0.0 &    0.0 &    0.0 &    0.0 &   24.7 &   21.7 &   19.9 &   18.8 &   18.1 &   20.5 &   21.2 &   23.3 &   26.5 &   30.2 &   40.7 &    0.0 &    0.0 &    0.0 &0.0 \\ 
0.0 &    0.0 &    0.0 &    0.0 &    0.0 &   27.5 &   23.1 &   22.1 &   19.5 &   19.3 &   23.0 &   22.6 &   24.1 &   27.2 &   30.6 &    0.0 &    0.0 &    0.0 &    0.0 &0.0 \\ 
0.0 &    0.0 &    0.0 &    0.0 &    0.0 &    0.0 &   26.8 &   23.0 &   21.5 &   20.9 &   23.5 &   24.7 &   26.0 &   31.6 &   37.8 &    0.0 &    0.0 &    0.0 &    0.0 &0.0 \\ 
0.0 &    0.0 &    0.0 &    0.0 &    0.0 &    0.0 &    0.0 &   26.8 &   23.7 &   21.9 &   24.7 &   25.9 &   30.3 &   36.3 &    0.0 &    0.0 &    0.0 &    0.0 &    0.0 &0.0 \\ 
0.0 &    0.0 &    0.0 &    0.0 &    0.0 &    0.0 &    0.0 &    0.0 &   29.0 &   26.2 &   28.2 &   30.4 &   34.6 &    0.0 &    0.0 &    0.0 &    0.0 &    0.0 &    0.0 &0.0 \\ 
0.0 &    0.0 &    0.0 &    0.0 &    0.0 &    0.0 &    0.0 &    0.0 &    0.0 &    0.0 &    0.0 &    0.0 &    0.0 &    0.0 &    0.0 &    0.0 &    0.0 &    0.0 &    0.0 &0.0 \\ 
\hline
\end{tabular}
}
\end{sidewaystable*}

\begin{sidewaystable*}
\tiny
\caption{Two-dimensional maps of Venus from data taken on September 17, 2009. Pixel size is 0.84 arcsec in X, and 0.85 arcsec in Y. }\label{sep16_table}
\centering
\scalebox{0.9}{
\begin{tabular}{c c c c c c c c c c c c c c c c c c c c}
\hline 
\multicolumn{20}{c}{Photometric map} \\ 
\hline 
45.7 &   64.8 &   98.7 &  148.9 &  224.4 &  327.6 &  441.3 &  571.8 &  675.2 &  719.2 &  698.6 &  622.2 &  514.2 &  399.4 &  286.8 &  199.6 &  134.3 &   88.8 &   59.1 &41.2 \\ 
60.0 &   89.7 &  138.4 &  210.0 &  310.0 &  446.2 &  654.8 &  881.3 & 1026.2 & 1073.1 & 1024.3 &  924.9 &  783.3 &  614.8 &  437.9 &  286.5 &  178.9 &  117.9 &   77.4 &50.5 \\ 
76.2 &  118.7 &  189.4 &  304.2 &  481.1 &  713.3 & 1017.4 & 1283.1 & 1440.3 & 1475.3 & 1413.9 & 1277.4 & 1071.9 &  837.3 &  599.0 &  393.5 &  248.2 &  158.1 &   97.3 &63.1 \\ 
91.5 &  143.2 &  232.7 &  384.9 &  621.4 &  977.2 & 1356.0 & 1639.1 & 1798.8 & 1821.4 & 1723.7 & 1531.0 & 1285.0 & 1009.0 &  725.1 &  471.9 &  295.4 &  172.5 &  102.4 &63.8 \\ 
117.7 &  192.1 &  323.8 &  545.2 &  866.5 & 1325.5 & 1773.6 & 2054.8 & 2167.1 & 2142.7 & 1999.2 & 1782.6 & 1502.7 & 1195.8 &  870.6 &  574.3 &  351.9 &  198.6 &  113.7 &69.9 \\ 
161.7 &  263.8 &  442.8 &  697.3 & 1038.9 & 1576.2 & 1982.5 & 2211.7 & 2286.2 & 2235.8 & 2077.5 & 1848.6 & 1556.9 & 1246.5 &  941.7 &  648.1 &  407.6 &  237.2 &  140.9 &88.3 \\ 
171.8 &  284.4 &  478.6 &  782.8 & 1239.6 & 1776.3 & 2179.1 & 2414.2 & 2487.4 & 2430.0 & 2267.1 & 2030.8 & 1727.7 & 1397.1 & 1054.8 &  723.8 &  464.4 &  282.0 &  169.6 &103.7 \\ 
169.0 &  284.9 &  491.1 &  837.8 & 1360.1 & 2012.6 & 2474.1 & 2707.7 & 2756.4 & 2663.3 & 2466.8 & 2194.3 & 1864.9 & 1488.5 & 1099.0 &  753.4 &  476.7 &  275.7 &  154.7 &89.8 \\ 
189.9 &  314.6 &  531.9 &  869.0 & 1344.1 & 1871.3 & 2282.8 & 2491.9 & 2538.2 & 2460.2 & 2285.0 & 2044.6 & 1745.4 & 1422.2 & 1092.9 &  770.9 &  488.0 &  283.7 &  166.9 &101.2 \\ 
185.1 &  298.5 &  494.1 &  815.1 & 1248.7 & 1740.6 & 2140.6 & 2345.6 & 2395.6 & 2324.4 & 2158.8 & 1921.0 & 1634.9 & 1323.0 & 1028.5 &  744.8 &  475.5 &  289.3 &  174.7 &107.7 \\ 
156.2 &  254.0 &  421.5 &  697.1 & 1119.9 & 1623.2 & 2120.0 & 2356.2 & 2408.0 & 2329.6 & 2153.0 & 1904.2 & 1601.5 & 1264.2 &  962.9 &  672.8 &  416.1 &  235.1 &  135.6 &81.0 \\ 
141.6 &  221.6 &  356.5 &  582.3 &  901.1 & 1240.0 & 1671.8 & 1932.7 & 2036.8 & 2013.5 & 1884.9 & 1686.6 & 1438.0 & 1157.9 &  872.5 &  596.3 &  388.8 &  240.5 &  147.9 &93.0 \\ 
110.2 &  169.8 &  268.7 &  434.9 &  701.7 & 1031.2 & 1408.4 & 1658.8 & 1783.4 & 1788.2 & 1690.8 & 1521.4 & 1293.2 & 1025.5 &  724.3 &  487.7 &  305.0 &  186.6 &  114.0 &72.6 \\ 
84.4 &  130.3 &  201.6 &  319.1 &  505.2 &  755.4 & 1007.7 & 1278.9 & 1414.0 & 1436.5 & 1359.1 & 1206.9 & 1011.4 &  812.3 &  579.6 &  374.7 &  226.1 &  138.1 &   87.1 &55.8 \\ 
63.5 &   95.3 &  145.1 &  222.4 &  347.2 &  528.6 &  731.6 &  886.0 & 1011.7 & 1056.2 & 1020.1 &  915.3 &  771.8 &  614.5 &  431.7 &  276.4 &  170.5 &  104.7 &   66.8 &43.4 \\ 
27.9 &   42.9 &   66.3 &  103.6 &  163.1 &  256.0 &  383.0 &  532.3 &  664.5 &  724.7 &  721.5 &  661.2 &  538.6 &  393.8 &  261.6 &  163.8 &  100.1 &   62.4 &   40.5 &28.2 \\ 
14.1 &   20.7 &   29.1 &   42.3 &   63.2 &   94.1 &  135.5 &  182.0 &  219.5 &  236.9 &  227.2 &  194.2 &  150.0 &  106.2 &   72.5 &   50.6 &   36.1 &   25.9 &   18.4 &13.5 \\ 
\hline 
\multicolumn{20}{c}{Line-of-sight velocity map} \\ 
\hline 
0.0 &    0.0 &    0.0 &    0.0 &    0.0 &    0.0 &    0.0 &    0.0 &   54.7 &   40.5 &   23.6 &   -8.1 &  -84.7 &    0.0 &    0.0 &    0.0 &    0.0 &    0.0 &    0.0 &0.0 \\ 
0.0 &    0.0 &    0.0 &    0.0 &    0.0 &    0.0 &    0.0 &  110.2 &   56.7 &   36.7 &   16.8 &  -22.9 &  -73.3 & -113.2 &    0.0 &    0.0 &    0.0 &    0.0 &    0.0 &0.0 \\ 
0.0 &    0.0 &    0.0 &    0.0 &    0.0 &    0.0 &   93.8 &   72.7 &   49.5 &   36.3 &    1.3 &  -33.5 &  -88.2 & -136.1 & -185.7 &    0.0 &    0.0 &    0.0 &    0.0 &0.0 \\ 
0.0 &    0.0 &    0.0 &    0.0 &    0.0 &  154.3 &  112.4 &  104.3 &   58.5 &   35.4 &    2.1 &  -48.7 &  -79.0 & -121.0 & -165.2 &    0.0 &    0.0 &    0.0 &    0.0 &0.0 \\ 
0.0 &    0.0 &    0.0 &    0.0 &    0.0 &  136.8 &  109.0 &   77.9 &   50.8 &   26.7 &   -6.5 &  -45.9 &  -90.6 & -136.4 & -151.0 & -171.9 &    0.0 &    0.0 &    0.0 &0.0 \\ 
0.0 &    0.0 &    0.0 &    0.0 &    0.0 &  114.9 &   88.7 &   62.7 &   35.8 &   11.7 &  -11.7 &  -48.0 &  -89.3 & -139.6 & -158.2 & -175.6 &    0.0 &    0.0 &    0.0 &0.0 \\ 
0.0 &    0.0 &    0.0 &    0.0 &  141.6 &  100.6 &   79.0 &   56.4 &   25.4 &    4.1 &  -21.5 &  -57.2 &  -92.6 & -144.5 & -174.7 & -185.2 &    0.0 &    0.0 &    0.0 &0.0 \\ 
0.0 &    0.0 &    0.0 &    0.0 &  120.3 &  103.1 &   80.3 &   53.9 &   31.0 &    9.5 &  -22.5 &  -59.6 &  -83.2 & -137.2 & -162.2 & -166.2 &    0.0 &    0.0 &    0.0 &0.0 \\ 
0.0 &    0.0 &    0.0 &    0.0 &  120.1 &   94.6 &   62.2 &   40.5 &   23.1 &   -5.0 &  -42.0 &  -79.7 & -109.3 & -153.0 & -178.1 & -203.7 &    0.0 &    0.0 &    0.0 &0.0 \\ 
0.0 &    0.0 &    0.0 &    0.0 &  136.3 &  102.0 &   84.9 &   59.1 &   44.7 &    5.4 &  -23.3 &  -52.1 &  -92.1 & -138.2 & -178.6 & -199.7 &    0.0 &    0.0 &    0.0 &0.0 \\ 
0.0 &    0.0 &    0.0 &    0.0 &  154.7 &  113.1 &   94.7 &   83.8 &   54.4 &   22.6 &   -0.6 &  -37.9 &  -76.8 & -118.0 & -177.4 & -173.0 &    0.0 &    0.0 &    0.0 &0.0 \\ 
0.0 &    0.0 &    0.0 &    0.0 &    0.0 &   90.1 &  105.0 &   80.0 &   54.3 &   26.8 &   13.0 &  -26.2 &  -65.3 & -120.7 & -172.8 & -208.8 &    0.0 &    0.0 &    0.0 &0.0 \\ 
0.0 &    0.0 &    0.0 &    0.0 &    0.0 &  121.6 &  130.8 &  102.9 &   80.4 &   69.8 &   30.9 &  -14.9 &  -67.0 & -117.0 & -165.7 &    0.0 &    0.0 &    0.0 &    0.0 &0.0 \\ 
0.0 &    0.0 &    0.0 &    0.0 &    0.0 &    0.0 &  116.7 &  124.1 &   85.6 &   62.4 &   35.8 &   -5.9 &  -67.8 &  -89.3 & -145.4 &    0.0 &    0.0 &    0.0 &    0.0 &0.0 \\ 
0.0 &    0.0 &    0.0 &    0.0 &    0.0 &    0.0 &    0.0 &   81.9 &   73.0 &   58.3 &   33.8 &  -11.4 &  -73.8 & -150.7 &    0.0 &    0.0 &    0.0 &    0.0 &    0.0 &0.0 \\ 
0.0 &    0.0 &    0.0 &    0.0 &    0.0 &    0.0 &    0.0 &    0.0 &   -6.7 &   -2.1 &  -36.6 &  -74.3 &  -88.5 &    0.0 &    0.0 &    0.0 &    0.0 &    0.0 &    0.0 &0.0 \\ 
0.0 &    0.0 &    0.0 &    0.0 &    0.0 &    0.0 &    0.0 &    0.0 &    0.0 &    0.0 &    0.0 &    0.0 &    0.0 &    0.0 &    0.0 &    0.0 &    0.0 &    0.0 &    0.0 &0.0 \\ 
\hline 
\multicolumn{20}{c}{Velocity error map} \\ 
\hline 
0.0 &    0.0 &    0.0 &    0.0 &    0.0 &    0.0 &    0.0 &    0.0 &   16.3 &   15.5 &   15.6 &   16.4 &   19.2 &    0.0 &    0.0 &    0.0 &    0.0 &    0.0 &    0.0 &0.0 \\ 
0.0 &    0.0 &    0.0 &    0.0 &    0.0 &    0.0 &    0.0 &   16.6 &   14.4 &   13.6 &   13.9 &   14.8 &   17.3 &   19.1 &    0.0 &    0.0 &    0.0 &    0.0 &    0.0 &0.0 \\ 
0.0 &    0.0 &    0.0 &    0.0 &    0.0 &    0.0 &   15.3 &   14.0 &   12.7 &   12.3 &   12.6 &   13.4 &   15.1 &   16.4 &   20.2 &    0.0 &    0.0 &    0.0 &    0.0 &0.0 \\ 
0.0 &    0.0 &    0.0 &    0.0 &    0.0 &   16.2 &   13.9 &   13.1 &   11.8 &   11.8 &   12.1 &   13.0 &   14.3 &   15.5 &   18.2 &    0.0 &    0.0 &    0.0 &    0.0 &0.0 \\ 
0.0 &    0.0 &    0.0 &    0.0 &    0.0 &   14.6 &   12.7 &   11.9 &   11.3 &   11.3 &   11.7 &   12.5 &   13.4 &   14.7 &   16.5 &   20.2 &    0.0 &    0.0 &    0.0 &0.0 \\ 
0.0 &    0.0 &    0.0 &    0.0 &    0.0 &   13.9 &   12.5 &   11.7 &   11.3 &   11.3 &   11.6 &   12.3 &   13.3 &   14.4 &   15.8 &   18.8 &    0.0 &    0.0 &    0.0 &0.0 \\ 
0.0 &    0.0 &    0.0 &    0.0 &   15.1 &   13.3 &   12.1 &   11.4 &   11.1 &   11.1 &   11.4 &   12.0 &   12.9 &   13.8 &   15.3 &   17.9 &    0.0 &    0.0 &    0.0 &0.0 \\ 
0.0 &    0.0 &    0.0 &    0.0 &   14.3 &   13.1 &   11.8 &   11.2 &   10.9 &   11.0 &   11.3 &   11.9 &   12.8 &   13.6 &   15.1 &   17.9 &    0.0 &    0.0 &    0.0 &0.0 \\ 
0.0 &    0.0 &    0.0 &    0.0 &   14.6 &   13.1 &   11.7 &   11.2 &   11.0 &   11.1 &   11.4 &   11.9 &   12.8 &   13.7 &   15.5 &   17.9 &    0.0 &    0.0 &    0.0 &0.0 \\ 
0.0 &    0.0 &    0.0 &    0.0 &   14.9 &   13.3 &   11.8 &   11.3 &   11.1 &   11.0 &   11.4 &   12.1 &   13.1 &   13.7 &   15.4 &   18.2 &    0.0 &    0.0 &    0.0 &0.0 \\ 
0.0 &    0.0 &    0.0 &    0.0 &   15.5 &   13.4 &   12.3 &   11.5 &   11.0 &   11.1 &   11.5 &   12.1 &   13.2 &   13.9 &   16.3 &   18.6 &    0.0 &    0.0 &    0.0 &0.0 \\ 
0.0 &    0.0 &    0.0 &    0.0 &    0.0 &   14.1 &   13.0 &   11.8 &   11.5 &   11.5 &   11.9 &   12.2 &   13.5 &   14.4 &   17.3 &   19.4 &    0.0 &    0.0 &    0.0 &0.0 \\ 
0.0 &    0.0 &    0.0 &    0.0 &    0.0 &   15.2 &   13.8 &   12.4 &   11.8 &   11.5 &   12.0 &   12.8 &   14.2 &   14.9 &   17.5 &    0.0 &    0.0 &    0.0 &    0.0 &0.0 \\ 
0.0 &    0.0 &    0.0 &    0.0 &    0.0 &    0.0 &   15.3 &   13.4 &   12.7 &   12.1 &   12.7 &   13.4 &   15.0 &   16.0 &   19.6 &    0.0 &    0.0 &    0.0 &    0.0 &0.0 \\ 
0.0 &    0.0 &    0.0 &    0.0 &    0.0 &    0.0 &    0.0 &   15.4 &   14.5 &   13.8 &   14.1 &   15.2 &   17.3 &   19.2 &    0.0 &    0.0 &    0.0 &    0.0 &    0.0 &0.0 \\ 
0.0 &    0.0 &    0.0 &    0.0 &    0.0 &    0.0 &    0.0 &    0.0 &   15.1 &   14.1 &   14.2 &   15.5 &   15.8 &    0.0 &    0.0 &    0.0 &    0.0 &    0.0 &    0.0 &0.0 \\ 
0.0 &    0.0 &    0.0 &    0.0 &    0.0 &    0.0 &    0.0 &    0.0 &    0.0 &    0.0 &    0.0 &    0.0 &    0.0 &    0.0 &    0.0 &    0.0 &    0.0 &    0.0 &    0.0 &0.0 \\ 
\hline
\end{tabular}
}
\end{sidewaystable*}

\end{appendix}

\end{document}